\documentstyle[12pt,epsf,twoside]{article}
\def\journal#1, #2, #3, #4 { {\sl #1~}{\bf #2~} (#3) #4 }

\def\cmp{\journal Comm. Math. Phys., }

\def\np{\journal Nucl. Phys., }

\def\pl{\journal Phys. Lett., }

\catcode`\@=11
\def\marginnote#1{}
\newcount\hour
\newcount\minute
\newtoks\amorpm
\hour=\time\divide\hour by60
\minute=\time{\multiply\hour by60 \global\advance\minute
by-\hour}\edef\standardtime{{\ifnum\hour<12
\global\amorpm={am}%
        \else\global\amorpm={pm}\advance\hour by-12 \fi
        \ifnum\hour=0 \hour=12 \fi
        \number\hour:\ifnum\minute<10
0\fi\number\minute\the\amorpm}}
\edef\militarytime{\number\hour:\ifnum\minute<10
0\fi\number\minute}

\def\draftlabel#1{{\@bsphack\if@filesw {\let\thepage\relax
   \xdef\@gtempa{\write\@auxout{\string
      \newlabel{#1}{{\@currentlabel}{\thepage}}}}}\@gtempa
   \if@nobreak \ifvmode\nobreak\fi\fi\fi\@esphack}
        \gdef\@eqnlabel{#1}}
\def\@eqnlabel{}
\def\@vacuum{}
\def\draftmarginnote#1{\marginpar{\raggedright\scriptsize\tt#1}}
\def\draft{\oddsidemargin -.5truein
        \def\@oddfoot{\sl preliminary draft \hfil
        \rm\thepage\hfil\sl\today\quad\militarytime}
        \let\@evenfoot\@oddfoot \overfullrule 3pt
        \let\label=\draftlabel
        \let\marginnote=\draftmarginnote

\def\@eqnnum{(\theequation)\rlap{\kern\marginparsep\tt\@eqnlabel}%
\global\let\@eqnlabel\@vacuum}  }

\def\numberbysection{\@addtoreset{equation}{section}
        \def\theequation{\thesection.\arabic{equation}}}

\def\underline#1{\relax\ifmmode\@@underline#1\else
 $\@@underline{\hbox{#1}}$\relax\fi}

\catcode`@=12
\relax

\numberbysection
\def\beq{\begin{equation}}
\def\eeq{\end{equation}}
\def\beqa{\begin{eqnarray}}
\def\eeqa{\end{eqnarray}}
 \def\nnn{\nonumber \\}
\def\sqr#1#2{{\vcenter{\vbox{\hrule height.#2pt
\hbox{\vrule width.#2pt height#1pt \kern#1pt
\vrule width.#2pt}
\hrule height.#2pt}}}}

\def\hhat{{\widehat h}}

\def\Jhat{{\widehat J}}

\def\Chat{{\widehat C}}

\def\Dhat{{\widehat D}}

\def\mhat{{\widehat m}}
\def\nhat{{\widehat n}}

\def\Vhat{{\widehat V}}

\def\psihat{{\widehat \psi}}

\def\Mhat{{\widehat M}}

\def\Vhat{{\widehat V}}

\def\muhat{{\widehat \mu}}
\def\nuhat{{\widehat \nu}}

\def\rhat{{\widehat r}}
\def\shat{{\widehat s}}
\def\that{{\widehat t}}

\def\varpihat{{\widehat \varpi}}

\def\mhat{{\widehat m}}
\def\nhat{{\widehat n}}

\def\square{\mathchoice\sqr34\sqr34\sqr{2.1}3\sqr{1.5}3}
\def\qed{ $ \quad\square $ }

\begin{document}
\begin{titlepage}

\nopagebreak
\begin{flushright}
 
LPTENS--93/02, \\
hep-th/9302035,  \\
   February   1993
\end{flushright}
 
\vglue 1.5  true cm
\begin{center}
{\large \bf 
THE QUANTUM GROUP STRUCTURE OF \\
\medskip
 2D GRAVITY AND MINIMAL MODELS II:\\ 
\medskip
THE  GENUS-ZERO  CHIRAL  BOOTSTRAP }
\vglue 1.5 true cm
{\bf Eug\`ene CREMMER}\\
\medskip
{\bf Jean-Loup~GERVAIS}\\
\medskip
{\bf Jean-Fran\c cois ROUSSEL}\\
\medskip
{\footnotesize Laboratoire de Physique Th\'eorique de
l'\'Ecole Normale Sup\'erieure\footnote{Unit\'e Propre du
Centre National de la Recherche Scientifique,
associ\'ee \`a l'\'Ecole Normale Sup\'erieure et \`a
l'Universit\'e
de Paris-Sud.},\\
24 rue Lhomond, 75231 Paris CEDEX 05, ~France}.
\end{center}
\vfill
\begin{abstract}
\baselineskip .4 true cm
\noindent
{\footnotesize 
The F  and B  matrices  associated with
Virasoro null vectors are derived in closed form
  by making use of
the operator-approach suggested by the Liouville theory, where
the quantum-group symmetry is explicit.
It is found  that the entries of the  fusing and braiding
matrices are not simply equal to quantum-group symbols, but
involve additional coupling constants whose derivation
is one aim of the present work. Our explicit  formulae are new,
to our knowledge, in spite of the numerous studies of
this problem.
 The relationship between the
quantum-group-invariant (of IRF type) and quantum-group-covariant
(of vertex type) chiral operator-algebras
is fully clarified,
and  connected
with the transition  to the shadow world for quantum-group
symbols. The corresponding 3-j-symbol dressing
is shown to reduce to  the simpler
transformation of Babelon and one of the author (J.-L. G.)
in a suitable infinite limit defined by analytic continuation. 
The above two types  of   operators are found  to coincide when applied to
states with Liouville momenta going to $\infty$ in 
 a suitable way.     
The introduction of  quantum-group-covariant  
operators in the three dimensional picture   
gives a generalisation of the  quantum-group version of discrete  
three-dimensional gravity that includes tetrahedra associated with  
3-j symbols and universal R-matrix
elements. Altogether the present work gives the concrete 
realization of Moore and Seiberg's scheme that describes the
chiral operator-algebra of two-dimensional gravity and minimal
models. }
\end{abstract}
\vfill
\end{titlepage}
\section {INTRODUCTION} 
\markboth{ 1. Introduction }{ 1. Introduction }

Clearly, the chiral operator-algebras  of two dimensional 
(2D) gravity,
and minimal models have  already received much attention.
It may thus be surprising that the corresponding chiral 
operator  algebra 
is not yet fully explored. 
There are  several reasons  for this situation. 

In recent years the emphasis has been  put on what is  
considered as  
particularly simple and elegant in conformal theories. On the
one hand, Verlinde's beautiful ideas\cite{V} led to extensive
studies of the fusion rules where one only considers  the 
number of independent couplings --multiplicities-- 
$N_{ij}^k$ which are
non-negative integers.  One  aim here is, on the contrary,
 to study the fusion 
algebra, that is to determine the  coefficients  of  
Wilson's operator-product-expansion 
OPE completely (since we deal with irrational theories, the 
$N_{ij}^k$ are trivially known)\footnote{Warning! throughout the
article the word fusion is related with full-fledged 
 OPE, not with Verlinde algebras}. On the other hand, after the
topological nature of the gravity-dressed minimal models 
was recognized,   and since
the recent developments are connected with matrix models, 
most recent articles deal with topological conformal theories,
where the chiral OPE we want to study becomes somewhat hidden.

Most  previous studies of the operator-product algebra
(OPA) focused on the braiding matrices\cite{TK,K,F,FFK,FK,R,RS}, 
a few years  after 
ref.\cite{GN4} first determined the simplest ones. Fusion 
 was
only used to leading order  as a tool to 
deduce the higher braiding matrices from the simplest ones, 
recursively.   

One main  result of  our study is the explicit correspondence
between  q-6j symbols and the fusing (F)  and braiding (B) 
matrices of the basic set of chiral vertex operators $V_m^{(J)}$
 of the 
type $(1,2J+1)$ in the BPZ classification, with $2J$ a positive
integer. In  the notation of Moore and Seiberg (MS)\cite{MS}, 
we will show
that 
\beq
F_{{J_{23}},{J_{12}}}\!\!\left[^{J_1}_{J}
\,^{J_2}_{J_3}\right]
=
{g_{J_1J_2}^{J_{12}}\
g_{J_{12}J_3}^{J}
\over
g _{J_2J_3}^{J_{23}}\
g_{{J_1}J_{23}}^{J}
}
\left\{ ^{J_1}_{J_3}\,^{J_2}_{J}
\right. \left |^{J_{12}}_{J_{23}}\right\}.
\label{1.1}
\eeq
The F and B matrices are found to be connected by the MS 
relation\cite{MS} (the $\Delta$'s are conformal weights) 
\beq
B^{\pm}_{{J_{23}},{J_{12}}}\!\!\left[^{J_1}_{J} 
\,^{J_3}_{J_2}\right] 
=e^{\pm i \pi \left( \Delta_{J}+\Delta_{J_2}-\Delta_{J_{23}}
-\Delta_{J_{12}}\right )}
F_{{J_{23}},{J_{12}}}\!\!\left[^{J_1}_{J} 
\,^{J_2}_{J_3}\right].  
\label{1.2}
\eeq
 In the first equation, 
$\left\{ ^{J_1}_{J_3}\,^{J_2}_{J} 
\right. \left |^{J_{12}}_{J_{23}}\right\}$ is the q-6j symbol
associated with $U_q(sl(2))$, and we have introduced
quantities  noted 
 $g_{JK}^{L}$,  
whose    explicit
expression will be  given in section 2. 
The V operators are closely related with  
operators called  IRF-chiral vertex operator   
in ref.\cite{MR}, which   are associated with   
integrable models with   
solid-on-solid interactions around-the-face\cite{P}.    
In the context of Liouville theory, they were called\cite{B}  
 Bloch-wave 
operators since they diagonalise the monodromy matrix, and 
we shall use this name. 
The above  formula is  new to our knowledge, although  many 
other studies\cite{K,FFK,FK,MR,AGS,W,T,FGP,Ga}  
gave strong hints that indeed the fusion and braiding
matrices of the  chiral vertex operators 
 should be ``essentially given'' by q-6j symbols. 
 The 
basic difficulty in trying to relate OPA coefficients with 
quantum-group symbols 
is  that the latter  are trigonometric 
functions of the quantum-group parameters, while the 
 correlators  of Bloch-wave operators 
involve ordinary $\Gamma$ functions. Thus  equality is
impossible, although, both satisfy the same MS polynomial 
equations. The form of the above formulae is precisely 
such that this crucial property holds. An  important 
 remark must be made 
at this point. We have derived 	Eq.\ref{1.1}  with a  specific 
normalization of the Bloch-wave  operators, that is,  
that their matrix elements between highest-weight states 
are  equal to one. Since it leaves invariant the polynomial 
equations, the multiplicative factor 
$g_{J_1J_2}^{J_{12}}\
g_{J_{12}J_3}^{J}
\Bigl /
g _{J_2J_3}^{J_{23}}\
g_{{J_1}J_{23}}^{J}
$ 
is a gauge 
transformation,  from the three dimensional viewpoint, 
which may be re-absorbed by a change of this  normalization. 
However this does not eliminate the $g's$ from the 
OPE, as one may easily see, for instance at the level of 
primaries.  Indeed,  
 the fusing-matrix entries  are equal to  the coefficients of the 
OPA for primaries only if  the vertex operators are normalized to 
one  
 (this will be spelled out later on). Accordingly,    
 the difference between our result and 6-j symbols 
is genuine. Returning to our main line, let us first sketch 
how the operator algebra of the $V$ fields is determined in
section 2.    
 At first  
$F_{{J_{23}},{J_{12}}}\!\!\left[^{1/2}_{J} 
\,^{J_2}_{J_3}\right]$ is re-derived from the hypergeometric
differential equation due to the decoupling of the Virasoro 
null vector at the second level. Next,  the
result is put under the form Eq.\ref{1.1}. Remarkably, this 
determines the coupling constants $g_{JK}^{L}$ uniquely, 
assuming, as is always possible that $g_{JK}^{J+K}=1$.  Then 
the complete operator-algebra is established applying 
 Moore and Seiberg's 
general scheme. 
 
Another difficulty of the quantum-group picture 
is that formulae like Eqs.\ref{1.1}, \ref{1.2} only involve
quantum-group invariants, so that one does not see how the 
quantum group acts. Previously, this problem has been  
 solved  in two seemingly different ways. 
First the invariant operators were  ``dressed  with  3-j symbols'' 
\cite{P,MR}. In this approach, 
 the quantum group covariance seems somewhat 
artificial and redundant, even though this  construction 
allows\cite{P} to relate integrable models. of the IRF
(interaction around-the-face) and vertex types.  
Another method\cite{G1,B,B2,G3} was  directly inspired by  the 
operator approach\cite{GN3,GN4,GN5,GN7} to
Liouville theory which is explicitly quantum-group symmetric.
A set of chiral primary fields noted $\xi_M^{(J)}$ 
 was  constructed\footnote{There exists a 
 related derivation  of the
universal R matrix 
 in the Coulomb gas picture\cite{GoS}. However it seems not to
be so well suited for obtaining the complete fusion algebra.}. 
They are deduced from the Bloch-wave operators by equations 
of the form\cite{B,G1}
\beq
\xi_M^{(J)} := \sum_{-J\leq m \leq J}\vert
J,\varpi)_M^m \>  E_m^{(J)}(\varpi) 
\> V_m^{(J)},  \quad -J\leq M\leq J, 
\label{1.3}
\eeq
where $|J,\varpi)_M^m$ are q-hypergeometrix functions of
$e^{ih(\varpi+m)}$, $E_m^{(J)}(\varpi)$ are normalization 
factors, and $\varpi$ is the zero-mode. 
The braiding matrices of the $\xi$'s 
 coincide\cite{B,G1} with the universal 
R-matrix of
$U_q(sl(2))$. This construction is more econonomical than 
the dressing by 3-j symbols, since it does not involve any  redundant
quantum number. Thus we call it the 
{\it intrinsic transformation}. 
 The leading-order fusion coefficients of the $\xi$ fields  were
shown\cite{G1}  to coincide with the quantum 3-$j$ symbols, 
and it was stated\cite{G3} without proof that this is also true, 
up to
a coupling constant, for every order. 
It is another  
purpose of the present work to complete that picture. 	
At first, using the above  relationship between  
$\xi$ fields and  $V$ fields, 
  we
deduce, in section 3, that the fusion of the former are given
by 
$$
\xi ^{(J_1)}_{M_1}\,\xi^{(J_2)}_{M_2} =
 \sum _{J_{12}= \vert J_1 - J_2 \vert} ^{J_1+J_2}
g _{J_1J_2}^{J_{12}} (J_1,M_1;J_2,M_2\vert J_{12})\times 
$$
\beq
\sum _{\{\nu\}} \xi ^{(J_{12},\{\nu\})} _{M_1+M_2}
<\!\varpi _{J_{12}},\{\nu\} \vert V ^{(J_1)}_{J_2-J_{12}}
\vert \varpi_{J_2}\! >, 
\label{1.4}
\eeq
where $(J_1,M_1;J_2,M_2\vert J_{12})$ are the q-Clebsch-Gordan
coefficients. Apart from the fact that the right-hand side
involves one $\xi$ field and one $V$ field, this  
has the standard MS form\footnote{It already appears, without
the $g$ factor in ref.\cite{MR}.} 
(worldsheet variables are suppressed,  
the states  $|\varpi _J,\{\nu\}>$ span the 
Verma module with highest weight state $|\varpi _J>$).  
 It realizes our
expectation that there must be fields such that the fusion 
corresponds  to making q-tensor-products of representations. 
However there is  the additional factor  $g _{J_1J_2}^{J_{12}}$. 
Since this form  may be 
 considered as governed by the q-deformed Wigner Eckart theorem, 
the $g$'s will be called coupling constants.  It is thus clear
that the $\xi$ fields are quantum-group covariant. 
Next we use the
fact that the   right-hand side
involves one $\xi$ field and one $V$ field to relate
them operatorially.
This is found equivalent to 
 the dressing by 3-j symbols of
refs.\cite{P,MR}. Since the $\xi$ fields were defined by the 
intrinsic transformation\cite{B,G1} (see Eq.\ref{1.3}), 
 we are able to establish the relationship between the two 
viewpoints: the dressing by 3-j symbols has an additional 
magnetic quantum number, and is shown to reduce to the 
intrinsic transformation, when this number   tends to
$\infty$ after a suitable analytic continuation. Finally, we 
terminate section 3, by giving a general formula connecting the
coupling constants $g_{JK}^L$ with quantities introduced
earlier\cite{G1,G3} in the quantum-group context. 

In section 4, we further  develop the idea of understanding
the connection  between $V$  and $\xi$ fields from infinite
limits. Inspired by a recent article of  Witten\cite{W} we show
that the $V$ and $\xi$ fields coincide in the limit where the 
zero-mode $\varpi$ goes to $\infty$, after suitable analytic
continuation. This is explicitly proven
from the intrinsic transformation summarized above
(Eq.\ref{1.3}). The expressions  of the fusing and braiding
matrices of $V$ and $\xi$ fields in terms of q symbols then 
show that these symbols 
 should be related by the same limit, and this 
is explicitly verified. This sheds  light on the 
method followed in ref\cite{W} to construct covariant vertex
operators, although the conformal theory 
considered is different.  Another point of this section is to
show that the intrinsic transformation is the 
conformal analogue of the transition to the shadow
world described in ref.\cite{KR} for quantum group symbols, and
to display  its three-dimensional aspect where 3-j symbols and
R-matrix are 
represented by tetrahedra, at the same time as 6-j symbols.

Finally, in section 5 we take account of the fact that, for 
irrational theories,  there
are actually two quantum-group parameters, one for each 
screening charge. The above picture should  thus be extended. To avoid
lengthy discussion, we take a short cut and only  determine the 
coupling constants, for which a general formula is 
written down. The underlying  quantum group
structure was already unravelled before in refs.\cite{B,G3,CG}.

\section{THE  BRAIDING AND FUSING MATRICES} 
\markboth{2. Braiding and fusing matrices}
{2. Braiding and fusing matrices }

The present operator algebra may, in principle be considered on any
given Riemann surface. In a typical situation, one may  work on the
cylinder $0\leq
\sigma \leq 2\pi$,  $-\infty \leq \tau \leq \infty $ obtained by
an
appropriate mapping from one of the handles. However, the
present work deals with the case of genus zero only. 
In this discussion we 
concentrate on the holomorphic fields which are
functions of $x=\sigma-i\tau$ only.
The notations are exactly the same as in refs.\cite{G1,G3,G5}. The
basic concepts are recalled in appendix A for completeness. 
The starting point is the 
 differential equation\cite{GN3,G1},  satisfied by the two
spin $1/2$ fields $V_{\pm1/2}(x)$, which is 
 recalled in appendix A  (Eq.A.8). In the BPZ framework it
expresses the vanishing of the Virasoro null fields  at the
second level for  operators  of the type $(1,2)$ 
(Untill section 5 we concentrate on operators of the $(1,2J+1)$
type, that is on the family with  
one of the two quantum group parameters).   
It then follows that for any holomorphic primary operator 
$A_\Delta(x)$
with conformal weight $\Delta$, one has (see Eq.A.6 of
ref\cite{G1}):
$$
< \varpi_4 \vert \, V_{\pm 1/2}(x)\, A_\Delta(x')\,
 \vert \varpi_1> =
e^{ix'(\varpi_1^2-\varpi_4^2)h/4\pi}e^{i(x'-x)
(-1/2\mp\varpi_4)h/2\pi }$$
\beq
\times \bigl(1-e^{-i(x-x')}\bigr)^\beta \,
 F(a_\pm ,b_\pm ;c_\pm ;e^{i(x'-x)})
\label{2.1}
\eeq
$$a_\pm =\beta -{h\over 2\pi}\mp 
h({\varpi_4-\varpi_1 \over 2\pi});\quad
b_\pm =\beta  -{h\over 2\pi}\mp h({\varpi_4+\varpi_1 \over
2\pi});\quad c_\pm =1\mp{h\varpi_4 \over \pi};$$
\beq
\beta ={1\over 2} (1+h/\pi)\bigl(1-
\sqrt{1-{8h\Delta  \over 2\pi(1+h/\pi)^2}}\bigr);  
\label{2.2}
\eeq
where $F(a,b;c;z)$ is the standard hypergeometric function. 
Similarly, one verifies that 
$$
< \varpi_4 \vert \, A_\Delta(x')\,V_{\pm 1/2}(x)\, 
 \vert \varpi_1> =
e^{ix'(\varpi_1^2-\varpi_4^2)h/4\pi}e ^{i(x-x')
(-1/2\pm \varpi_1)h/2\pi }\times
$$
\beq
\bigl(1-e^{-i(x-x')}\bigr)^\beta \,
 F(a'_\pm ,b'_\pm ;c'_\pm;\,  e^{-i(x-x')})
\label{2.3}
\eeq
\beq
a'_\pm =\beta -{h\over 2\pi}\mp h({\varpi_1-\varpi_4 \over
2\pi});\quad
b'_\pm =\beta  -{h\over 2\pi}\mp h({\varpi_1+\varpi_4 \over
2\pi});\quad c'_\pm =1\mp{h\varpi_1 \over \pi}. 
\label{2.4}
\eeq
As already mentioned in the introduction, $\varpi$ denotes the
zero mode, and $|\varpi >$ is the corresponding highest-weight
state. The two fields $V_{\pm 1/2}$ have the same conformal
weight, but shift $\varpi$ in opposite directions.   
These are the basic equations for establishing  the 
operator algebra. They were already  used\cite{G1} to determine the
operator-product algebra (OPA) to leading order in the singularity.
This showed that there are operators $V_m^{(J)}$, with  $2J$ 
a positive
integer, and $-J\leq m \leq J$, 
(also denoted $V^{\mu,\, \nu}$, with
$2J=\mu+\nu$, and $2m=-\mu+\nu$)  and conformal weights 
\beq 
\Delta_J=-hJ(J+1)/\pi -J. 
\label{2.5}
\eeq
  The above formulae were also  used in order  
to determine the braiding properties of the $V$ operators.    
We shall only  deal with the case of genus zero, so
that 
 it is appropriate, following ref.\cite{G5} to change the
normalization of the operator.  In this connection, let
us recall that a  primary field $A(z, z^*)$ of weight
$\Delta$, $\overline \Delta$  transforms so that
$A(z, z^*) (dz)^\Delta (dz^*)^{\overline \Delta}$ is
invariant\cite{GS}. Given $A(z,z^*)$,  and a conformal
map $z\to Z(z)$, it is thus convenient to define
\begin{equation}
 A (Z,Z^*) =  
\Bigl ({dz\over dZ}\Bigr )^\Delta
\Bigl ({dz^*\over dZ^*}\Bigr ) ^{\overline \Delta}\> 
U_{\{Z\}}A(z,z^*)  U_{\{Z\}}^{-1},  
\label{2.6}
\eeq
where $U_{\{Z\}}$ is the operator that realizes  the transformation of
the states on which $A$ acts. It  is a functional of $Z(z)$. 
We are using coordinates such that the Riemann  sphere  is
described by the complex variable
 $z=\exp (i x)$. If we denote by $A(\sigma, \tau)$ a general 
primary field on the cylinder, and by $A(z,z^*)$ its transformed 
on the sphere, one has 
\begin{equation}
A(z,z^*)= z^{-\Delta} z^{*\> - {\overline
\Delta}} \> U_{\{z\}}A(\sigma, \tau)U_{\{z\}}^{-1}.
\label{2.7}
\end{equation}
 $A(\sigma,\tau)$, which was used before ref\cite{G5}, 
is such that  its  Fourier expansion in  $\sigma$ has
integer coefficients. The transformation   rule
just recalled gives  the standard
definition for the Laurent expansion of $A(z,z^*)$ as  series in
$z^{n-\Delta}$, and $(z^*)^{m-\overline \Delta}$,  $n$, $m$
integers\footnote{The main practical 
advantage of using $z$, $z^*$ is that 
 $A(z,z^*)$  has a simple behaviour  under 
M\"obius transformations}. 
Using the formulae just recalled, one may see that 
$$<\varpi_4 | V_{\pm 1/2}^{(1/2)}(z_3) V_{m_2}^{(J_2)}(z_2) 
|\varpi_1> =z_2^{-\Delta_{J_2}-\Delta(\varpi_1)} \> 
z_3^{-\Delta_{1/2}+\Delta(\varpi_4)} 
\times 
$$
\beq 
 \left ( {z_2 \over
z_3}\right )^{\Delta(\varpi_4\pm 1) }  \left ( 1-{z_2 \over
z_3}\right )^{-hJ_2 /\pi} F(a_\pm ,b_\pm ;c_\pm ; {z_2\over z_3});
\label{2.8}
\eeq
$$a_\pm ={h\over 2\pi}\left [-1-2J_2  \mp (\varpi_4-\varpi_1 )\right
]
;\quad
b_\pm = {h\over 2\pi}\left [-1-2J_2  \mp (\varpi_4+\varpi_1 ) 
\right ]
;
$$
\beq
 c_\pm =1\mp{h\varpi_4 \over \pi}. 
\label{2.9}
\eeq
The symbols $|\varpi >$ represent highest-weight states, with 
Virasoro weights $\Delta(\varpi)$ given by Eq.A.11. 
According to Eqs.A.12, $V_m^{(J)}$ shifts $\varpi$ by $2m$,
thus   $\varpi_1=\varpi_4+2m_2\pm 1$. 
In order to avoid clumsy formulae, we do not explicitly write down  
the operators  $U_{\{z\}}$, and $U_{\{z\}}^{-1}$ that appear when 
Eq.\ref{2.7} is used. There should be no confusion. 
One may consider Eq.\ref{2.8} for arbitrary $\varpi_1$. 
However, in the framework of standard conformal theory, a
basic assumption is that all highest-weight states are
generated by applying a primary field to the $sl(2,C)$
invariant vacuum. 
 As
discussed, in
ref.\cite{G5}, a highest-weight state $|\varpi_J >$, is created
from
the $sl(2,C)$-invariant state $|\varpi_0 >$ (
$\varpi_0=1+\pi/h$)
by the limit
\beq
|\varpi_J > =\lim_{z\to 0} V^{(J)}_{-J}(z)  |\varpi_0 >,\quad
\hbox {with} \> \varpi_J=\varpi_0+2J.
\label{2.10}
\eeq
We shall thus consider matrix elements between states with
momenta of this type. According to Eq.A.17, the 
overall normalization of
Eq.\ref{2.1}, \ref{2.3}, and \ref{2.8} is fixed by assuming that 
the operators $V_m^{(J)}$ are such that  
\beq
<\varpi_{J_1}|V_m^{(J_2)}(1)|\varpi_{J_3}> 
=\delta_{m,\,  J_3 -J_1}.
\label{2.11}
\eeq
Similarly, one also deduces from Eq.\ref{2.3} that 
$$<\varpi_4 |  V_{m_2}^{(J_2)}(z_2)
V_{\pm 1/2}^{(1/2)}(z_3)
|\varpi_1> =z_2^{-\Delta_{J_2}+\Delta(\varpi_4)} \> z_3^
{-\Delta_{1/2}-\Delta(\varpi_1) }
\times
$$
\beq
 \left ( {z_2 \over
z_3}\right )^{\Delta(\varpi_1\mp  1) }  \left ( 1-{z_2 \over
z_3}\right )^{-hJ_2 /\pi} F(a'_\pm ,b'_\pm ;c'_\pm ; {z_3\over z_2})
\label{2.12}
\eeq
$$a'_\pm ={h\over 2\pi}\left [-1-2J_2  \mp (\varpi_1-\varpi_4 )\right
]
;\quad
b'_\pm = {h\over 2\pi}\left [-1-2J_2  \mp (\varpi_1+\varpi_4 )
\right ]
;
$$
\beq
 c'_\pm =1\mp{h\varpi_1 \over \pi}.
\label{2.13}
\eeq
Next using the well known relation 
$$F(a,b;c;x)=
{ \Gamma(c)\Gamma(c-b-a) \over \Gamma(c-a) \Gamma (c-b) }
\> F(a,b;a+b-c+1;1-x)+
$$
\beq 
 {\Gamma(c)\Gamma(a+b-c) \over \Gamma(a)
\Gamma (b) }\> 
(1-x)^{c-a-b} F(c-a,c-b;c-a-b+1;1-x),
\label{2.14}
\eeq
one deduces that 
$$<\varpi_4 | V_{\pm 1/2}^{(1/2)}(z_3) V_{m_2}^{(J_2)}(z_2)
|\varpi_1> =
z_2^{2(\Delta_{1/2}+\Delta(\varpi_4)-\Delta(\varpi_1))} \> 
z_3^{-2\Delta_{1/2}}
\times
$$
\beq
\sum_{\epsilon=\pm 1} f_{\pm 1,\, \epsilon}(J_2,m_2;\varpi)\,  
<\varpi_1 |  V_{m_4}^{(J_4)}(z_2)
V_{-\epsilon /2}^{(1/2)}\left (z_2(1-z_2/z_3)\right )
|\varpi_2>,  
\label{2.15}
\eeq
where  $m_4=(\varpi_2-\varpi_1-\epsilon)/2$, and 
$$f_{1, \,  1}= {\Gamma\left ( 1-\varpi_4 h/\pi \right ) 
\, \Gamma\left ( 1+\left (2J_2+1\right )  h/\pi \right ) \over 
\Gamma\left ( 1+\left (-\varpi_4+J_2-m_2\right )  h/\pi \right ) 
\,  \Gamma\left ( 1+\left (J_2+m_2+1 \right )  h/\pi \right )},
$$
$$ f_{-1, \,  -1}= {\Gamma\left ( 1+\varpi_4 h/\pi \right )
\, \Gamma\left ( -1-\left (2J_2+1\right )  h/\pi \right ) \over
\Gamma\left ( -\left (J_2+m_2\right )  h/\pi \right )
\,  \Gamma\left ( \left (\varpi_4-J_2+m_2-1 \right )  h/\pi \right
)},
$$
$$f_{1, \,  -1}= {\Gamma\left ( 1-\varpi_4 h/\pi \right )    
\, \Gamma\left ( -1-\left (2J_2+1\right )  h/\pi \right ) \over  
\Gamma\left ( -\left (J_2-m_2\right )  h/\pi \right )   
\,  \Gamma\left ( -\left (\varpi_4+J_2+m_2+1 \right )  h/\pi \right )}  ,
$$  
\beq
f_{-1, \,  1}= {\Gamma\left ( 1+\varpi_4 h/\pi \right ) 
\, \Gamma\left ( 1+\left (2J_2+1\right )  h/\pi \right ) \over 
\Gamma\left ( 1+\left (\varpi_4+J_2+m_2\right )  h/\pi \right ) 
\,  \Gamma\left ( 1+\left (J_2-m_2+1 \right )  h/\pi \right
)}. 
\label{2.16}
\eeq
Following ref.\cite{MS} we re-transform the right-hand side of
Eq.\ref{2.15}, in order to see the operator-product expansion 
more clearly. Denote by $|\varpi, \{\nu\}>$ an arbitrary vector 
of the Verma module corresponding to the highest weight
$\Delta(\varpi)$. The notation $\{\nu\}$ represents 
a multi-index. 	 We shall use a basis such that  
 the  metric of inner products  
$ {\cal G}_{\{\nu\},\,
\{\nu'\}}$ is equal to $\delta_{\{\nu\},\, 
\{\nu'\}}$.  
It is convenient to write 
$$<\varpi_1 |  V_{m_4}^{(J_4)}(z_2)
V_{-\epsilon /2}^{(1/2)}\left (z_2(1-z_2/z_3)\right )
|\varpi_2>
=
$$
\beq
\sum_{\{\nu\}}
<\varpi_1 |  V_{m_4}^{(J_4)}(z_2) 
|\varpi_\epsilon , \{\nu\}> 
<\varpi_\epsilon , \{\nu'\}| V_{-\epsilon /2}^{(1/2)}
\left (z_2(1-{z_2\over z_3})\right )
|\varpi_2>, 
\label{2.17}
\eeq
where $\varpi_\epsilon=\varpi_2+\epsilon$. 
According to Eq.\ref{2.11}, 
 one has 
\beq
<\varpi_1 |  V_{m_4}^{(J_4)}(z_2)
|\varpi_\epsilon , \{0\}> \equiv 
<\varpi_1 |  V_{m_4}^{(J_4)}(z_2)
|\varpi_\epsilon > = z_2^{\Delta_1+\Delta_4-\Delta_\epsilon}, 
\label{2.18}
\eeq
where 
\beq
\Delta_1=\Delta(\varpi_1),\quad 
\Delta_4=\Delta(\varpi_4),\quad 
\Delta_\epsilon=\Delta(\varpi_\epsilon).
\label{2.19}
\eeq
On the other hand, 
\beq
<\varpi_4 |  V_{m_2\pm 1/2}^{(J_\epsilon)}(z_2)
|\varpi_1> 
 = z_2^{-\Delta_1+\Delta_4+\Delta_\epsilon},
\label{2.20}
\eeq
and we get 
$$<\varpi_1 |  V_{m_4}^{(J_4)}(z_2)
|\varpi_\epsilon >= z_2^{2(\Delta_1-\Delta_4)} 
<\varpi_4 |  V_{m_2\pm 1/2}^{(J_\epsilon)}(z_2) 
|\varpi_1>. 
$$
It is thus consistent to let 
\beq 
<\varpi_1 |  V_{m_4}^{(J_4)}(z_2) 
|\varpi_\epsilon, \{\nu \} >= z_2^{2(\Delta_1-\Delta_4)}  
<\varpi_4 |  V_{m_2\pm 1/2}^{(J_\epsilon, \{\nu \})}(z_2)  
|\varpi_1>, 
\label{2.21}
\eeq
 where the right-hand side involves the descendant fields, noted 
$V_{m}^{(J, \{\nu \})}$ in general.  
This expresses  the symmetry
between the three
legs (sphere with three punctures). 
The last equation allows us to
re-write Eq.\ref{2.15} as 
$$<\varpi_4 | V_{\pm 1/2}^{(1/2)}(z_3) V_{m_2}^{(J_2)}(z_2)
|\varpi_1> = \left ( {z_2\over z_3}\right ) ^{2\Delta_{1/2}}
\sum_{\epsilon=\pm 1} f_{\pm 1,\, \epsilon}(J_2,m_2;\varpi_4) 
\times        
$$
\beq 
\sum_{\{\nu\}}
 <\varpi_4 |  V_{m_2\pm 1/2}^{(J_\epsilon, \{\nu \})}(z_2)
|\varpi_1> 
 <\varpi_\epsilon , \{\nu\}| V_{-\epsilon /2}^{(1/2)}
\left (z_2(1-{z_2\over z_3})\right )
|\varpi_2>.
\label{2.22}
\eeq
This formula is not explicitly invariant under $z$ translation.
This apparent contradiction is resolved in appendix B, by
performing a suitable M\"obius transformation. 
At this point it is useful to establish a closer contact with
the Moore Seiberg  (MS) scheme\cite{MS}.  There is a
 Verma module
${\cal H}_J$ for each $\varpi_J$ (defined in Eq.\ref{2.10}).  The MS
chiral vertex-operators connect three specified Verma modules
and are thus  of the form
$\phi_{J_3,J_1}^{J_2}$. The Bloch-wave 
operators $V_m^{(J)}$, on the
contrary,
act in the direct sum
${\cal H}=\oplus_J {\cal H}_J$. According
to
Eq.\ref{2.11},  the two are
related by the projection operator ${\cal P}_{J}$:
\beq
{\cal P}_{J} {\cal H} ={\cal H}_J, \quad
{\cal P}_{J_3} V_m^{(J_2)} \equiv \phi_{J_3,J_3+m}^{J_2}.
\label{2.23}
\eeq

In the MS formulation, 
the fusing and braiding  algebras   of the V operators, 
including the explicit contribution
of the descendants,  read (from now on we omit the 
worldsheet variables):
$$
{\cal P}_J
V^{(J_1)}_{m_1}\,V^{(J_2)}_{m_2} =
{\cal P}_J
\sum _{J_{12}= \vert J_1 - J_2 \vert} ^{J_1+J_2}
F_{J+m_1,J_{12}}\!\!\left[
^{J_1}_J
\>^{\quad J_2}_{J+m_1+m_2 }
\right]\times
$$
\beq
\sum _{\{\nu_{12}\}} V ^{(J_{12},\{\nu_{12}\})} _{m_1+m_2}
<\!\varpi_{J_{12}},{\{\nu_{12}\}} \vert
V ^{(J_1)}_{J_2-J_{12}} \vert \varpi_{J_2} \! >,
\label{2.24}
\eeq
\beq
{\cal P}_{J}
V^{(J_1)}_{m_1}\,V^{(J_2)}_{m_2} =
{\cal P}_{J}
\sum _{m'_1,m'_2,m'_1+m'_2=m_1+m_2}
B_{J+m_1,J+m'_2}^\pm \!\!\left[
^{J_1}_J
\>^{\quad J_2}_{J+m_1+m_2 }
\right]
V^{(J_2)}_{m'_2}\,V^{(J_1)}_{m'_1}.
\label{2.25}
\eeq
As already recalled in the introduction,  
the F and B matrices should be related by the 
equations \cite{MS} 
\beq 
B^{\pm}_{{J_{23}},{J_{12}}}\!\!\left[^{J_1}_{J} 
\,^{J_3}_{J_2}\right] 
=e^{\pm i \pi \left( \Delta_{J}+\Delta_{J_2}-\Delta_{J_{23}} 
-\Delta_{J_{12}}\right )}
F_{{J_{23}},{J_{12}}}\!\!\left[^{J_1}_{J}
\,^{J_2}_{J_3}\right].
\label{2.26}
\eeq
Comparing Eq.\ref{2.24} with Eq.\ref{2.22}, one concludes 
that\footnote{we slightly change the choice of indices for 
later convenience}
$$
F_{J+\epsilon_1/2,J_2+\epsilon_2/2}\!
\left[^{1/2}_{J}\,^{J_2}_{J_3}\right]=
f_{\epsilon_1,\epsilon_2}(J_2,J_3-J-\epsilon_1 /2;\varpi_{J})
=
$$
$$
{\Gamma((1-\epsilon_1)-\epsilon_1(2J+1)h/\pi)
\over
\Gamma(1+(\epsilon_2-\epsilon_1)/2+(J_3+\epsilon_2
J_2-\epsilon_1J+(1-\epsilon_1+\epsilon_2)/2)h/\pi)
}
\times
$$
\beq
{\Gamma(\epsilon_2+\epsilon_2(2J_2+1)h/\pi)
\over
\Gamma((\epsilon_2-\epsilon_1)/2+(-J_3+\epsilon_2
J_2-\epsilon_1J+(-1-\epsilon_1+\epsilon_2)/2)h/\pi)}. 
\label{2.27}
\eeq
Using the explicit form of the braiding algebra of ref.\cite{GN4} 
(or  ref.\cite{G1}), one easily cheks that Eq.\ref{2.26} 
holds for $J_1=1/2$. 

Next we want to deduce the complete F and B matrices  from the
ones  just written using  the polynomial equations\cite{MS} 
that express the associativity of the OPA, and the fact that
the order between fusing and braiding should be irrelevant. 
It is well known that the q-6j symbols with appropriate 
entries give a solution. However they only involve 
trigonometric functions of $h$, while the above partial
solution Eq.\ref{2.27} involves $\Gamma$ functions. 
   At this point we remark
that, given a solution of the polynomial equations, one may 
get a family of other solutions if one multiplies by
 factors whose products is equal to one in every 
equation. With this in mind, we consider the
ansatz 
\beq
F_{{J_{23}},{J_{12}}}\!\!\left[^{J_1}_{J}
\,^{J_2}_{J_3}\right]
=
{g_{J_1J_2}^{J_{12}}\
g_{J_{12}J_3}^{J}
\over
g _{J_2J_3}^{J_{23}}\
g_{{J_1}J_{23}}^{J}
}
\left\{ ^{J_1}_{J_3}\,^{J_2}_{J}
\right. \left |^{J_{12}}_{J_{23}}\right\}.
\label{2.28}
\eeq
The  symbol\footnote{since all symbols are q deformed, 
we omit 
the letter q most of the time.} 
$\left\{ ^{J_1}_{J_3}\,^{J_2}_{J}
\right. \left |^{J_{12}}_{J_{23}}\right\}$  represents the 6-j
coefficient wich is not completely  
tetra\-hed\-ron-symmetric,  
as the notation indicates. This is  
in contrast with the Racah-Wigner 6-j symbol, noted 
$\left\{ ^{J_1}_{J_3}\,^{J_2}_{J}
\,^{J_{12}}_{J_{23}}\right\}$. 
The form of the $gg/gg$-factor is precisely such that the
left-hand side is also a solution of the polynomial equations,
(if the B matrix is defined by Eq.\ref{2.26}) for 
arbitrary $g$'s. We will give a very simple proof of this fact
later on, following the argument indicated in the
introduction. 
 For the time being, let us show that the consistency of the
ansatz just written, with the value of $F$ for $J_1=1/2$ 
(Eq.\ref{2.27}) fixes  $g_{JK}^L$   
for arbitrary $J$, $K$, $L$ so that 
F matrix is completely determined. 
 This is of
course due to the very special form of the proportionality
factor which is dictated by the polynomial equations (below 
 we shall  see   
how the polynomial equations arise from associativity of the 
OPA in our case).     
In order to determine the $g$'s, 
 we  introduce 
\beq
A\!\left[ ^{J_1}_{J_3}\,^{ J_2}_{J}
\,^{J_{12}}_{ J_{23}}\right ]\equiv 
{F_{{J_{23}},{J_{12}}}\!\!\left[^{J_1}_{J}
\,^{J_2}_{J_3}\right]
\over
\left\{ ^{J_1}_{J_3}\,^{J_2}_{J}
\right. \left |^{J_{12}}_{J_{23}}\right\}
}.
\label{2.29} 
\eeq
Next we determine $A\!\left[ ^{1/2}_{J_3}\,^{ J_2}_{J}
\,^{J_{12}}_{ J_{23}}\right]$ by using Eq.\ref{2.27}, together
with the explicit expression of the 6-j symbols (see, e.g.
ref.\cite{KR} and Eq.\ref{2.55} below).
After some calculations, one obtains
 $$
A\!\left[ ^{1/2}_{J_3}\,^{ J_2}_J
\,^{J_2+\epsilon_2/2}_{\ J+\epsilon_1/2}\right]
=\sqrt{
F(\epsilon_2+\epsilon_2(2J_2+1)h/\pi)\over
F((1+(\epsilon_2-\epsilon_1)/2)
+(\epsilon_2J_2-\epsilon_1J+J_3+(1-\epsilon_1+\epsilon_2)/2)h/\pi)}
$$
\beq
\times \sqrt{
F((1-\epsilon_1)-\epsilon_1(2J+1))h/\pi)
\over
F((\epsilon_2-\epsilon_1)/2+(\epsilon_2J_2
-\epsilon_1 J-J_3+(-1-\epsilon_1+\epsilon_2)/2)h/\pi)},
\label{2.30}
\eeq
\beq
\hbox{with}\qquad
F(z)\equiv
\Gamma(z)\bigl / \Gamma(1-z).
\label{2.31}
\eeq 
The obvious property $F(z)F(1-z)=1$ is important for the
following. At this point  our ansatz gives 
\beq
{
g_{J_2+\epsilon_2/2,\, J_3}^{J}
\over
g _{J_2J_3}^{J+\epsilon_1/2}
}
= {g_{1/2,\, J+\epsilon_1/2}^{J}\over 
g_{1/2,\, J_2}^{J_2+\epsilon_2/2}} 
A\!\left[ ^{1/2}_{J_3}\,^{ J_2}_{J}
\,^{J_2+\epsilon_2/2}_{ J+\epsilon_1/2}\right].
\label{2.32}
\eeq
Clearly, once 
 we know the expression  of $g_{1/2\>J}^{J\pm
1/2}$,
this last relation  will allow us  to determine all of the
$g_{J_1 \, J_2}^{J_{12}}$
by a double recursion on the indices $J_1$ and $J_{12}$ (for any
$J_2$). The next basic point is that the integrability condition 
for this recurrence relation, completely fixes
$g_{1/2\>J}^{J\pm
1/2}$, as we  now  show. As a preliminary remark,  
 note that  
$g_{1/2\>J}^{J+1/2}$  can be taken  equal
to one, without loss of generality. For, 
if it  were  not, it would be  possible to define
$$
\tilde g_{J_1J_2}^{J_{12}}\equiv
{\alpha_{J_1}\alpha_{J_2}
\over \alpha_{J_{12}}}
g_{J_1J_2}^{J_{12}}
\qquad
\hbox{such\ that}
\qquad
\tilde g_{1/2\ J}^{J+1/2}\equiv
{\alpha_{1/2}\alpha_J
\over \alpha_{J+1/2}}
g_{1/2\ J}^{J+1/2}
=1, 
$$
and the fusion coefficients $F$ defined by
Eq.\ref{2.28} with $\tilde g$ are the same as the ones
defined with $g$.  Returning to our main line, 
it is useful  to consider Eq.\ref{2.32} as defining a  
connection on a two
dimensional lattice, for fixed $J_3$, by letting 
\beq
 U^{(J_3)}_{-\epsilon_1,\epsilon_2}
(J+{\epsilon_1\over 2},J_2)\equiv 
{g_{1/2,\, J+\epsilon_1/2}^{J}\over
g_{1/2,\, J_2}^{J_2+\epsilon_2/2}}
A\!\left[ ^{1/2}_{J_3}\,^{ J_2}_{J}
\,^{J_2+\epsilon_2/2}_{ J+\epsilon_1/2}\right].
\label{2.33}
\eeq
 This connection is
flat, since Eq.\ref{2.32} is equivalent to 
\beq
g_{J_2+\epsilon_2/2,\, J_3}^{J+\epsilon_1/2}
= 
U^{(J_3)}_{\epsilon_1,\epsilon_2}(J,J_2)\>     
g _{J_2J_3}^{J}.
\label{2.34}
\eeq
 There are two types  of flatness conditions. 
First,  the  inverse relations  
$$U^{(J_3)}_{-\epsilon_1,-\epsilon_2}
(J+{\epsilon_1\over 2},J_2+{\epsilon_2\over 2})\> 
U^{(J_3)}_{\epsilon_1,\epsilon_2}(J,J_2)=1. 
$$
Second,  the  elementary-plaquette relations   
$$ 
U^{(J_3)}_{\eta_1,\eta_2}(J+{\epsilon_1\over 2},
J_2+{\epsilon_2\over 2})\> 
U^{(J_3)}_{\epsilon_1,\epsilon_2}(J,J_2)=
U^{(J_3)}_{\epsilon_1,\epsilon_2}
(J+{\eta_1\over 2},J_2+{\eta_2\over 2})\> 
U^{(J_3)}_{\eta_1,\eta_2}(J,J_2).
$$
Each of the two types is enough to determine
$g_{1/2\>J}^{J-1/2}$, and we shall  only detail the first 
one for $\epsilon_1=\epsilon_2=1$. One has 
\beq
U^{(J_3)}_{1,1}(J,J_2)=
A\!\left[ ^{1/2}_{J_3}\,^{\quad J_2}_{J
+1/2}\,^{J_2+1/2}_{ J}\right], 
\label{2.35}
\eeq
\beq
U^{(J_3)}_{-1,-1}(J+1/2,J_2+1/2)=
{g_{1/2, \,J+1/2}^{ \quad J}
\over
g _{1/2, \,J_2+1/2}^{\quad J_2}}
A\!\left[ ^{1/2}_{J_3}\,^{J_2+1/2}_{ J}
\,^{J_2}_{J+1/2}\right].
\label{2.36}
\eeq
The inverse relation gives 
\beq
g_{1/2\>J_{2}+1/2}^{\quad J_{2}}
\ /\
g_{1/2\>J+1/2}^{\quad J}
=
A\!\left[ ^{1/2}_{J_3}\,^{\quad J_2}_{J
+1/2}\,^{J_2+1/2}_{ J}\right]\> 
A\!\left[ ^{1/2}_{J_3}\,^{J_2+1/2}_{ J}
\,^{J_2}_{J+1/2}\right].
\label{2.37}
\eeq
So, the r.h.s. must be of a very specific form:
it must be the
quotient of a function of $J$ by the same function
of $J_{2}$.
Thanks to the particular form of the A coefficients,
this is true, and one finds by explicit calculations that 
$$
g_{1/2\>J_{2}+1/2}^{\quad J_{2}} 
\ /\
g_{1/2\>J+1/2}^{\quad J}
=
\sqrt{
F(1+(2J_2+1)h/\pi)F(-1-(2J_2+2)h/\pi)
\over
F(1+(2J+1)h/\pi)F(-1-(2J+2)h/\pi)
}. 
$$
   The expression  of $g_{1/2\>J}^{J-1/2}$
is thereby determined 
up to an irrelevant constant factor (it does not
affect $F$):
\beq
g_{1/2\ J}^{J-1/2}
=g_0
\sqrt{F(1+2Jh/\pi)F(-1-(2J+1)h/\pi)}. 
\label{2.38}
\eeq
It is straightforward to verify that the other flatness 
conditions are satisfied\footnote{This only makes use of the
condition $F(z) F(1-z)=1$, and this construction is more
general than the particular structure we are discussing.}. 
Finally, we solve the recurrence relations, and compute 
the general expression for $g_{J_1\, J_2}^{J_{12}}$. Clearly,
there
exists an integer number p such that $(J_1-p/2)+J_2=J_{12}+p/2$, 
that is $p=J_1+J_2-J_{12}$. Thus, by repeated multiplications
through $U_{-1,+1}^{(J_2)}$, we may relate $g_{J_1\,
J_2}^{J_{12}}$ to 
$$
g_{(J_1-J_2+J_{12})/2 , \, J_2}^{(J_1+J_2+J_{12})/2}=1.
$$
This leads to 
$$g_{J_1\, J_2}^{J_{12}}=
\Pi_{n=1}^{J_1+J_2-J_{12}} U_{-1,+1}^{(J_2)}(J_{12}+n/2, J_1-n/2).
$$
Eqs.\ref{2.30}, and \ref{2.38} give   
$$
U_{-1,+1}^{(J_2)}(J_{12}, J_1)=
g_0
\sqrt{ F(1+(2J_1+1)h/\pi) F(-1-(2J_{12}+1)h/\pi)
\over F(1+(J_1+J_2-J_{12})h/\pi) 
F((J_1-J_2-J_{12}-1)h/\pi)}. 
$$
After some re-arrangement, one finds 
$$
g_{J_1J_2}^{J_{12}}
= (g_0)^{J_1+J_2-J_{12}} 
\prod_{k=1}^{J_1+J_2-J_{12}}
$$
\beq
\sqrt{
F(1+(2J_1-k+1)h/\pi)
F(1+(2J_2-k+1)h/\pi)
F(-1-(2J_{12}+k+1)h/\pi)
\over
F(1+kh/\pi)
}. 
\label{2.39}
\eeq
The result is symmetric in $J_1$, $J_2$. The lack of symmetry
between $J_1$ or $J_2$ and $J_{12}$ will be explained later
on.

So far we only delt with the fusing and braiding for $J_1=1/2$,    
and considered the matrix elements  between highest-weight states. 
Next we show that   
the generalization to the full OPE  follows  from 
the associativity  conditions, if we assume that Eq.\ref{2.22} 
holds for any matrix element, according to the MS scheme. 
Let us rederive, for pedagogical purpose,  the associated 
pentagonal relation  for the F matrices. 
We fuse $<\!\varpi_J\vert V^{(J_1)}_{m_1}\,
V^{(J_2)}_{m_2}V^{(J_3)}_{m_3}\vert
\varpi_J+2m_1+2m_2+2m_3\!>$ in two different ways,  beginning
from the left,
and  from the right, and identify the
coefficients of the resulting operator
$V^{(J_{123},\{\nu_{123}\})}_{m_1+m_2+m_3}$.
This gives
$$
\sum_{J_{12}}
F_{J+m_1+m_2,J_{123}}
\!\!\left[
_J^{J_{12}}\,
^{\qquad J_3}_{J+m_1+m_2+m_3}
\right]\,
F_{J+m_1,{J_{12}}}\!\!\left[^{J_1}_J
\,^{\quad J_2}_{J+m_1+m_2}\right]\,\times
$$
$$
\sum_{\{\nu_{12}\}}
<\!\varpi_{J_{123}},\{\nu_{123}\}|
V^{J_{12},\{\nu_{12}\}}_{J_3-J_{123}}|\varpi_{J_3}\!>
<\!\varpi_{J_{12}},\{\nu_{12}\}|
V^{J_1}_{J_2-J_{12}}|\varpi_{J_2}\!>
=
$$
$$
\sum_{J_{23}}
F_{J+m_1,{J_{123}}}\!\left[^{J_1}_J
\,^{\qquad J_{23}}_{J+m_1+m_2+m_3}\right]\,
F_{J+m_1+m_2,{J_{23}}}\!\left[^{\ J_2}_{J
+m_1}\,^{\qquad J_3}_{J+m_1+m_2+m_3}\right]\times
$$
\beq
\sum_{\{\nu_{23}\}}
<\!\varpi_{J_{123}},\{\nu_{123}\}|
V^{J_1}_{J_{23}-J_{123}}|\varpi_{J_{23}},{\{\nu_{23}\}}\!>
<\!\varpi_{J_{23}},\{\nu_{23}\}|
V^{J_2}_{J_3-J_{23}}|\varpi_{J_3}\!>. 
\label{2.40}
\eeq
On the r.h.s. we use
\beq
\sum_{\{\nu_{23}\}}\ |\varpi_{J_{23}},{\{\nu_{23}\}}\!>
<\!\varpi_{J_{23}},\{\nu_{23}\}|\
=\ {\cal P}_{{J_{23}}}, 
\label{2.41}
\eeq
and obtain the factor
$<\!\varpi_{J_{123}},\{\nu_{123}\}|
V^{J_1}_{J_{23}-J_{123}}\
V^{J_2}_{J_3-J_{23}}|\varpi_{J_3}\!>$.
These  last operators are fused in their turn, obtaining,
$$
\sum_{J_{23}} F_{J+m_1+m_2,{J_{23}}}\!
\left[^{\ J_2}_{J+m_1}
\,^{\qquad J_3}_{J+m_1+m_2+m_3}\right]
F_{J+m_1,{J_{123}}}\!\left[^{J_1}_J
\,^{\qquad J_{23}}_{J+m_1+m_2+m_3}\right]\,
F_{{J_{23}},{J_{12}}}\!\left[^{J_1}_{J_{123}}
\,^{J_2}_{J_3}\right]
$$
\beq
=\>F_{J+m_1,{J_{12}}}\!\left[^{J_1}_J
\,^{\quad J_2}_{J+m_1+m_2}\right]\,
F_{J+m_1+m_2,{J_{123}}}\!\left[^{J_{12}}_J
\,^{\qquad J_3}_{J+m_1+m_2+m_3}\right]. 
\label{2.42}
\eeq
On the other hand, the q-6j symbols are well known to satisfy
the same relation\cite{KR}, that is 
$$
\sum_{J_{23}}
\left\{ ^{\qquad J_2}_{J+m_1+m_2+m_3}
\,^{\ J_3}_{J+m_1}\right. \left |^{\quad
J_{23}}_{J+m_1+m_2}\right\}
\left\{ ^{\qquad J_1}_{J+m_1+m_2+m_3}
\,^{J_{23}}_{J}\right. \left |^{\ J_{123}}_{J+m_1}\right\}
\left\{ ^{J_1}_{J_3}\,^{J_2}_{J_{123}}
\right. \left |^{J_{12}}_{J_{23}}\right\}
$$
\beq
=
\left\{ ^{\quad J_1}_{J+m_1+m_2}
\,^{J_2}_{J}\right. \left |^{\ J_{12}}_{J+m_1}\right\}
\left\{ ^{\qquad J_{12}}_{J+m_1+m_2+m_3}
\,^{J_3}_{J}\right. \left |^{\quad J_{123}}_{J+m_1+m_2}\right\}.
\label{2.43}
\eeq
These two relations are used for a proof of Eq.\ref{2.28}, by
recurrence on $J_1$, starting from our previous derivation of
the case $J_1=1/2$. Assume that Eq.\ref{2.28} holds for 
$J_1\leq K$, and write Eqs.\ref{2.42}, and \ref{2.43} 
for $J_1\leq K$ and $J_2\leq K$. In Eq.\ref{2.42}, 
the last term, that is
$F_{J+m_1+m_2,{J_{123}}}\!\left[^{J_{12}}_J 
\,^{\qquad J_3}_{J+m_1+m_2+m_3}\right]$ may have $J_{12}>K$, and
then it is the only one which is not known from our hypothesis.
Combining this equality with the similar one for 6-j symbols
one immediately sees that
$F_{J+m_1+m_2,{J_{123}}}\!\left[^{J_{12}}_J  
\,^{\qquad J_3}_{J+m_1+m_2+m_3}\right]$ is also given by
Eq.\ref{2.28}. Thus this relation holds for all $J$'s. \qed 

It is simple to see, at once 
 that all the MS conditions will be  satisfied since 
we may define new chiral vertex operators $\widetilde V$, by 
\beq
{\cal P}_{J_{12}} {\widetilde V}_{J_2-J_{12}}^{J_1} 
\equiv g_{J_1\, J_2}^{J_{12}} 
{\cal P}_{J_{12}}  V_{J_2-J_{12}}^{J_1},  
\label{2.44}
\eeq
so that the fusing and braiding matrices  simply become 
\beq
{\widetilde F}_{{J_{23}},{J_{12}}}\!\!\left[^{J_1}_{J}
\,^{J_2}_{J_3}\right]
=
\left\{ ^{J_1}_{J_3}\,^{J_2}_{J}
\right. \left |^{J_{12}}_{J_{23}}\right\}, \quad 
{\widetilde B}^{\pm}_{{J_{23}},{J_{12}}}\!\!\left[^{J_1}_{J}
\,^{J_3}_{J_2}\right]
=e^{\pm i \pi \left( \Delta_{J}+\Delta_{J_2}-\Delta_{J_{23}}
-\Delta_{J_{12}}\right )}
\left\{ ^{J_1}_{J_3}\,^{J_2}_{J}
\right. \left |^{J_{12}}_{J_{23}}\right\}. 
\label{2.45}
\eeq
It is well known\cite{KR} that the q-6j symbols satisfy the polynomial
equations. For instance, 
we already used the pentagonal relation Eq.\ref{2.43}. 
Let us stress, however, as already mentioned in the introduction, 
that {\bf this does not mean that our fusing and 
braiding matrices are equivalent to  
Eqs.\ref{2.45}  from the viewpoint of 
conformal theory}. Indeed, the new fusing equations  read 
$$
{\cal P}_{J}
{\widetilde V}^{(J_1)}_{m_1}\, {\widetilde V}^{(J_2)}_{m_2} =
{\cal P}_{J}
\sum _{J_{12}= \vert J_1 - J_2 \vert} ^{J_1+J_2}
{\widetilde F}_{J+m_1,J_{12}}\!\!\left[
^{J_1}_J
\>^{\quad J_2}_{J+m_1+m_2 }
\right]\times
$$
\beq
\sum _{\{\nu_{12}\}}  {\widetilde V}^{(J_{12},\{\nu_{12}\})} _{m_1+m_2}
<\!\varpi_{J_{12}},{\{\nu_{12}\}} \vert
 {\widetilde V}^{(J_1)}_{J_2-J_{12}} \vert \varpi_{J_2} \! >. 
\label{2.46}
\eeq
The matrix element of ${\widetilde V}$ on the right-hand side is a
book-keeping device to recover the coefficients of the OPE. 
There  the normalization of ${\widetilde V}$ appears explicitly. 
According to Eqs.\ref{2.44}, Eq.\ref{2.11} is replaced by
\beq
<\varpi_{J_2}|{\widetilde V}_m^{(J_1)}(1)|\varpi_{J_{12}}>
= g_{J_1 \, J_2}^{J_{12}}\> 
\delta_{m,\,  J_{12} -J_2}.
\label{2.47}
\eeq
For instance, at the level of primaries we have 
\beq 
{\cal P}_{J}   
{\widetilde V}^{(J_1)}_{m_1}\, {\widetilde V}^{(J_2)}_{m_2} =
{\cal P}_{J}
\sum _{J_{12}= \vert J_1 - J_2 \vert} ^{J_1+J_2} 
g_{J_1 \, J_2}^{J_{12}}\> 
{\widetilde F}_{J+m_1,J_{12}}\!\!\left[
^{J_1}_J 
\>^{\quad J_2}_{J+m_1+m_2 }
\right]  
 {\widetilde V}^{(J_{12})}
_{m_1+m_2} \> +\cdots . 
\label{2.48}
\eeq
The $g$'s have re-appeared. There is no way to get rid of them at the
level of the two dimensional  OPA.

Of course, 
it is easy to verify that 
  the $g$ factors of Eq.\ref{2.28}
 cancel in pairs, in every polynomial
equation.   
Consider, for instance,  the Yang-Baxter equation:
$$
\sum_{J_{134}}
B_{{J_{234}},{J_{134}}}\!\!\left[^{J_{1}}_{J_{1234}}
\,^{J_{2}}_{J_{34}}\right]
B_{{J_{34}},{J_{14}}}\!\!\left[^{J_{1}}_{J_{134}}
\,^{J_{3}}_{J_{4}}\right]
B_{{J_{134}},{J_{124}}}\!\!\left[^{J_{2}}_{J_{1234}}
\,^{J_{3}}_{J_{14}}\right]
$$
\beq
=\sum_{J_{24}}
B_{{J_{34}},{J_{24}}}\!\!\left[^{J_{2}}_{J_{234}}
\,^{J_{3}}_{J_{4}}\right]
B_{{J_{234}},{J_{124}}}\!\!\left[^{J_{1}}_{J_{1234}}
\,^{J_{3}}_{J_{24}}\right]
B_{{J_{24}},{J_{14}}}\!\!\left[^{J_{1}}_{J_{124}}
\,^{J_{2}}_{J_{4}}\right], 
\label{2.49}
\eeq
which is obtained from  operator-braidings  in
$<\!\varpi_{J_{1234}}\vert V^{(J_1)}\,V^{(J_2)}\,
V^{(J_3)}\vert \varpi_{J_4}\!>$.
After cancelling out the g-factors, there  remains 
$$
\sum_{J_{134}}
e^{i\pi\epsilon
(\Delta_{J_{1234}}-\Delta_{J_{124}}
-\Delta_{J_{134}}-\Delta_{J_{234}})}
\left\{ ^{J_{1}}_{J_{2}}\,^{J_{34}}_{J_{1234}}
\right. \left |^{J_{134}}_{J_{234}}\right\}
\left\{ ^{J_{1}}_{J_{3}}\,^{J_{4}}_{J_{134}}
\right. \left |^{J_{14}}_{J_{34}}\right\}
\left\{ ^{J_{2}}_{J_{3}}\,^{J_{14}}_{J_{1234}}
\right. \left |^{J_{124}}_{J_{134}}\right\}
$$
\beq
=\sum_{J_{24}}
e^{i\pi\epsilon
(\Delta_{J_{4}}-\Delta_{J_{14}}-\Delta_{J_{24}}-\Delta_{J_{34}})}\left\{
^{J_{2}}_{J_{3}}\,^{J_{4}}_{J_{234}}
\right. \left |^{J_{24}}_{J_{34}}\right\}
\left\{ ^{J_{1}}_{J_{3}}\,^{J_{24}}_{J_{1234}}
\right. \left |^{J_{124}}_{J_{234}}\right\}
\left\{ ^{J_{1}}_{J_{2}}\,^{J_{4}}_{J_{124}}
\right. \left |^{J_{14}}_{J_{24}}\right\}, 
\label{2.50} 
\eeq
which  is a relation satisfied by the 6-j coefficients \cite{KR}.

As a conclusion of this section, 
 we give the final  form of the 
operator-algebra for the $V$ fields. 
In the MS form, and  according
to 
Eqs.\ref{2.24} and  \ref{2.28}, the fusing 
algebra reads 
$$
<\!\varpi_{J_{123} }, \{\nu_{123}\}|
V^{(J_1)}_{J_{23}-J_{123}} V^{(J_2)}_{J_{3}-J_{23}}
|\varpi_{J_{3 }}, \{\nu_3\} \! >
=\sum _{J_{12}= \vert J_1 - J_2 \vert} ^{J_1+J_2}
{g_{J_1J_2}^{J_{12}}\
g_{J_{12}J_3}^{J_{123}}
\over
g _{J_2J_3}^{J_{23}}\
g_{{J_1}J_{23}}^{J_{123}}
}
\left\{ ^{J_1}_{J_3}\,^{J_2}_{J_{123}}
\right. \left |^{J_{12}}_{J_{23}}\right\} \times
$$
\beq
\sum _{\{\nu_{12}\}}
<\! \varpi_{J_{123} }, \{\nu_{123}\}|
V ^{(J_{12},\{\nu_{12}\})}_{J_3-J_{123}} |\varpi_{J_{3 }},
\{\nu_3\} \! >
<\!\varpi_{J_{12}},{\{\nu_{12}\}} \vert
V ^{(J_1)}_{J_2-J_{12}} \vert \varpi_{J_2} \! >.
\label{2.51}
\eeq
In the operator formalism of the Liouville theory, this may be
rewritten as
$$
V^{(J_1)}_{m_1} V^{(J_2)}_{m_2} =
\sum _{J_{12}= \vert J_1 - J_2 \vert} ^{J_1+J_2}
{g_{J_1J_2}^{J_{12}}\> 
g_{J_{12}\, ( \varpi-\varpi_0+2m_1+2m_2)/2} 
^{\> (\varpi-\varpi_0)/2}
\over 
g_{J_{2}\, ( \varpi-\varpi_0+2m_1+2m_2)/2}
^{ \> (\varpi-\varpi_0+2m_1)/2} \> 
g_{J_{1}\, ( \varpi-\varpi_0+2m_1)/2}
^{ \> (\varpi-\varpi_0)/2} } \times 
$$
$$   
\left\{ ^{\quad \quad  J_1}_{( \varpi-\varpi_0 +2m_1+2m_2)/2}
\> ^{\quad J_2}_{( \varpi-\varpi_0)/2}
\right. 
\left |^{\quad J_{12}}_{( \varpi-\varpi_0+2m_1)/2}\right\} 
\times
$$
\beq
\sum _{\{\nu_{12}\}}
V ^{(J_{12},\{\nu_{12}\})}_{m_1+m_2}
<\!\varpi_{J_{12}},{\{\nu_{12}\}} \vert
V ^{(J_1)}_{J_2-J_{12}} \vert \varpi_{J_2} \! >.
\label{2.52}
\eeq
In this last formula, $\varpi$ is an operator. It is easy to
check that this operator-expression 
 is equivalent to Eq.\ref{2.51}, by computing
the matrix element between the states  $<~\varpi_{J_{123} },
\{\nu_{123}\}|$, and $|\varpi_{J_{3 }}, \{\nu_3\} \! >$. Then,
the additional spins of Eq.\ref{2.51}, as compared with 
Eq.\ref{2.52}, are given by
\begin{eqnarray}
J_{123}&=&(\varpi-\varpi_0)/2,  \nnn
J_{23}&=&(\varpi-\varpi_0+2m_1)/2,  \nnn 	
J_{3}&=&(\varpi-\varpi_0+2m_1+2m_2)/2.
\label{2.53}
\eeqa

The explicit formula  for the  Racah-Wigner 6-j coefficients,
which have the tetrahedral symmetry,
are given in
\cite{KR} by
$$
\left\{ ^{a\ b\ e}_{d\ c\ f}\right\}=
\big (
\lfloor z-a-b-e \rfloor \! !
\lfloor z-a-c-f \rfloor \! !
\big )^{-1/2}
(-1)^{a+b-c-d-2e}
\left\{ ^{a\ b}_{d\ c}\right .\left | ^e_f\right\}=
$$
$$
=\Delta(a,b,e)\Delta(a,c,f)\Delta(c,e,d)
\Delta(d,b,f)\times
$$
$$
\sum_{z\hbox{ integer}}
(-1)^z
\lfloor z+1 \rfloor \! !
\bigg[
\lfloor z-a-b-e \rfloor \! !
\lfloor z-a-c-f \rfloor \! !
\lfloor z-b-d-f \rfloor \! !\times
$$
\beq
\lfloor z-d-c-e \rfloor \! !
\lfloor a+b+c+d-z \rfloor \! !
\lfloor a+d+e+f-z \rfloor \! !
\lfloor b+c+e+f-z \rfloor \! !
\bigg]^{-1}
\label{2.54}
\eeq
with
$$
\Delta(l,j,k)=
\sqrt{\lfloor -l+j+k \rfloor \! !
\lfloor l-j+k \rfloor \! !
\lfloor l+j-k \rfloor \! !
\over
\lfloor l+j+k+1 \rfloor \! !}
$$
and
\beq
\lfloor n \rfloor \! ! \equiv
\prod_{r=1}^n \lfloor r \rfloor
\qquad \lfloor r \rfloor \equiv {\sin (hr)\over \sin h}.
\label{2.55}
\eeq

For completeness, we also summarize the formulae for the
braiding\footnote{an
equivalent form is given in ref\cite{B2},
without connection to the 6-j.}.
In the MS form, Eqs.\ref{2.25}, and \ref{2.26} show
that it is given by
$$
<\varpi_{J_{123}},\{\nu_{123}\}|  V_{J_{23}-J_{123}}^{(J_1)} 
 V_{J_{3}-J_{23}}^{(J_2)} | \varpi_{J_{3}},\{\nu_3\}>=
\sum_{J_{13}} e^{\pm i\pi (\Delta_{J_{123}}+\Delta_{J_3}
-\Delta_{J_{23}}-\Delta_{J_{13}})} \times
$$
\beq
{g_{J_1 J_3}^{J_{13}} g_{J_{13} J_2}^{J_{123}} \over
g_{J_2 J_3}^{J_{23}} g_{J_1 J_{23}}^{J_{123}}}
\left\{
^{J_1}_{J_2}\,^{J_3}_{J_{123}}
\right. \left |^{J_{13}}_{J_{23}}\right\}
<\varpi_{J_{123}},\{\nu_{123}\} | V_{J_{13}-J_{123}} ^{(J_2)} 
 V_{J_{3}-J_{13}}^{(J_1)} | \varpi_{J_{3}},\{\nu_3\}>
\label{2.56}.
\eeq
The equivalent operator-form is given by
$$
V^{(J_1)}_{m_1} \>  V^{(J_2)}_{m_2} =
\sum_{n_1+n_2=m_1+m_2} 
e^{\mp ih (2m_1m_2+m_2^2-n_2^2+\varpi(m_2-n_2))}
\times
$$
$$
\left\{ ^{J_1\quad }_{J_2\quad }
\> ^{( \varpi-\varpi_0 +2m_1+2m_2)/2}
_{ \quad \quad (\varpi-\varpi_0)/ 2}
\right.
\left |^{\quad ( \varpi-\varpi_0 +2n_2)/2}
_{\quad ( \varpi-\varpi_0+2m_1)/2}\right\} \times 
$$
\beq
{ g_{J_{1}\, ( \varpi-\varpi_0+2n_1+2n_2)/2}
^{ \> (\varpi-\varpi_0+2n_2)/2} \>
g_{J_{2}\, ( \varpi-\varpi_0+2n_2)/2} 
^{ \> (\varpi-\varpi_0)/2}
\over 
g_{J_{2}\, ( \varpi-\varpi_0+2m_1+2m_2)/2} 
^{ \> (\varpi-\varpi_0+2m_1)/2} \>
g_{J_{1}\, ( \varpi-\varpi_0+2m_1)/2} 
^{ \> (\varpi-\varpi_0)/2}} \quad 
V^{(J_2)}_{n_2}\>  V^{(J_1)}_{n_1} 	
\label{2.57}. 
\eeq
The correspondence table is again given by Eq.\ref{2.53}, with,
in addition, 
\beq
J_{13}= (\varpi-\varpi_0 +2n_2)/2. 
\label{2.58}
\eeq
In the operator-forms Eqs.\ref{2.52}, \ref{2.57}, one sees that
the fusion and braiding matrices involve the operator
$\varpi$, and thus do not commute with the V-operators 
(see Eq.A.12). Such is the 
general situation of the operator-algebras in the MS formalism.
This is in contrast with, for instance,  the braiding relations
for quantum group representations.  In the coming section, and
completing the results of refs.\cite{B,G1}, we will change
basis to the  holomorphic operators $\xi$ which are 
 such that these  $\varpi$
dependences  of the fusing and braiding matrices disappear. 
After the transformation,  one is in the same situation as 
for quantum group, and its structure becomes  more transparent. 

Our last point of the section will concern the treatment of the
end points in correlators.
Following ref.\cite{G5}, one has 
\begin{eqnarray}
\lim_{z\to 0}   V_{m\, \mhat}^{(J\, \Jhat )}(z)
\vert \varpi_0 > = \delta_{m+J,\,0}
\> \delta_{\mhat+\Jhat,\,0} \vert \varpi_{J,\,\Jhat} >
\nonumber \\
\lim_{z\to \infty }
z^{2\Delta_{J\Jhat}} 	<-\varpi_0 \vert
V_{m\, \mhat}^{(J\, \Jhat )}( z)
=   \delta_{m+J,\,0}\>
\delta_{\mhat+\Jhat,\,0}
< -\varpi_{J,\,\Jhat} \vert,
\label{2.59}
\end{eqnarray}
where $\Delta_{J\Jhat}$ is the conformal weight
(see Eq.A.15)
and $\varpi_{J,\,\Jhat}=\varpi_0+2J+2\Jhat\,\pi/h$.
At this point we consider  shortly  the general fields $V_{m\,
\mhat}^{(J\, \Jhat )}$, since, keeping  the $V_m^{(J)}$
fields only  
would give  vanishing two-point functions.  
General correlators, which are $sl(2,C)$ invariant, 
 may be described by the operator 
matrix-elements 
\beq
<-\varpi_0 | V_{m_n\, \mhat_n}^{(J_n\, \Jhat_n )}(z_n) 
\, \cdots \, V_{m_2\, \mhat_2}^{(J_2\, \Jhat_2 )}(z_2)
\, V_{m_1\, \mhat_1}^{(J_1\, \Jhat_1 )}(z_1)
|\varpi_0 >. 
\label{2.60}
\eeq
If we are on the sphere, where the points $0$ and
$\infty$ are equivalent to any other points,  the limit 
$z_1\to 0$ must be finite, and the limit $z_n\to \infty$ must 
also become finite after transforming  $z_n \to 1/z_n$.  
Thus, according to Eqs.\ref{2.59}, one should only consider 
correlators with 
\beq
m_n+J_n=\mhat_n+\Jhat_n=m_1+J_1=\mhat_1+\Jhat_1=0. 
\label{2.61}
\eeq
Then the question arises whether the operator algebra just 
summarized has a  consistent restriction, namely, whether,
conditions Eq.\ref{2.61} 
 are preserved under fusing and braiding. Of course,
this is only possible since the F and B matrices are functions
of $\varpi$. We shall only discuss the conditions  arising from
the limit $z_1\to 0$ of the right-most operator, for the fields
with $J_1+m_1=\Jhat_1+ \mhat_1=0$, explicitly. 

Consider first the fusion. One applies Eq.\ref{2.51}, or
equivalently, Eq.\ref{2.52}, for the highest-weight state 
with $\varpi_{J_3} =\varpi_0$, so 
that $J_3=0$. Recall the defining relation of the 
6-j symbols:
$$
\sum_{J_{23}}
\bigl(J_2,\mu_2;J_3,\mu_3 \vert J_{23}\bigr)
\bigl(J_1,\mu_1;J_{23},\mu_{23} \vert J_{123}\bigr)
\left \{^{J_1}_{J_3}\,^{J_2}_{J_{123}}\right.
\! \left |^{J_{12}}_{J_{23}}
\right \}=
$$
\beq
\bigl(J_1,\mu_1;J_2,\mu_2 \vert J_{12}\bigr)
\bigl(J_{12},\mu_{12};J_{3},\mu_{3} \vert J_{123}\bigr).
\label{2.62}
\eeq
The symbols $\bigl(J_k,\mu_k;J_\ell,\mu_\ell \vert J_{k\ell}\bigr)$ 
denote the  q Clebsch-Gordan (3-j)  coefficients\footnote{We use a
condensed notation
$(J_1,\mu_1;J_2,\mu_2|J_{12})$ instead of
$(J_1,\mu_1;J_2,\mu_2|J_{12}, \mu_1+\mu_2)$.}.
In our present
situation, we have $J_3=0$, and thus $\mu_3=0$. Obviously  
one has $J_{123}=J_{12}$, and $J_{23}=J_3$, and one may easily
verify, from the last relation,  that 
\beq 
\left\{ ^{ J_1}_0
\> ^{\> J_2}_{ J_{123}}
\right.
\left |^{J_{12}}_{J_{23}}\right\}= 
\delta_{J_{123},\, J_{12}} \delta_{J_{23},\, J_{2}}. 
\label{2.63}
\eeq
It follows from Eqs.\ref{2.53} 
 that, for $J_3=0$, the fusing relation vanishes
unless $m_2+J_2=0$, that is, unless the restriction condition 
Eq.\ref{2.61} is satisfied. Then, 
the $g$-factor becomes  one, and one  gets
\beq
V^{(J_1)}_{m_1} V^{(J_2)}_{m_2} |\varpi_0> 
=\delta_{ J_2+m_2,\, 0}\> 
\sum _{\{\nu_{12}\}} V ^{(J_{2}-m_1,\{\nu_{12}\})}_{m_1-J_2}
|\varpi_0> <\!\varpi_{J_{2}-m_2},{\{\nu_{12}\}} \vert
V ^{(J_1)}_{J_2-J_{12}} \vert \varpi_{J_2} \! >.
\label{2.64}
\eeq
A similar discussion shows that 
$$  
V^{(J_1)}_{m_1} V^{(J_2)}_{m_2} |\varpi_0> 
= \delta_{ J_2+m_2,\, 0}\> 
e^{\mp ih [ (\varpi_0+2J_2-2m_1)^2+\varpi_0^2 -
(\varpi_0+2J_2)^2 -(\varpi_0+2J_2)^2 ]/4} \times 
$$
\beq
V^{(J_2)}_{m_1+J_1-J_2} V^{(J_1)}_{-J_1}  |\varpi_0>. 
\label{2.65}
\eeq
Thus, the above restriction conditions  are  preserved by fusing
and braiding.  Note that one may equivalently describe the 
correlators by 
\[
<\varpi_0 | V_{m_n\, \mhat_n}^{(J_n\, \Jhat_n )}(z_n)
\, \cdots \, V_{m_2\, \mhat_2}^{(J_2\, \Jhat_2 )}(z_2)
\, V_{m_1\, \mhat_1}^{(J_1\, \Jhat_1 )}(z_1)
|-\varpi_0 >. 
\]
 Then, conditions Eq.\ref{2.61} are 
replaced by 
\[
m_n-J_n=\mhat_n-\Jhat_n=m_1-J_1=\mhat_1-\Jhat_1=0.
\]
The two formalisms are completely equivalent.

\section{ THE COVARIANT OPERATOR ALGEBRA} 
\markboth{3. Covariant operator algebra} 
{3. Covariant operator algebra} 

In the preceding section, the quantum numbers  $J$ and $m$ of the
$V_m^{(J)}$ operators should be regarded as 
 quantum-group invariant. Indeed, 
the $J$'s appear as total spins in q-6j symbols, and the $m$'s 
are given by differences of $J$'s, as is clear on Eq.\ref{2.11}, 
for instance. Thus the quantum group does not act on the $V$'s. 
 In refs.\cite{B,G1}, other  fields $\xi_M^{(J)}$ 
were   defined  
which are   quantum-group covariant. Following ref.\cite{G1}, this is 
done in two steps. A first change of field is  performed by
introducing  $\psi$ fields of the form 
\beq
\psi^{(J)}_m\equiv E^{(J)}_m (\varpi) 
\> V_m^{(J)}, 
\label{3.1}
\eeq  such
that
the braiding matrix and the fusing 
coefficients  for $\psi^{(1/2)}_\alpha \psi^{(J)}_m\to
\psi^{(J+1/2)}_{m+\alpha}$ become trigonometric. The
definition of $E^{(J)}_m(\varpi)$ (which in the notations 
of ref.\cite{G1} is equal to 
$C^{J-m,J+m} D^{J-m,J+m}(\varpi)$)    is recalled
in Eqs.A.16-A.25.  Then the $\xi$ fields are defined by expressions 
of the form 
\beq
\xi_M^{(J)} = \sum_{-J\leq m \leq J}\vert
J,\varpi)_M^m \>
\psi_m^{(J)},  \quad -J\leq M\leq J.
\label{3.2}
\eeq
The explicit form of the coefficients $| J,\varpi)_M^m$
is given in Eq.\ref{3.35}, below. The formulae 
just written  allow us
to deduce the F and B matrices of the $\xi$ fields from those of the 
$V$ fields derived in the previous section. Indeed, it was
already shown  
in ref\cite{G1} that the braiding matrix of the $\xi$ field
coincides  with the universal $R$-matrix of $U_q(sl(2))$. 
Concerning the fusing matrices, 
we shall establish an explicit connection   
later on, by first relating the coefficients 
$| J,\varpi)_M^m$ to a limit of q-Clebsch-Gordan coefficients. 
 At the present stage of the discussion, it is more enlightening
to proceed in another way. We  shall first transform the fusing 
matrix for the OPE of $V_{m_1}^{(1/2)}$ and $V_{m_2}^{(J_2)}$,
which was the starting point of last section, and after, generalize
the result using the associativity of the OPA. 

Consider, thus Eq.\ref{2.24}, with $J_1=1/2$, 
and make use of Eq.\ref{2.27}. Taking Eq.A.12
into account, one sees that
it is appropriate to multiply both sides  by
$ E^{(1/2)}_{\pm 1/2} \!
(\varpi)
\> E^{(J)}_m\! (\varpi\pm 1)$. 
Using the recurrence relations
satisfied by $C$ and $D$ (Eqs.A.15, and
A.17 of ref.\cite{G1}) one  thereby derives  
 the fusing relations   
$$
\psi_{\pm 1/2}^{(1/2)} \psi_{m}^{(J)}=
\sum_{\{\nu\}} {\lfloor \varpi\mp J+m\rfloor \over \lfloor
\varpi \rfloor}
\psi_{m \pm 1/2}^{(J+1/2, \{\nu\})}
<\varpi_J+1,\{\nu\} | V_{-1/2}^{(1/2)} |\varpi_J>.
$$
$$+ \sum_{\{\nu\}} 
{\lfloor  \mp J+m\rfloor \over \lfloor \varpi \rfloor}
\Gamma(1+2Jh/\pi)\Gamma(-1-(2J+1)h/\pi) \times
$$
\beq
\psi_{m \pm 1/2}^{(J-1/2, \{\nu\})}
<\varpi_J-1,\{\nu\} | V_{1/2}^{(1/2)} |\varpi_J>
\label{3.3}
\eeq
The
original motivation for introducing the $\psi$
fields\cite{G1,G3}  was that
the braiding matrix, and the leading-order fusing coefficients 
for them are trigonometrical. The last
equation written shows that this is not true for    the other fusing 
coefficients $\psi^{(1/2)}_\alpha \psi^{(J)}_m\to
\psi^{(J-1/2)}_{m+\alpha}$.  
As a result,
the OPE is not associative if one forgets the contribution
of the
secondaries (more about this below).  
Next, comparing with the expression Eq.\ref{2.38} 
of the $g$'s, one sees
that the last equation is naturally rewritten as 
\beq
\psi_{\pm 1/2}^{(1/2)} \psi_{m}^{(J)}=
\sum_{\epsilon, \{\nu\}}
g_{1/2,\, J}^{J+\epsilon/2}
N\!\!\left | ^{1/2}_{\pm  1/2} \,  ^{ J;}_{m;}
\,^ {J+\epsilon /2}_{m\pm 1/2}
;\varpi \right |
\psi_{m \pm 1/2}^{(J+\epsilon/2, \{\nu\})}
<\varpi_J+{\epsilon} ,\{\nu\} |
V_{-\epsilon/2}^{(1/2)} |\varpi_J>
\label{3.4}
\eeq
where,
$$
N\!\!\left | ^{1/2}_{\pm  1/2} \,  ^{ J;}_{m;}
\,^ {J+1/2}_{m\pm 1/2}
;\varpi \right |=
{\lfloor \varpi\mp J+m\rfloor \over \lfloor \varpi \rfloor},
$$
\beq
N\!\!\left | ^{1/2, \,   J;\,   J-1/2}
_{\pm  1/2,  \,  m; \, m\pm 1/2};\varpi \right |=
{i \pi g_0^{-1} \over \sqrt{ \sin (2hJ) \sin ((2J+1)h)}}
{\lfloor  \mp J+m\rfloor \over \lfloor \varpi \rfloor}
\label{3.5}.
\eeq
One sees that the non-trigonometric part is entirely contained
in the $g$'s. It will be shown at the end of the section  
 that Eq.\ref{3.4}  is a particular case of the
general fusing algebra 
\beq
\psi_{m_1 }^{(J_1)} \psi_{m_2}^{(J_2)}=
\sum_{J, \{\nu\}}
g_{J_1,\, J_2}^{J}
N\!\!\left | ^{J_1}_{m_1} \,  ^{ J_2;}_{m_2;}
\,^ {\> J}_{m_1+m_2}
;\varpi \right |\>
\psi_{m_1+m_2}^{(J, \{\nu\})}
<\varpi_J ,\{\nu\} | V_{J_2-J}^{(J_1)} |\varpi_{J_2}>.
\label{3.6}
\eeq
In addition,
 $ N\!\!\left | ^{J_1}_{m_1} \,  ^{ J_2;}_{m_2;}
\,^ {\> J}_{m_1+m_2}
;\varpi \right | $ will be
related to a
q-6j symbol.
For the time being we transform  Eq.\ref{3.4} further, in order
to derive the fusion of the $\xi$ fields. Their definition Eq.\ref{3.2},
together
with the shift properties of the $V$ fields 
(see Eq.\ref{2.11}, or Eq.A.12), are such that    
\beq
\xi_\alpha ^{(1/2)}\xi_M^{(J)} =\sum_{\eta=\pm 1/2,\, m}
|1/2,\varpi)_\alpha^\eta\>  | J,\, \varpi+2\eta)_M^m
\> \psi_\eta^{(1/2)} \psi_m^{(J)}.
\label{3.7}
\eeq
It will be shown in appendix C that, if we choose 
$g_0=2\pi $, we have 
$$\sum_{\eta=\pm 1/2,\, m ,\>   (\eta+m =n)}
|1/2,\varpi)_\alpha^\eta | J,\,
\varpi+2\eta)_M^m\>
N\!\!\left | ^{1/2}_{\eta} \,  ^{ J;}_{m;}
\,^ {J+\epsilon /2}_{\eta+ m}
;\varpi \right |
$$
\beq
=(1/2,\alpha;J,M|J+\epsilon/2)\  | J+\epsilon/2,\,
\varpi)_{M+\alpha}^n.
\label{3.8}
\eeq
(recall that $(1/2,\alpha;J,M|J+\epsilon/2)$ denotes the
3-j  symbols).   
 Eq.\ref{3.4} becomes
$$  
\xi_{\pm 1/2}^{(1/2)} \xi_{m}^{(J)}=
\sum_{\epsilon=\pm 1} 
g_{1/2,\, J}^{J+\epsilon/2}
(1/2,\alpha;J,M|J+\epsilon/2) \times  
$$
\beq
\sum_{\{\nu\}}
\xi_{m \pm 1/2}^{(J+\epsilon/2, \{\nu\})}
<\varpi_J+\epsilon/2 ,\{\nu\} | V_{-\epsilon/2}^{(1/2)}
|\varpi_J>.
\label{3.9}
\eeq

Our next task is to generalize this last fusing identity.
Note an important feature of this  equation. As expected, 
it expresses the
OPE of two $\xi$ fields in terms of one $\xi$ fields, (and its 
descendants),  that is $\xi_{m \pm 1/2}^{(J+\epsilon/2,
\{\nu\})}$. However, the coefficients of this OPE 
 are proportional to the
matrix element of  a $V$ field. 
The basic reason is that the $V$ matrix element,
of the fusing equation for the $V$ fields (Eq.\ref{2.24})
contains no $m_1$  or $m_2$  dependence, and is thus 
 unchanged when going from the $\psi$ to the $\xi$ fields. 
 Thus we shall start from the general ansatz
$$
{\cal P}_J
\xi ^{(J_1)}_{M_1}\,\xi^{(J_2)}_{M_2} =
{\cal P}_J
\sum _{J_{12}= \vert J_1 - J_2 \vert} ^{J_1+J_2}
{\cal F}(J_1,M_1,J_2,M_2,J_{12},M_{12},\varpi_J)\times
$$
\beq
\sum _{\{\nu\}} \xi ^{(J_{12},\{\nu\})} _{M_1+M_2}
<\!\varpi _{J_{12}},{\{\nu\}} \vert
V ^{(J_1)}_{J_2-J_{12}} \vert \varpi_{J_2}\! >.
\label{3.10}
\eeq
Our next task is to prove
that the fusion coefficients
are proportional to the CG coefficients 
and independent of $\varpi_J$:
\beq
{\cal F}(J_1,M_1,J_2,M_2,J_{12},M_{12},\varpi_J)=
g _{J_1J_2}^{J_{12}} (J_1,M_1;J_2,M_2\vert J_{12}).
\label{3.11}
\eeq
To do this, we have to write the associativity equations
for the $\xi$ fields.
The demonstration is the same as the one for
the V operators (Eq.\ref{2.42}),
this is why we will not give all
the details right now (more about it soon, however).
There is a difference yet, that we have to emphasize.
In the demonstration of Eq.\ref{2.42}, we  pointed out that only
four of the five fusing  coefficients
really came from the fusion of the operators considered,
as the other one --- the third one on the left-hand side of 
Eq.\ref{2.42} ---
came from the fusion of the operators in the matrix
elements, which had been restored as operators
thanks to the closure relation.
So,  in the case of the $\xi$ operators,
the four fusion coefficients will become coefficients
of the $\xi$, but  {\bf the other one will remain a fusing coefficient
of V-fields}. Accordingly, we get a pentagonal relation of the
form 
\beq
\sum {\cal F}\, {\cal F}\,  F\, =\ {\cal F}{\cal F}.
\label{3.12}
\eeq
Next, using  the explicit expression Eq.\ref{3.10}, 
let us show that Eq.\ref{3.11} is a solution of the 
associativity condition. Indeed, with this
ansatz, the pentagonal relation becomes
$$
\sum_{J_{23}}
\bigl(J_2,M_2;J_3,M_3 \vert J_{23}\bigr)
\bigl(J_1,M_1;J_{23},M_{23} \vert J_{123}\bigr)
\left \{^{J_1}_{J_3}\,^{J_2}_{J_{123}}\right.
\! \left |^{J_{12}}_{J_{23}}
\right \}=
$$
\beq
\bigl(J_1,M_1;J_2,M_2 \vert J_{12}\bigr)
\bigl(J_{12},M_{12};J_{3},M_{3} \vert J_{123}\bigr).
\label{3.13}
\eeq
This is the basic identity that defines 
the q-6j  coefficients\cite{KR}, as already recalled (see
Eq.\ref{2.62}). 
Once we know that the equation \ref{3.11} gives an associative
algebra, it is easy to derive it by recursion from the
particular case $J_1=1/2$, $J_2$ arbitrary 
(Eq.\ref{3.9}) using a recurrence proof which is completely
parallel to the one we gave for the $V$ fields, in the
previous section. We do not go through it again, and consider 
Eq.\ref{3.11} as established. \qed

An important feature of the result
is that the fusing matrix does not depend upon $\varpi_J$. 
Thus the projector of Eq.\ref{3.10} does not serve any purpose
and may be removed. The fusion algebra finally reads
$$
\xi ^{(J_1)}_{M_1}\,\xi^{(J_2)}_{M_2} =
\sum _{J_{12}= \vert J_1 - J_2 \vert} ^{J_1+J_2}
g _{J_1J_2}^{J_{12}} (J_1,M_1;J_2,M_2\vert J_{12})\times
$$
\beq
\sum _{\{\nu\}} \xi ^{(J_{12},\{\nu\})} _{M_1+M_2}
<\!\varpi _{J_{12}},{\{\nu\}} \vert
V ^{(J_1)}_{J_2-J_{12}} \vert \varpi_{J_2}\! >.
\label{3.14}
\eeq
For the coming discussion, 
we shall actually  need  the following generalization of 
this last relation 
$$ 
\xi ^{(J_1,\{\gamma\})}_{M_1}\,\xi^{(J_2,\{\mu\})}_{M_2} = 
 \sum _{J_{12}= \vert J_1 - J_2 \vert} ^{J_1+J_2} 
g _{J_1J_2}^{J_{12}} (J_1,M_1;J_2,M_2\vert J_{12}) \times
$$ 
\beq 
\sum _{\{\nu\}} \xi ^{(J_{12},\{\nu\})} _{M_1+M_2} 
<\!\varpi_{J_{12}},\{\nu\} \vert V ^{(J_1, \{\gamma\})}_{J_2-J_{12}} 
\vert \varpi_{J_2}\!, \{\mu\} >,  
\label{3.15} 
\eeq 
which holds  according to   the general 
principles  of ref.\cite{MS}. 

We have just shown  that the complete fusion rule
of two $\xi$ fields is most naturally expressed in terms of
one $\xi$ field  and one $V$ field. Now, we use   this fact   to
operatorially relate the fusing  and braiding properties  of
V fields with
those of the $\xi$ fields. This will provide the general
identities which relate their fusing and braiding matrices. 
The method is to apply the fusion algebra Eq.\ref{3.15} 
repeatedly to  the OPE of 
several $\xi$ fields. In fact, the forthcoming calculation 
will  explicitly verify some of the polynomial equations 
of the OPE of the $\xi$ fields, and may also be regarded as 
a pedagogical explanation of the arguments given above to
derive Eq.\ref{3.15}. For that purpose, one should of course 
deal with the descendant matrix-elements, and this is 
why Eq.\ref{3.15} is needed. Consider the matrix element 
$<\varpi_J,\{\alpha\}|\xi_{M_1}^{(J_1)}
\xi_{M_{2}}^{(J_{2})}
\xi_{M_{3}}^{(J_{3},\{\rho\})}|\varpi_K,\{\beta\}>$. We apply
Eq.\ref{3.15}  twice:  $\xi_{M_{3}}^{(J_{3},\{\rho\})}$ 
and  $\xi_{M_{2}}^{(J_{2})}$ are fused  first, and the result 
is then fused with $\xi_{M_1}^{(J_1)}$. One gets 
$$ <\varpi_J,\{\alpha\}|\xi_{M_1}^{(J_1)} 
\xi_{M_{2}}^{(J_{2})}
\xi_{M_{3}}^{(J_{3},\{\rho\})}|\varpi_K,\{\beta\}>=
$$
$$ \sum _{ J_{23}  \{\nu\}} 
g _{J_2J_{3}}^{J_{23}} \> (J_2,M_2;J_{3},M_{3}\vert J_{23}) 
\sum _{J_{123}  \{\mu\}}
g _{J_1J_{23}}^{J_{123}} \> (J_1,M_1;J_{23},M_2+M_3\vert J_{123})\times
$$
$$
<\varpi_J,\{\alpha\}|
\xi ^{(J_{123},\{\mu\})} _{M_1+M_2+M_3}|\varpi_K,\{\beta\}>\times 
$$
$$
<\varpi_{J_{123}},\{\mu\} | V^{(J_1)}_{J_{23}-J_{123}} 
| \varpi_{J_{23}},\{\nu\}>
<\varpi_{J_{23}}, \{\nu\} | V^{(J_2)}_{J_{3}-J_{23}}
 | \varpi_{J_{3}}, \{\rho\}>. 
$$
Performing the sum over $\{\nu\}$ gives 
$$
<\varpi_J,\{\alpha\}|\xi_{M_1}^{(J_1)}
\xi_{M_{2}}^{(J_{2})}
\xi_{M_{3}}^{(J_{3},\{\rho\})}|\varpi_K,\{\beta\}>=
$$
$$
 \sum _{J_{123} J_{23}  \{\nu\}}
g _{J_1J_{23}}^{J_{123}} \> (J_1,M_1;J_{23},M_2+M_3\vert J_{123})
g _{J_2J_{3}}^{J_{23}} \> (J_2,M_2;J_{3},M_{3}\vert J_{23})
\times $$
\beq
<\varpi_J,\{\alpha\}|
\xi ^{(J_{123},\{\nu\})} _{M_1+M_2+M_3}|\varpi_K,\{\beta\}>
<\varpi_{J_{123}},\{\nu\} | V^{(J_1)}_{J_{23}-J_{123}} 
V^{(J_2)}_{J_{3}-J_{23}}  
  | \varpi_{J_{3}}, \{\rho\}>. 
\label{3.16}
\eeq
 At this point it is convenient to consider the last
equation as follows. The two fields $\xi_{M_1}^{(J_1)}$ and 
$\xi_{M_{2}}^{(J_{2})}$ on the left-hand side  
have been converted into two 
$V$ fields, that is 
$V^{(J_1)}_{J_{23}-J_{123}}$, and $V^{(J_2)}_{J_{3}-J_{23}}$, 
on the right-hand side. 
The third field $\xi_{M_{3}}^{(J_{3},\{\rho\})}$
plays the role of a background field which allows us to
operatorially relate $\xi$ fields to $V$ fields by successive
fusions. Its  quantum numbers  $J_3$ and $\{\rho \}$  specify 
which matrix element of $V$ operators will come out at the end. 
Its quantum number $M_3$ is arbitrary and does not appear in the
final matrix element of the two $V$ fields. We shall come back to it
later on.  
 It is easily seen  that this procedure may be repeated for more
than three  
$\xi$ fields, and that the structure is similar. The right-most 
$\xi$ field which is the only one not converted into a $V$ field 
is to be considered as a background field. In fact, 
each $V$ field is multiplied by a 3-j symbols, and this method 
naturally leads to the transformation through dressing 
by 3-j symbols of refs.\cite{P,MR}, as we see next. 
Clearly, fusing or
braiding the $\xi$ fields and the corresponding $V$ fields on each 
side of Eq.\ref{3.16} 
will directly relate their $F$ and $B$ matrices. Since this
relation is, to begin  with,  different from the one which comes out 
from the connection through the $|J,\varpi)_M^m$ coefficients, we 
next go through the derivations. 

First consider
the fusion.   It follows from Eq.\ref{2.51}  that the  right 
member of Eq.\ref{3.16} may be rewritten as
$$
 \sum _{J_{123} J_{23}}
(J_1,M_1;J_{23},M_2+M_3\vert J_{123})
(J_2,M_2;J_{3},M_{3}\vert J_{23}) \left\{
^{J_1}_{J_3}\,^{J_2}_{J_{123}}
\right. \left |^{J_{12}}_{J_{23}}\right\}
g_{J_1J_{2}}^{J_{12}} g_{J_{12}J_{3}}^{J_{123}}
\times
$$
$$
\sum_{\{\nu\}}<\varpi_J,\{\alpha\}|
\xi ^{(J_{123},\{\nu\})} _{M_1+M_2+M_3}|\varpi_K,\{\beta\}>
\times
$$
\beq
\sum_{\{\gamma\}}
<\varpi_{J_{123}},\{\nu\} | V_{J_{3}-J_{123}}^{(J_{12},\{\gamma\})} 
 | \varpi_{J_{3}},
\{\rho\}>
<\varpi_{J_{12}},\{\gamma\} | 
V_{J_{2}-J_{12}}^{(J_1)} | \varpi_{J_{2}}>.
\label{3.17}
\eeq
On the left-hand side, we fuse $\xi^{(J_1)}_{M_1}$, and
$\xi^{(J_2)}_{M_2}$, making use of Eq.\ref{3.15}. This gives  
$$
\sum_{J_{12}} g_{J_1J_{2}}^{J_{12}} 
(J_1,M_1;J_{2},M_{2}\vert J_{12}) 
\sum_{\{\gamma\}} 
<\varpi_J,\{\alpha\}|\xi_{M_1+M_2}^{(J_{12}, \{\gamma\})}
\xi_{M_{3}}^{(J_{3},\{\rho\})}|\varpi_K,\{\beta\}> \times
$$
\beq
 <\varpi_{J_{12}},\{\gamma\} |
V_{J_{2}-J_{12}}^{(J_1)} | \varpi_{J_{2}}>. 
\label{3.18}
\eeq
Next the remaining two $\xi$ fields are fused in their turn, and one
gets 
$$ 
\sum_{J_{12}, J_{123}} g_{J_1J_{2}}^{J_{12}} 
(J_1,M_1;J_{2},M_{2}\vert J_{12})
g_{J_{12}J_{3}}^{J_{123}} 
(J_{12},M_1+M_2;J_{3},M_{3}\vert J_{123})\times
$$
$$ \sum_{\{\gamma\}\, \{\nu\}} 
<\varpi_J,\{\alpha\}|\xi_{M_1+M_2+M_3}^{(J_{123},
\{\nu\})} 
|\varpi_K,\{\beta\}>\times 
$$
\beq
<\varpi_{J_{123}},\{\nu\} | 
V_{J_{3}-J_{123}}^{(J_{12},\{\gamma\})} | \varpi_{J_{3}},\{\rho \}>
<\varpi_{J_{12}},\{\gamma\} | 
V_{J_{2}-J_{12}}^{(J_1)} | \varpi_{J_{2}}>.
\label{3.19}
\eeq
Comparing this last expression with Eq.\ref{3.17}, one sees that they
coincide if the following relation holds:
$$
\sum_{J_{23}}
\bigl(J_2,M_2;J_3,M_3 \vert J_{23}\bigr)
\bigl(J_1,M_1;J_{23},M_2+M_3 \vert J_{123}\bigr)
\left \{^{J_1}_{J_3}\,^{J_2}_{J_{123}}\right.
\! \left |^{J_{12}}_{J_{23}}
\right \}=
$$
\beq
\bigl(J_1,M_1;J_2,M_2 \vert J_{12}\bigr)
\bigl(J_{12},M_1+M_2;J_{3},M_{3} \vert J_{123}\bigr).
\label{3.20}
\eeq	
This is indeed the defining relation of the 6-j symbols already
recalled in Eq.\ref{2.62}. 
  Thus Eq.\ref{3.16} 
does establish the correct correspondence between the fusion
properties of the $V$ and $\xi$ fields. 

Consider,  next,  the braiding.   
For  the V fields, it is   given by 
Eq.\ref{2.56}, or  \ref{2.57}. 
Concerning the $\xi$ fields, the  braiding properties were derived 
in ref.\cite{G1}. One has
\beq
\xi_{M_1}^{(J_1)}\,\xi_{M_2}^{(J_2)}=
\sum_{-J_1\leq N_1\leq J_1;\> -J_2\leq N_2\leq J_2}\>
(J_1,J_2)_{M_1\, M_2}^{N_2\, N_1}\, \xi_{N_2}^{(J_2)}
 \,\xi_{N_1}^{(J_1)}.
\label{3.21}
\eeq
The symbol $(J_1,J_2)_{M_1\, M_2}^{N_2\, N_1}$ denotes the following
matrix
element of the universal R-matrix: 
\beq
(J_1,J_2)_{M_1\, M_2}^{M'_2\, M'_1}=\Bigl(<\! <J_1,M_1\vert \otimes <\! <J_2,M_2
\vert\Bigr)\> {\bf R}
\>\Bigl(\vert J_1,M'_1>\! > \otimes \vert J_2,M'_2>\! >\Bigr),
\label{3.22}
\eeq
where $|J,M>\! >$ are group theoretic states which span the
representation of spin $J$ of $U_q(sl(2))$. 
The universal R-matrix  $R$  is given by 
\beq
{\bf R}= e^{-2ihJ_3 \otimes J_3}
\sum_{n=0}^\infty \,
{{(1-e^{2ih})^{n}\,e^{ihn(n-1)/2} \over
\lfloor n \rfloor \! !}} e^{-ihnJ_3}(J_+)^n \otimes
e^{ihnJ_3}(J_-)^n,
\label{3.23}
\eeq
 $J_\pm$, and $J_3$ are the quantum-group generators.
For later use we recall that the $R$-matrix-elements may be 
simply written in terms of CG coefficients, since the latter
are ``twisted'' eigenvectors, namely, 
\beq
\sum_{N_1 N_2} (J_2,N_2;J_1,N_{1}\vert
J_{12})\,  (J_1,J_2)_{M_1\, M_2}^{N_2\, N_1}	
=
e^{i\pi ( \Delta_{J_{12}}-\Delta_{J_1}-\Delta_{J_2})}
\,  (J_1,M_1;J_2,M_{2}\vert 
J_{12}) .
\label{3.24}
\eeq
It follows from the orthogonality of the CG coefficients that 
\beq 
 (J_1,J_2)_{M_1\, M_2}^{N_2\, N_1} =
\sum_{J_{12}} (J_2,N_2;J_1,N_{1}\vert 
J_{12})\,  e^{i\pi ( \Delta_{J_{12}}-\Delta_{J_1}-\Delta_{J_2})} 
\, (J_1,M_1;J_2,M_{2}\vert  
J_{12})  .
\label{3.25} 
\eeq 

Returning to our main line, we follow the same 
 procedure  as for fusion. We shall skip details since
the present discussion goes in close parallel.  
  One exchanges  the  first two
$\xi$ fields on the left-hand side of Eq.\ref{3.16}, and  the
two $V$ fields on its right-hand side. Comparing the results, one 
derives   the consistency condition 
$$
\sum_{N_1 N_2} (J_1,N_1;J_3,M_{3}\vert
J_{13})(J_2,N_2;J_{13},N_1+M_{3}\vert J_{123},M_{123})
(J_1,J_2)_{M_1\, M_2}^{N_2\, N_1} =
$$
$$
\sum_{J_{23}} (J_2,M_2;J_3,M_{3}\vert
J_{23})(J_1,M_1;J_{23},M_2+M_{3}\vert J_{123}) 
\times $$ 
\beq
e^{i\pi (\Delta_{J_{123}}+\Delta_{J_3}
-\Delta_{J_{23}}-\Delta_{J_{13}})}
\left\{ 
^{J_1}_{J_2}\,^{J_3}_{J_{123}}  
\right. \left |^{J_{13}}_{J_{23}}\right\}.
\label{3.26}
\eeq
This last relation may be easily proven using Eq.\ref{2.50}
with $J_4=0$, Eq.\ref{2.62}, and Eq.\ref{3.25}. 
This defines the 6-j coefficient of the second type as was
introduced in 	
  ref.\cite{HHM}.

The outcome of the present  discussion
 is that the defining relations for 
the two types of 6-j symbols (Eqs.\ref{3.20}
and \ref{3.26}) may be considered as relating the 
braiding and fusing matrices  
of the $V$ and $\xi$ fields.  Thus the connection is established
in a way where the quantum-group meaning is transparent. 
Clearly, Eqs.\ref{3.20}, and \ref{3.26} show that the
connection is established via 
3-j symbols. As a matter of fact,  we have effectively 
re-derived 
the transformation
through dressing by 3-j symbols of refs.\cite{P,MR}. 
This is   in contrast with the intrinsic transformation  of 
refs.\cite{B,G1},  recalled in Eq.\ref{3.2}, which 
 uses  the $|J,  \varpi )_M^m$ coefficients. Our next point is
to establish the connection between these two transformations. 
The q-CG symbols involve $5$ independent quantum numbers, 
and the $|J,  \varpi )_M^m$ coefficients only $4$. In this
connection, we have remarked that, in Eq.\ref{3.16}, $M_3$ 
does not appear in the $V$-matrix element. It only appears 
in the two CG coefficients. The first one, that is 
$(J_1,M_1;J_{23},M_2+M_3\vert J_{123})$, (resp. the second 
one, that is $(J_2,M_2;J_{3},M_{3}\vert J_{23})$) only involve 
J-quantum-numbers of the field $V^{(J_1)}_{J_{23}-J_{123}}$, 
(resp. $V^{(J_2)}_{J_{3}-J_{23}}$). These two sets of quantum
numbers are treated on the same footing, but this is not 
the case for the $M$-quantum numbers, 
 however, since 
the first CG coefficient contains  $M_1$, $M_2+M_3$, and the 
second $M_2$, $M_3$. To motivate the coming mathematical
derivation, we may remark that the symmetry is restored if 
$M_2+M_3 \sim M_3$, that is if $M_3$ is very large compared 
with $M_2$. One   way to achieve  this is to keep $J_3$ finite and to 
continue the quantum group for $M_3 > J_3$. This may be done
rigourously, as we shall next show, since the q-3j symbols are 
given by q-deformed hypergeometric functions\cite{KR,G3}.
In this limit one quantum number of the q-CG symbol drops 
out, and we will be  left with the right number to identify 
the result with a 
 $|J,  \varpi )_M^m$ coefficient.	

We start from the explicit expression of the CG
coefficients\cite{KR,G3}, that is,  
$$\bigl(J_1,M_1;J_2,M_2 \vert J\bigr)= B(J_1,J_2,J,M_1) 
\times $$
$$\sqrt{{\lfloor  J_2-M_2 \rfloor\! !\,\lfloor J_2+M_2 \rfloor\! !\,
\lfloor  J-M_1-M_2 \rfloor\! !\,\lfloor J+M_1+M_2 \rfloor\!
!}}\times$$
$$e^{ihM_2 J_1}\>
\sum_{\mu=0}^{J_1+J_2-J}\>\Bigl\{{e^{-ih\mu(J+J_1+J_2+1)}
\,(-1)^\mu \over
\lfloor \mu \rfloor\! !\,\lfloor  J_1+J_2-J-\mu \rfloor\! !}\times
 $$
\beq
{1\over \lfloor J_1-M_1-\mu \rfloor\! !\,\lfloor J-J_2+M_1+\mu
\rfloor\! !\,
\lfloor  J_2+M_2-\mu \rfloor\! !\,\lfloor J-J_1-M_2+\mu \rfloor\!
!}\Bigr\},
\label{3.27}
\eeq
where 
$$
B(J_1,J_2,J,M_1)= 
e^{ih(J_1+J_2-J)(J_1+J_2+J+1)/2} e^{-ihM_1J_2}
 \sqrt{\lfloor  J_1-M_1 \rfloor\! !\,\lfloor J_1+M_1 \rfloor\!
!}  \times
$$
\beq
\sqrt {\lfloor 2J+1\rfloor}\>
\sqrt{{\lfloor  J_1+J_2-J \rfloor\! !\,\lfloor -J_1+J_2+J
\rfloor\! !\,
\lfloor J_1-J_2+J \rfloor\! !\over \lfloor J_1+J_2+J+1\rfloor\!
!}}.
\label{3.28}
\eeq
The terms in $M_2$ are
conveniently  rewritten as
\beq
\sqrt{ \lfloor  J_2-M_2 \rfloor\! !  \over \lfloor J-J_1+\mu -M_2
\rfloor\!  !} 
\sqrt{ \lfloor  J-M_1-M_2 \rfloor\! ! \over
J-J_1+\mu -M_2
\rfloor\!  !}  
\sqrt{ \lfloor  J_2+M_2 \rfloor\! ! 
\over \lfloor  J_2-\mu+M_2 \rfloor\! !} 
\sqrt{ \lfloor J+M_1+M_2 \rfloor\! ! \over 
\lfloor  J_2-\mu+M_2 \rfloor\! !} .
\label{3.29}
\eeq
We shall take the limit by giving an imaginary part to 
$M_2$, thus we have to continue the above formulae in this 
variable. The last formula  contains all the $M_2$ dependence. 
It has been written as a product of square roots of ratios 
of q-deformed factorials. Consider each term one-by-one. 
the differences between the arguments  of numerators and 
denominators are 
$$ \left \{ \begin{array}{ccc}
J_2-M_2-(J-J_1+\mu -M_2)&=&J_1+J_2-J-\mu \\
J-M_1-M_2-(J-J_1+\mu -M_2)&=&J-M_1-\mu\\
J_2+M_2 -(J_2-\mu+M_2)&=&\mu\\
J+M_1+M_2-(J_2-\mu+M_2)&=&J+M_1-J_2+\mu.
\end{array}\right. 
$$
The right members are independent from $M_2$.   
In the expression Eq.\ref{3.27}, the actual range of summation is
dictated by the fact  that a factorial with negative argument is
infinite, so that each factorial may only have a  non-negative 
argument. This immediately shows that the right-hand sides 
of the last set  of equations are non-negative 
integers. 
 As in ref.\cite{G3}, let us introduce 
($\nu$ is a positive integer, and $a$ arbitrary)
\beq
\lfloor a\rfloor _\nu \equiv 
\lfloor a \rfloor \lfloor a+1 \rfloor \cdots 
\lfloor a+\nu -1 \rfloor={\lfloor a+\nu-1\rfloor !\over 
\lfloor a-1\rfloor !}.
\label{3.30}
\eeq
Eq.\ref{3.29} may be rewritten as 
$$
\sqrt{ \lfloor J-J_1+\mu -M_2+1 \rfloor_{J_1+J_2-J-\mu}\,
\lfloor J-J_1+\mu -M_2+1 \rfloor _{J-M_1-\mu}}\times
$$
\beq	
\sqrt{ \lfloor J_2-\mu+M_2+1 \rfloor_{\mu} \,
\lfloor J_2-\mu+M_2+1 \rfloor_{J+M_1-J_2+\mu}}.
\label{3.31}
\eeq
According to the definition Eq.\ref{3.30}, each term involves a 
number of factors  which is independent from $M_2$, so that 
the last expression makes sense for arbitrary complex $M_2$.
The limit is taken  with an imaginary part, since, otherwise, 
the  functions  $\sin [h(M_2+\alpha)]$, $\alpha$  
constant, which appear in Eq.\ref{3.31},  
 would not have a well defined limit. Of course, with the 
imaginary part, one exponential of trigonometric functions   blows up 
  while the other vanishes.   
The choice of  sign is such that 
\beq 
\lim_{M_2\to \infty} 
\lfloor  \alpha\pm M_2 \rfloor = \mp  {1\over 2i \sin h} 
e^{-ih ( M_2 \pm \alpha)}
\label{3.32}
\eeq
\beq
{\lfloor  \alpha\pm M_2 \rfloor \! ! \over 
\lfloor  \beta\pm M_2 \rfloor \! !} \sim
(-1)^{(\alpha-\beta)(1\pm 1)/2} (2i \sin h)^{\beta-\alpha} 
e^{ih(\beta-\alpha)M_2} e^{\mp ih(\alpha-\beta)(\alpha+\beta+1)/2}.
\label{3.33}
\eeq
Substitute into Eq.\ref{3.27}, one gets
$$\bigl(J_1,M_1;J_2,M_2 \vert J\bigr)\sim 
\>e^{ih[-M_1(J+1/2)+(J_2^2-J_1^2-J^2+2J_1J_2)/2+(J_1+J_2-J)/2]}\times $$
$$ B(J_1,J_2,J,M_1)
{e^{i\pi(J-J_2+M_1)/2}\over (2i\sin h)^{J_1}} 
\sum_{\mu=0}^{J_1+J_2-J}\>\Bigl\{{e^{-2ih\mu(J+J_2+1)}
 \over
\lfloor \mu \rfloor\! !\,\lfloor  J_1+J_2-J-\mu \rfloor\! !}\times
 $$
\beq
{1\over \lfloor J_1-M_1-\mu \rfloor\! !\,\lfloor J-J_2+M_1+\mu
\rfloor\! !\,
}\Bigr\}.
\label{3.34}
\eeq
Note that the limit is perfectly finite, since all exponentials
in $M_2$ cancell out. 
 On the other hand, the explicity expression of 
$\vert j,\varpi)_M^m$ is\cite{G1}\footnote{
${n\choose p}$ denotes the q-deformed binomial coefficients
 ${n \choose p} \equiv {\lfloor n \rfloor \! !
 /  \lfloor p \rfloor \! !\lfloor n-p \rfloor \! !}$}   
$$\vert j,\varpi)_M^m\,=
\sqrt{\hbox{$ {2j \choose j+M} $ }} \>e^{ihm/2}\times 
$$
\beq
\sum_t\, {e^{iht(\varpi
+m)} \lfloor j-M\rfloor \! ! \lfloor j+M\rfloor \! ! \over 
\left \lfloor {j-M+m-t\over 2}\right\rfloor \! !  
\left \lfloor {j-M-m+t\over 2}\right\rfloor\! !
\left \lfloor {j+M+m+t\over 2}\right\rfloor\! !
\left \lfloor {j+M-m-t\over 2}\right\rfloor\! !}
\label{3.35}
\eeq
and, letting $\mu =(j+M-m-t)/2$, 
$$\vert j,\varpi)_M^m\,= 
\sqrt{\hbox{$ {2j \choose J+M} $ }} \>e^{ihm/2} 
e^{ih((\varpi+m)(j+M-m)} \times  
$$
\beq
\sum_\mu\, {e^{-2ih\mu (\varpi 
+m)} \lfloor j-M\rfloor \! ! \lfloor j+M\rfloor \! ! \over    
\left \lfloor {\mu}\right\rfloor\! !  
\left \lfloor {j-m-\mu }\right\rfloor\! ! 
\left \lfloor { m-M+\mu }\right\rfloor\! ! 
\left \lfloor {j+M-\mu }\right\rfloor\! !} .
\label{3.36}
\eeq    
Comparing Eqs.\ref{3.34}, and \ref{3.36}, one sees that the
variables should be related by   
\beq
\varpi=\varpi_0+2J_2,\>  J_1=j,\>  M=-M_1,\> 
m=J-J_2.
\label{3.37}
\eeq
One gets altogether,
$$\bigl(J_1,M_1;J_2,M_2 \vert J\bigr) \sim
\sqrt { \lfloor 2J+1\rfloor \over {2J_1\choose J_1+J_2-J} 
\lfloor J_2+J-J_1+1\rfloor _{2J_1+1}} \times 
$$
\beq
e^{i\pi [(J-J_2+M_1)/2 +J+M_1-J_1-J_2]} 
{e^{ih(M_1+J_2-J)/2}\over (2i \sin h)^{J_1}} 
\vert J_1,\,  \varpi_0+2J_2
) _{-M_1}^{J-J_2}. 
\label{3.38}
\eeq
Recall that one has\cite{G3}
\beq
\sum_{M=-j}^j (-1)^{j-M}\,e^{ih(j-M)}\,\vert j,\,\varpi)_M^m
\,\vert j,\,\varpi+2p)_{-M}^{-p}=\delta_{m,\,p}\,
C_m^{(j)}(\varpi),
\label{3.39}
\eeq
where
\beq
C_m^{(j)}(\varpi)=(-1)^{j-m}\, (2i \sin h )^{2j} e^{ihj} {2j
\choose j-m}\,
{\lfloor \varpi-j+m \rfloor_{2j+1}\over \lfloor \varpi+2m
\rfloor}.
\label{3.40}
\eeq 
Eq.\ref{3.38}
 may be re-written as 
\beq
\bigl(J_1,M_1;J_2,M_2 \vert J\bigr) \sim 
\sqrt{ e^{ih( J_1+J_2-J+M_1)} e^{i\pi (J_1+M_1)}\over 
C_{J-J_2}^{(J_1)}(\varpi_0+2J_2)} \vert J_1,\, \varpi_0+2J_2
) _{-M_1}^{J-J_2}. 
\label{3.41}
\eeq
This is consistent with the orthogonality of the C.G.
coefficients:
$$\sum_{M_1} \Bigl \{ 
\bigl(J_1,M_1;{\varpi-\varpi_0\over 2} ,M_2 \vert 
{\varpi+2m-\varpi_0\over 2}\bigr)\times 
$$\beq
\bigl(J_1,M_1;{\varpi-\varpi_0\over 2} ,M_2 \vert 
{\varpi+2p-\varpi_0\over 2}\bigr) \Bigr \}
=\delta_{m,p}.
\label{3.42}
\eeq
In the limit $M_2\to \infty$ this gives 
\beq
\sum_{M_1 }(-1)^{J_1+M_1}\,e^{ih(J_1+M_1)}\,\vert J_1,\,\varpi)_{-M_1}^m
\,\vert J_1,\,\varpi)_{-M_1}^{p}=\delta_{m,\,p}\,
C_m^{(J_1)}(\varpi) e^{ihm}.
\label{3.43}
\eeq  
This is equivalent to Eq.\ref{3.39} since 
$\vert J_1,\,\varpi+2p)_{-M_1}^{-p} = 
e^{-ihp} \vert J_1,\,\varpi)_{M_1}^{p}$. 	
Finally we use the defining  relation for the  6-j symbols 
(Eq.\ref{2.62}, or \ref{3.13}), that is, 
$$
\sum_{j_{23}} 
\bigl(j_2,\mu_2;j_3,\mu_3 \vert j_{23}\bigr) 
\bigl(j_1,\mu_1;j_{23},\mu_{23} \vert j_{123}\bigr)
\left \{^{j_1}_{j_3}\,^{j_2}_{j_{123}}\right.  
\! \left |^{j_{12}}_{j_{23}}
\right \}=
$$
\beq
\bigl(j_1,\mu_1;j_2,\mu_2 \vert j_{12}\bigr)
\bigl(j_{12},\mu_{12};j_{3},\mu_{3} \vert j_{123}\bigr). 
\label{3.44}
\eeq
For the present purpose, it is convenient to let
$$ j_2=J_1,\> \mu_2=-M_1,\> j_3={\varpi-\varpi_0\over 2}, 
\> j_{23}= {\varpi-\varpi_0+2m_1 \over 2} ,
$$
\beq
j_1=J_2,\> \mu_1=-M_2,\>   
\> j_{123}= {\varpi-\varpi_0+2m_1+2m_2 \over 2}.
\label{3.45}
\eeq
In the limit one gets
$$ \sum_{m_1+m_2=m_{12} }
\vert J_1,\, \varpi )_{M_1}^{m_1}
\vert J_2,\, \varpi+2m_1 )_{M_2}^{m_2}
\sqrt{  e^{ih( J_1-m_1+J_2-m_2)}
\over
C_{m_1}^{(J_1)}(\varpi) C_{m_2}^{(J_2)}(\varpi+2m_1)} \times
$$
$$ e^{i\pi (J_1+J_2)/2}
\left \{ \begin{array} {cc} 
J_2 & J_1  \\ 
{\varpi-\varpi_0\over 2}  & 
{\varpi+2m_1+2m_2-\varpi_0\over 2}
\end{array} \right. \left|\begin{array}{c} J_{12} \\
{\varpi+2m_1-\varpi_0\over 2}\end{array}
\right \} =
$$
\beq
\bigl(J_1, M_1;J_2,M_2 \vert J_{12}\bigr)
\vert J_{12},\, \varpi )_{M_1+M_2}^{m_{12}}\sqrt{
e^{ih( J_{12}-m_{12})} e^{i\pi
J_{12}}
\over
C_{m_{12}}^{(J_{12})}(\varpi) }
\label{3.46}
\eeq
where we used that fact that 
\beq
\bigl(J_2,-M_2;J_1,-M_1 \vert J_{12}\bigr) 
=\bigl(J_1,M_1;J_2,M_2 \vert J_{12}\bigr). 
\label{3.47}
\eeq
This relation may be verified on 
the explicit expression Eq.\ref{3.27}. 
Eq.\ref{3.46} is the generalization of Eq.\ref{3.8}, derived in 
Appendix C. Indeed, starting from the fusing algebra of the $\xi$
fields (Eq.\ref{3.15}), 
and taking account of the relationship between $\xi$ and
$\psi$ fields (Eq.\ref{3.2}), one concludes that 
the fusing of the $\psi$
fields is of the form\footnote{already announced on
Eq.\ref{3.16}.} 
\beq
\psi_{m_1 }^{(J_1)} \psi_{m_2}^{(J_2)}=
\sum_{J, \{\nu\}}
g_{J_1,\, J_2}^{J}
N\!\!\left | ^{J_1}_{m_1} \,  ^{ J_2;}_{m_2;}
\,^ {\> J}_{m_1+m_2}
;\varpi \right |\>
\psi_{m_1+m_2}^{(J, \{\nu\})}
<\varpi_J ,\{\nu\} | V_{J_2-J}^{(J_1)} |\varpi_{J_2}>,
\label{3.48}
\eeq
where 
$$ N\!\!\left | ^{J_1}_{m_1} \,  ^{ J_2;}_{m_2;}
\,^ {\> J_{12}}_{m_1+m_2}
;\varpi \right |\>
=\sqrt{ e^{i\pi( J_{12}-J_1-J_2)}  C_{m_{12}}^{(J_{12})}(\varpi) \over 
C_{m_1}^{(J_1)}(\varpi) C_{m_2}^{(J_2)}(\varpi+2m_1)} 
\times
$$
\beq
e^{ih(J_1+J_2-J_{12} )/2} 
\left \{ \begin{array} {cc}  
J_2 & J_1  \\
{\varpi-\varpi_0\over 2}  & 
{\varpi+2m_1+2m_2-\varpi_0\over 2} 
\end{array} \right. \left|\begin{array}{c} J_{12} \\
{\varpi+2m_1-\varpi_0\over 2}\end{array} 
\right \}.
\label{3.49}
\eeq

Our last point is a cross check of the whole section. 
We     re-derive the fusing matrix of
the $\psi$ fields by starting  from that of the 
$V$ fields (Eq.\ref{2.51}),
 and applying   
the transformation  Eq.\ref{3.1}. The result reads
$$
<\!\varpi_{J_{123} } \{\nu_{123}\}| 
\psi^{(J_1)}_{J_{23}-J_{123}} \psi^{(J_2)}_{J_{3}-J_{23}} 
|\varpi_{J_{3 }} \{\nu_3\} \! >=
$$
$$
\sum _{J_{12}= \vert J_1 - J_2 \vert} ^{J_1+J_2} 
{g_{J_1J_2}^{J_{12}}\
g_{J_{12}J_3}^{J_{123}}
\over
g_{J_2J_3}^{J_{23}}\
g_{{J_1}J_{23}}^{J_{123}}}
{E^{(J_1)}_{J_{23}-J_{123}} (\varpi_{J_{123}}) \> 
E^{(J_2)}_{J_{3}-J_{23}} (\varpi_{J_{23}}) 
\over  E ^{(J_{12})}_{J_2-J_{123}} (\varpi_{J_{123}})}
\left\{ ^{J_1}_{J_3}\,^{J_2}_{J_{123}}
\right. \left |^{J_{12}}_{J_{23}}\right\} \times 
$$
\beq
\sum _{\{\nu_{12}\}} 
<\! \varpi_{J_{123} } \{\nu_{123}\}|
V ^{(J_{12},\{\nu_{12}\})}_{J_3-J_{123}} |\varpi_{J_{3 }}
\{\nu_3\} \! >
<\!\varpi_{J_{12}},{\{\nu_{12}\}} \vert
V ^{(J_1)}_{J_2-J_{12}} \vert \varpi_{J_2} \! >.
\label{3.50}
\eeq
Comparing with Eq.\ref{3.49}, we conclude that 
$$
{
g_{J_{12}J_3}^{J_{123}}
\over  
g _{J_2J_3}^{J_{23}}\
g_{{J_1}J_{23}}^{J_{123}}
}= {e^{-ih(J_1+J_2-J_{12} )/2} 
\sqrt{  C_{J_{3}-J_{123}}^{(J_{12})}(\varpi_{J_{123})} 
e^{i\pi (J_{12}-J_1-J_2)}
\over
 C_{J_{23}-J_{123}}^{(J_1)}(\varpi_{J_{123}}) \> 
C_{J_{3}-J_{23}}^{(J_2)}(\varpi_{J_{23}})}}\times 
$$
\beq
{ E^{(J_{12})}_{J_{3}-J_{23}} (\varpi_{J_{23}}) \over 
E^{(J_1)}_{J_{23}-J_{123}} (\varpi_{J_{123}}) \> 
E^{(J_2)}_{J_{3}-J_{23}} (\varpi_{J_{23}})}.     
\label{3.51}
\eeq
Thus, we should have 
\beq
g _{J_2J_3}^{J_{23}}= 
e^{-ih(J_2+J_3-J_{23} )/2} 
\sqrt{ e^{i\pi (J_{23}-J_2-J_3)} 
C_{J_{3}-J_{23}}^{(J_2)}(\varpi_{J_{23}})} \> 
E^{(J_2)}_{J_{3}-J_{23}} (\varpi_{J_{23}}) 
\beta_{J_2J_3}^{J_{23}}, 
\label{3.52}
\eeq
where $\beta$ is a solution of the equation 
\beq
\beta_{J_{12}J_3}^{J_{123}} 
=\beta_{J_2J_3}^{J_{23}} \beta_{J_1J_{23}}^{J_{123}}
\label{3.53}
\eeq
that ensures that it disappears from Eq.\ref{3.51}. In this last
relation, no summation over $J_{23}$ is understood. Since the
left-hand side is independent of $J_1$, this must be true for the
right-hand side, and it follows that, in general,  
$\beta_{L\,J}^{K} $ is independent from  the first lower
index $L$. This shows that the general solution of
Eq.\ref{3.53} is  of the form 
\beq
\beta_J^K=
f(K)/f(J)
\label{3.54}
\eeq
The remaining unknown function $f(I)$ is determined from 
the condition $g _{J_2J_3}^{J_{2}+J_{3}}=1$, and one finally gets
$$   
g _{J_2J_3}^{J_{23}}=
e^{ih(J_{23}-J_2-J_3)/2} 
\sqrt{ e^{i\pi (J_{23}-J_2-J_3)} 
C_{J_{3}-J_{23}}^{(J_2)}(\varpi_{J_{23}})
C_{-J_2}^{(J_2)}(\varpi_{J_2})
\over
C_{-J_3}^{(J_3)}(\varpi_{J_3})
} \> 
\times
$$
\beq
{E^{(J_2)}_{J_{3}-J_{23}} (\varpi_{J_{23}})
E_{-J_2}^{(J_2)}(\varpi_{J_2})
\over
E_{-J_3}^{(J_3)}(\varpi_{J_3})
}.
\label{3.55}
\eeq
It is straightforward, but a bit lengthy, to verify that this is
equivalent to our previous expression Eq.\ref{2.39}.

\section{THE GENERAL (3D) STRUCTURE}
\markboth{4. General (3D) structure}
{4. General (3D) structure} 
In the first subsection we discuss the general structure of the bootstrap 
equations. We shall not give all details, but rather establish 
the connection with
earlier works\cite{P,MR,KR,R} where the $U_q(sl(2))$ quantum group 
structure was discussed in other contexts, and show the generalization 
brought about by the introduction of $\xi$ fields.  Moreover, in the 
second part, we will introduce,  by 
analytic continuation,  a suitable limit of the general scheme spelled out 
in the first part, where  the  $V$ and $\xi$ operator algebras 
coincide.  	  
\subsection{Pictorial representations}
In ref.\cite{KR}, quantum-group diagrams were introduced which 
involve two different ``worlds'': the ``normal'' one and 
the ``shadow'' one.  Adopting  this terminology from now on,    
we are going to verify  that the OPA of the $V$ and $\xi$ fields is 
in exact correspondence with these diagrams,  if the  $\xi$ and $V$  OPE's    
are associated with the normal and shadow worlds respectively. 
At the same time we shall discuss the associated three dimensional 
aspect. For the $V$ fields  it is already known, since it 
 corresponds 
to the quantum-group version of the Regge-calculus approach to the discrete 
three-dimensional gravity\cite{DisGr}
   or to the discussion of ref.\cite{W},  for instance.
This case will serve as an introduction to the novel structure that 
comes  out when $V$ and $\xi$ fields are considered together. 

In the pictorial representations, we omit the $g$ coefficients.
Thus,  we actually make use of the    
operator-algebra  expressed in terms of the $\widetilde V$ fields  defined
by Eq.\ref{2.44}, that is ${\cal P}_{J_{12}}
{\widetilde V}_{J_{2}-J_{12}}^{J_1}
\equiv g_{J_1\, J_2}^{J_{12}}
{\cal P}_{J_{12}}  V_{J_{2}-J_{12}}^{J_1}$, instead of the $V$'s.
Though of great importance for operator-product expansion,
the coupling constants $g$  define a pure gauge for the  polynomial equations
or knot-theory viewpoints.
We could draw other figures including $g$ coefficients,
to show how they cancel out of those equations,
but this  would be cumbersome. The basic fusing and braiding operations   
on the $\widetilde V$ operators  have three equivalent representations 
$$
\epsffile{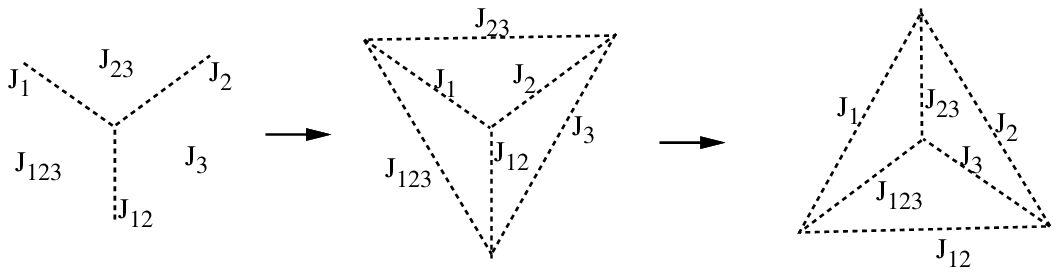}
$$
\beq
=\left\{ ^{J_1}_{J_3}\,^{J_2}_{J_{123}}
\right. \left |^{J_{12}}_{J_{23}}\right\}
\hbox{  (fusion of $\widetilde V$ operators)},
\label{4.1}
\eeq
$$
\epsffile{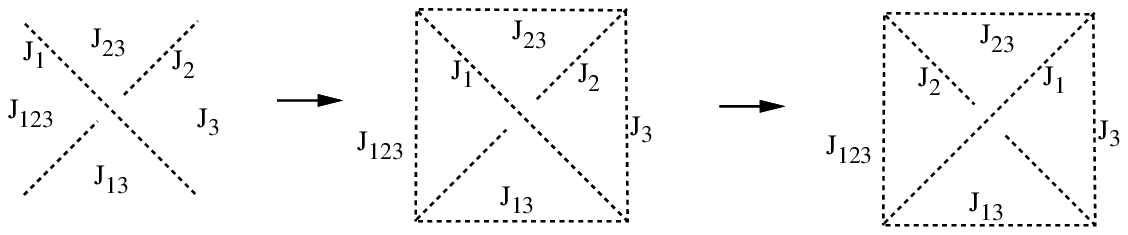}
$$
\beq
= e^{i\pi (\Delta_{J_{123}}+\Delta_{J_3}
-\Delta_{J_{23}}-\Delta_{J_{13}})}
\left\{
^{J_1}_{J_2}\,^{J_3}_{J_{123}}
\right. \left |^{J_{13}}_{J_{23}}\right\}
\hbox{  (braiding of $\widetilde V$ operators)}. 
\label{4.2}
\eeq
First consider the left diagrams, and the associated Eqs.\ref{2.51}, 
and \ref{2.56}. 
Apart from the  $\widetilde V$-matrix element 
on the right-hand side 
of the fusing relation, which has no specific representative, each  
operator  
$\widetilde V^{(J)}_m$ is represented by a dashed line carrying the 
label $J$.    
 The spins on the faces display the zero-modes of the 
Verma modules on which the  $\widetilde V^{(J)}_m$ operators  act. 
Thus the $m$'s are differences between the spin-labels of the two 
neighbouring faces. For the braiding  diagram, the spins on the 
edges are unchanged at crossings, and, for given $J_1$, $J_2$,  the braiding 
diagram has the form of  a  vertex of 
an interaction-around-the-face (IRF) model.   
These diagrams  are two-dimensional (2D). The appearence of spins on the faces 
reflect the fact that    
the  fusion and braiding properties
depend upon the Verma module on which the operators  act. 
It is easily seen that, when they are used as building blocks,   the above 
 drawings  
generate diagrams which 
  have the same structure as the quantum-group  ones of  ref.\cite{KR}
  in the shadow world\footnote{we used 
 dashed lines to agree with the conventions of ref.\cite{KR}.}.  
The polynomial equations can be viewed as  
link-invariance conditions.
For instance, 
\begin{equation}
\epsffile{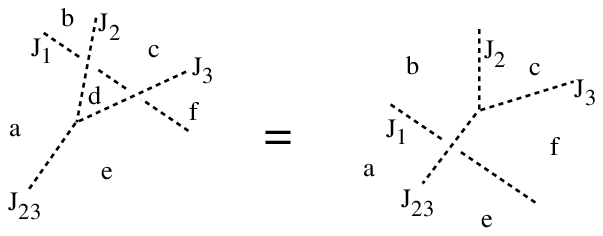}
\label{4.3}
\end{equation}
gives the pentagonal relation
Eq.\ref{2.43} after cancellation of the phases
(with a change of indices).

The middle diagrams of figures \ref{4.1}, and \ref{4.2} 
 are obtained from the left ones  (first arrow) 
by  enclosing the 2D  figures with  extra dashed 
lines  carrying  the spin labels 
which were previously on the faces. In this way, one gets three-dimensional (3D) 
tetrahedra, with spin labels only on the edges. 
The right figures are obtained from the middle ones 
by  dualisation:
the  face, surrounded by the  edges $J_a,J_b,J_c$,
becomes the  vertex where  the  edges $J_a,J_b,J_c$ join,
and conversely, a vertex becomes a face.
An edge joining two vertices becomes    
the  edge between the two dual  faces.  
There is one triangular face
for each $ \widetilde V$ field, including the $\widetilde V$
matrix-element of the
fusing relation Eq.\ref{2.51}.
On the dualised polyhedra,
the triangular inequalities give the addition rules for spins.
The main point of the middle and right diagrams is that, as a consequence 
of  the basic MS 
properties 
of the OPA,     they are really 2D projections of 
 three-dimensional diagrams which may be rotated at  
essentially no cost\footnote{We use
6-j symbols which do not have the full tetrahedral symmetry,
so that two edges should be distinguished.
This will be discussed at the end of this section.}.
For instance,  
 the 
relation Eq.\ref{2.26} between fusing and braiding matrices 
simply corresponds  to the fact that they are 
represented by tetrahedra which may be identified  after  a  
rigid 3D  
rotation.  We shall illustrate the general properties of the 
3D diagrams on the example of the pentagonal relation.  
In the same way as 
 we closed the basic figures in Eq.\ref{4.1}, \ref{4.2},
the rule to go to 3D  is to close the composite figure \ref{4.3}.
It gives a polyhedron
 which  has vertices with three edges only, which we call type  V3E.
The two-dimensional Eq.\ref{4.3} now simply corresponds to 
viewing the V3E  polyhedron  from two different angles: 
\begin{equation}
\epsffile{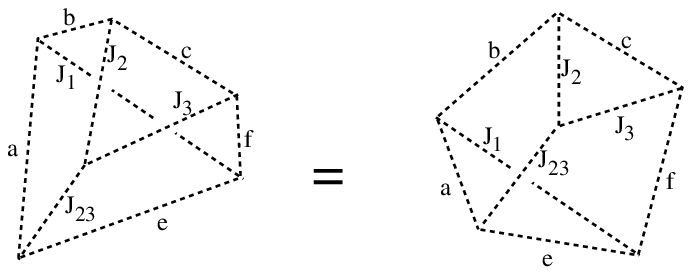}
\end{equation}
Dualisation gives a polyhedron,
with  only triangular faces, which we call F3E. 
The polynomial equations are recovered 
by decomposing a F3E polyhedron  into  tetrahedra   	
(this correspondence only works with  F3E 
polyhedra, this is the reason for  dualisation).
In parallel with the  two different fusing-braiding 
decompositions of each side of 
Eq.\ref{4.3},   
there are  two  3D  decompositions of the F3E  polyhedron. 
This is represented in split view on the next figure, 
where the internal 
faces are  hatched for clarity.                
\begin{equation}
\epsffile{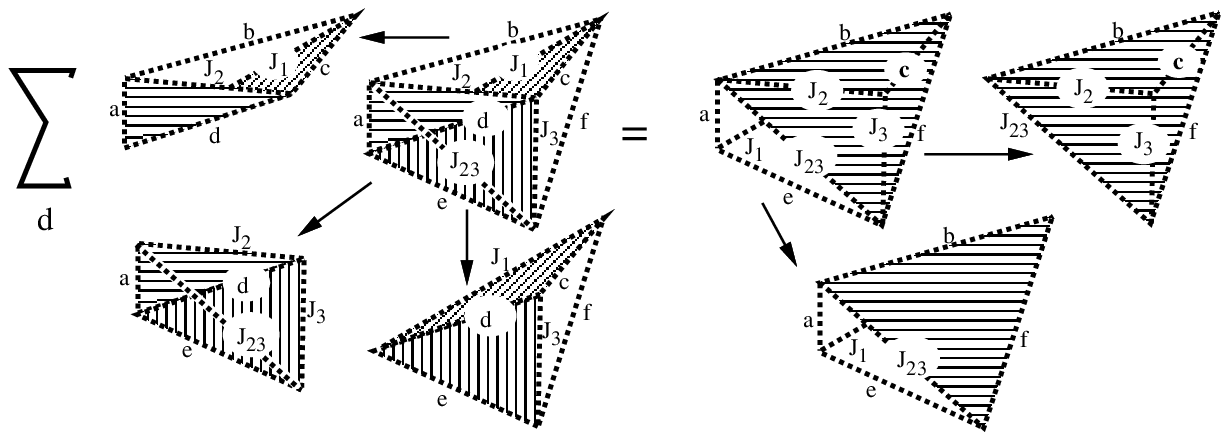}
\label{4.5}
\end{equation}
In general,  the rule is to take a polyhedron with triangular faces
and to decompose it in tetrahedra in different ways.
Substituting the associated 6-j symbols  yields the polynomial 
identities\footnote{we restrict ourselves to 
polyhedra which are orientable surfaces.}.

In quantum-group diagrams\cite{KR}, a second world was introduced -- the 
normal one -- which is represented by solid lines.                
In this world, the quantum numbers
are spins $J$  and  magnetic numbers $M$ both on the lines. 
The label $M$  changes  at the crossings.
We now show how our $\xi$ operator algebra is a realisation
of this normal world. Indeed, the fusing and braiding matrices  of the 
fields $\xi_M^{(J)}$ do not depend upon the Verma module on which 
they act, so that it is consistent that 
the quantum numbers $J$ and $M$  be attached to the 
corresponding line. Comparing Eqs.\ref{3.14}, and \ref{3.21} with the 
formulae given in ref.\cite{KR}, one sees that one has the following 2D 
representation\footnote{For  Eq.\ref{3.14}, the factor $g_{J_1 J_2}^{J_{12}}$ 
is absorbed by going from $V$ to $\widetilde V$ matrix element 
on the right-hand side.} 
\begin{equation}
\begin{array}{c}
\epsffile{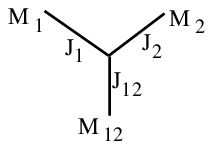}
\end{array}
=(J_1,M_1;J_2,M_2\vert J_{12})
\hbox{     =  fusion of $\xi$ operators}, 
\label{4.6}
\end{equation}
\begin{equation}
\begin{array}{c}
\epsffile{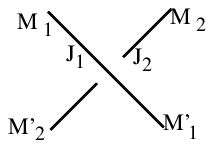}
\end{array}
=
(J_1,J_2)_{M_1\, M_2}^{M'_2\, M'_1}
\hbox{      =   braiding of $\xi$ operators}, 
\label{4.7}
\end{equation}
which coincides with the corresponding quantum-group vertices of ref.\cite{KR}. 
Clearly, the braiding diagram Eq.\ref{4.7} should be regarded as   an
interaction  of  
the vertex type.  
In the same way as for the $\widetilde V$ fields, the fusing vertex has only 
three legs so that the $\widetilde V$ matrix element is not represented per se. 
On the other hand, we have shown, using repeated fusions 
(Eqs.\ref{3.16}-\ref{3.26}) that the fusing equation also provides the 
transitions between $\widetilde V$ and $\xi$ operator-product
algebras. From this viewpoint, the result of the 
discussion just recalled may be pictorially   represented  by introducing 
transitions  between the two worlds based on the graph 
\begin{equation}
\begin{array}{c}
\epsffile{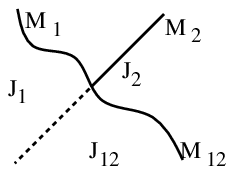}
\end{array}
=(J_1,M_1;J_2,M_2\vert J_{12})
\hbox{     (from normal to shadow world)},
\label{4.8}
\end{equation}
which coincides with the one introduced in ref.\cite{KR}. 
 
The polynomial equations involving $\widetilde V$, and/or $\xi$ fields are 
summarized by the link-invariance of the diagrams constructed out of the 
building bocks just given. For instance, the example previously given for the 
$\widetilde V$ operators becomes, in the normal world, 
\beq
\epsffile{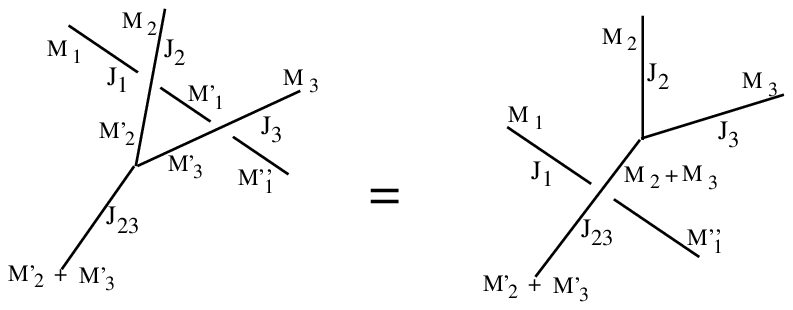} 
\label{4.9}
\eeq
Let us establish  a three-dimensional representation  involving 
$\xi$ operators. We propose the following  
\begin{equation}
\epsffile{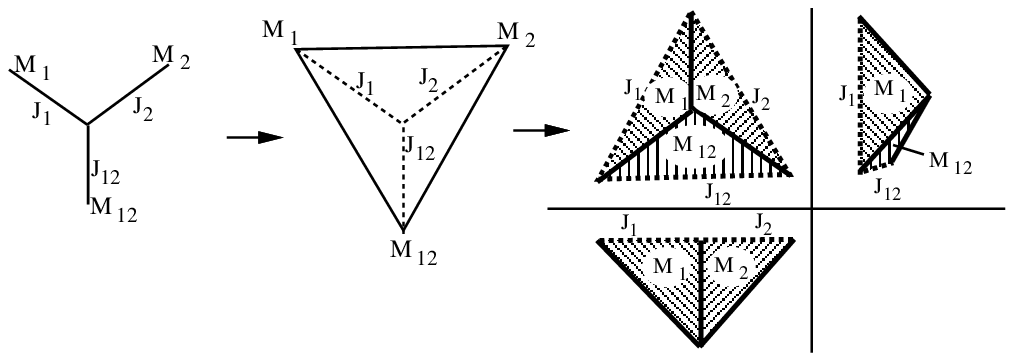} 
\label{4.10}
\end{equation}
\begin{equation}
\epsffile{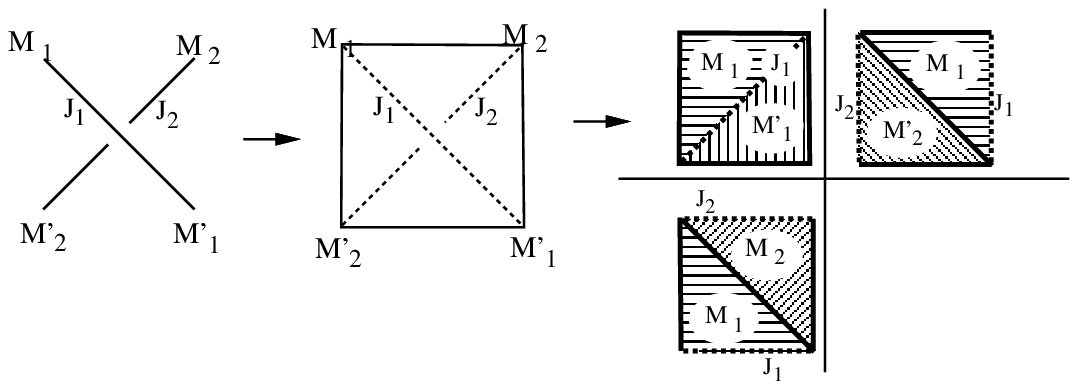}
\label{4.11}
\end{equation}
When the left diagrams are enclosed, we put the $M$'s at the edges. The 
surrounding lines are drawn as solid, while the lines which 
already existed 
become dashed. This ensures  consistency with 
the tetrahedral representation of the $\widetilde V$ operator-product 
algebra  given above,  
since dashed lines have a $J$ label 
in agreement with the previous convention  -- contrary to the solid ones. 
In the dualisation, the dashed lines are transformed as before, while the 
$M$'s naturally go on the faces.  These come out  of two types.
In the fusion, 	there is one  face  which has 
no $M$ and is  surrounded by three
dashed lines. It represents  the $\widetilde V$ field which appears
in Eq.\ref{3.14}.  All other faces   in the two above diagrams 
are similar: 
 surrounded by two solid  lines
with no label, and  a dashed line with a $J$ label,
they  carry the corresponding magnetic number  
 $M$.  Each of them represents a $\xi$ field. 
In the dualised diagrams,  the face associated with $\widetilde V$ 
fields is  drawn as transparent, while the $\xi$ faces  are dashed.  
For them we also give a top-view drawn on the lower-left quarter plane, 
and a left-view drawn on the upper-right quarter plane. 
The tetrahedral representation of the diagram Eq.\ref{4.8} is chosen so that 
its  3D representations agrees with the one of the fusion up to a 
rotation since they are given by the same 3-j symbols. Thus we let   
\begin{equation}
\epsffile{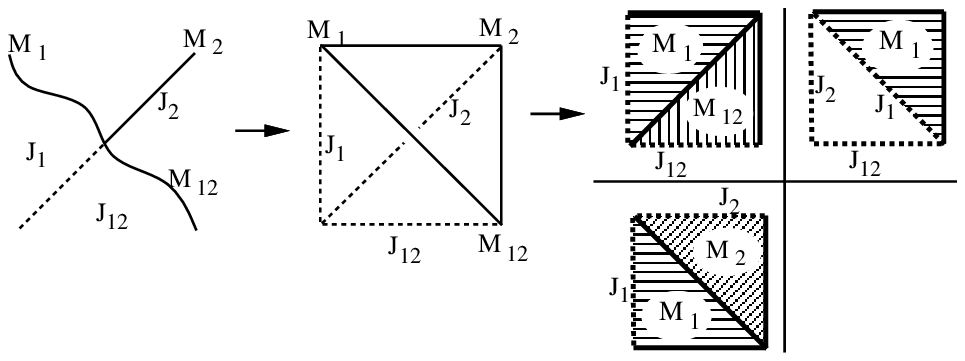}
\label{4.12}
\end{equation}

The 3D aspect now summarizes the general polynomial equations. 
The relation displayed by the figure \ref{4.9} is transformed into
\beq
\epsffile{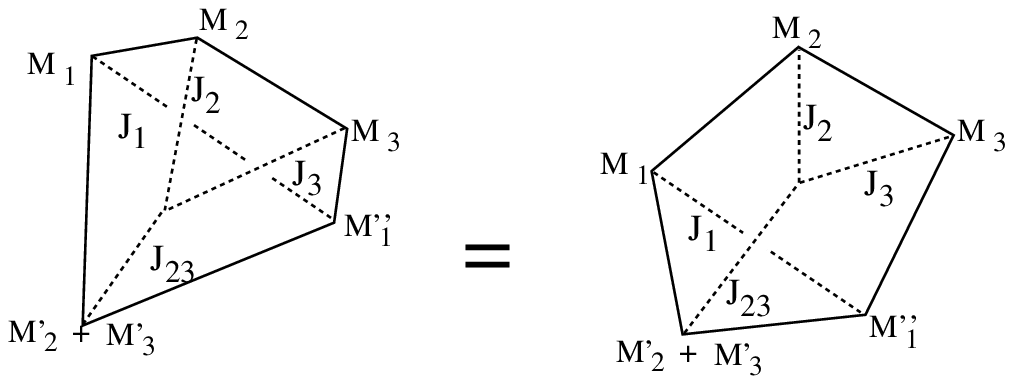}
\eeq
The dualised polyhedron is
\beq
\epsffile{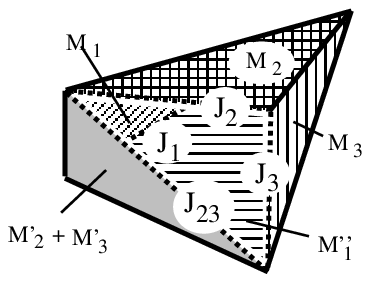}
\eeq
where all the faces are dashed  except
the face $J_2,J_3,J_{23}$, of the $\widetilde V$ type, which is transparent, 
according with the general convention.
We next draw its  decomposition, altering the general convention for   
visibility: the external quantum number are not written, and  
some external faces are made transparent
\beq
\epsffile{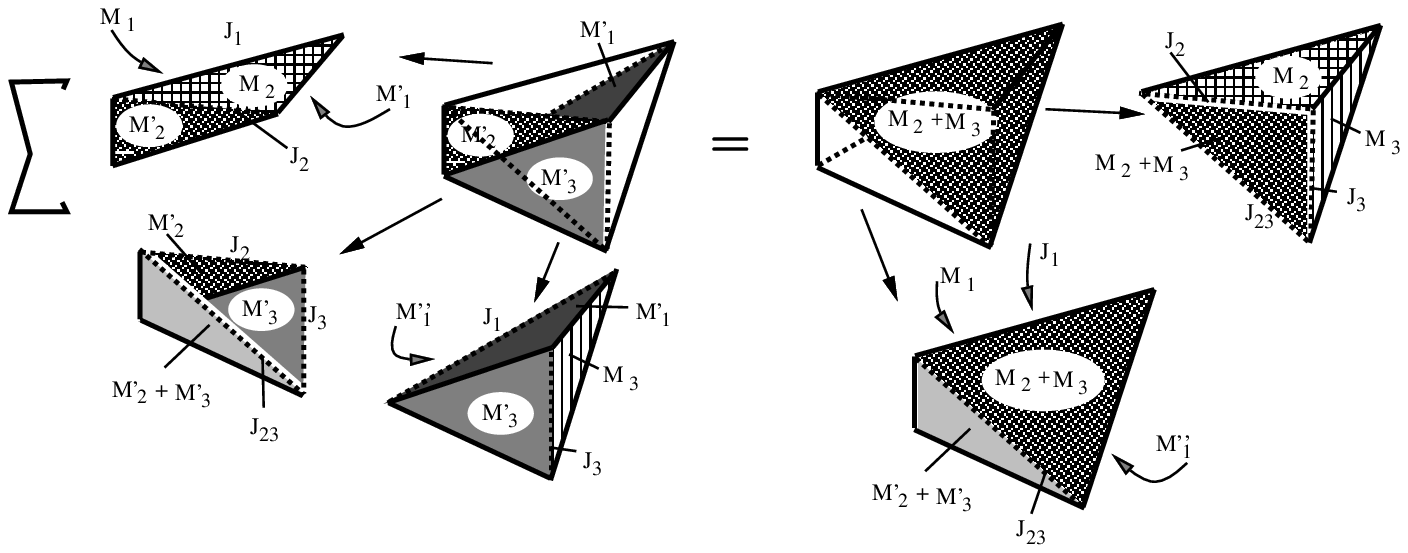} 
\label{4.15}
\eeq
Consider in general a higher 2D diagram with one  separation between a shadow
and a real part. The enclosure proceeds as follows. In each world the 
rule is as indicated above. Concerning the separation line, one follows the 
prescription suggested by figure \ref{4.12}, namely, the separation line 
becomes solid, and thus carries no label.   As a result, the higher 
3D diagram before dualisation is again of the V3E type, with    any number of 
dashed lines, and   one closed loop of solid lines. Thus there are only two types 
of vertices:  with three dashed lines, or  
  with one dashed and two solid  lines. After dualisation, one may obtain    
 any polyhedron of the F3E type with  
the two kinds of faces introduced above: faces of the $\widetilde V$ type 
(three dashed lines,  each with a   $J$,    around a -- transparent -- face), 
and  faces of the $\xi$ type 
(two solid  lines with no number,  a dashed line with a $J$,  and
an $M$ on the face -- which is  dashed).
A $J$ has to be interpreted as the length of the corresponding edge,
and an $M$ as the difference of length of the two surrounding solid edges.
This gives all the spin addition rules and relations between  $M$'s  as
triangular inequalities.
Of course, the $J$ value  of a dashed line is common to the two adjacent faces. 
The polynomial equations are derived by 
 splitting a general F3E polyhedron in tetrahedra.
Since there are two types of faces,  
 one can only obtain three types of tetrahedra.
There is a first type of  tetrahedron, with
three  $\widetilde V$   and one $\xi$  faces,
(see figure \ref{4.10} or \ref{4.12}),
its value is the corresponding Clebsch-Gordan coefficient. 
The tetrahedra of the second type  have  
four $\xi$ faces,
(see figure \ref{4.11}),
the  values are the corresponding R-matrix elements.
The third type tetrahedron has 
four $\widetilde V $  faces
(see figure \ref{4.1} or \ref{4.2}),
its value is the corresponding 6-j coefficient.
In the case of the third type tetrahedron, changing 
the orientation yields an extra phase as we already mentioned. 

For completeness,
we have to add that the
6-j, R-matrix or Clebsch-Gordan
have less symmetries than the tetrahedra by
which they are represented, and therefore that
these symmetries
must be broken
by adding extra caracteristics to the tetrahedra,
so that the correspondence be one-to-one.
We give them briefly.
First,
the faces of the tetrahedra must be
either 'incoming' or 'outgoing'.
Each tetrahedron has two incoming faces
(the faces $J_1$, $J_{23}$, $J_{123}$ and
$J_2$, $J_3$, $J_{23}$ in figures \ref{4.1} and \ref{4.2},
the faces
$M_1$ and $M_2$ in figures \ref{4.10}, \ref{4.11} and \ref{4.12})
and two outgoing ones (the two other ones).
When splitting a composite polyhedron in tetrahedra,
the internal faces, common to two tetrahedra,
are outgoing for
one tetrahedron and incoming for the other one.
This rule allows to single out the two particular $J$
of the non-symmetric non-RW 6-j (the $J$ between the
two incoming faces, and the $J$ between the two outgoing ones),
to distinguish $M_1$ and $M_2$ from $M_{12}$ for the C-G,
and to distinguish between the $M_i$'s and $M'_i$'s
of the R-matrix.
It is clear that the incoming faces represent the operators
on which the fusing or braiding are performed,
and the outgoing faces the resulting operators.
Secondly,
the $M_i$'s on the faces are oriented quantities.
Like the $m_i$'s,
they should be considered as differences of lengths
of the two solid lines surrounding the face $M_i$,
which thus must be supplemented by an ordering.
This ordering must always be from left to right (for instance)
on the 2D projection.
If we rotate the tetrahedron of figure \ref{4.11}
representing $(J_1,J_2)_{M_1\, M_2}^{M'_2\, M'_1}$,
by $\pi$ around a vertical axis in the plane of the page,
this exchanges 1 and 2, but the orderings as well.
The signs of the $M_i$'s must therefore be changed to restore the
left-right ordering of the 2D projection.
The resulting value is then  $(J_2,J_1)_{-M_2\, -M_1}^{-M'_1\, -M'_2}$,
which is indeed equal to  $(J_1,J_2)_{M_1\, M_2}^{M'_2\, M'_1}$.
So, these tetrahedra with two outgoing and two incoming faces,
and oriented $M_i$'s,
are in one-to-one correspondence with
the 6-j, R-matrix or Clebsch-Gordan coefficients.

\subsection{$\xi$ as limit of ${\widetilde V}$}

In part 4.1, we showed that the $V$ braiding diagrams could
be interpreted as IRF model vertices and the $\xi$ braiding
diagrams as
vertex model interactions.
Witten showed that one could get vertex models
as limit of IRF models\cite{W},
when letting the spins on the faces go to infinity
with fixed differences.
In this part we shall apply this method directly to our operators.
The spins on the faces are the ones corresponding to
zero-modes.
We shall denote by $I_\varpi$ the left-most one (it is defined
by $\varpi\equiv\varpi_0+2I_\varpi$),
and let $\varpi$ go to infinity.
Hence, getting a vertex model as limit
of an IRF model is equivalent to obtaining $\xi$ operators
as limits of ${\widetilde V}$ operators.
More precisely, let us prove that
\beq
\lim_{\varpi\to\infty}
{\cal P}_{I_\varpi}\
{\widetilde V}^{(J)}_m\,
/\,\beta_{I_\varpi+m}^{I_\varpi}
=
(2i)^m
e^{-ihm}
\xi^{(J)}_m
\label{4.16}
\eeq
The role of the $\beta$ coefficients
is to remove the $\Gamma$ functions
which would have no well defined limit.
They will cancel out of the braiding or fusing
identities thanks to their property Eq.\ref{3.53}.
${\cal P}_{I_\varpi}$ is the projector on the Verma module
of spin $I_\varpi$ defined in Eq.\ref{2.23}.
Moreover,
we have to give an imaginary part to $I_\varpi$ (or $\varpi$)
so that the limit  $e^{\pm ih\varpi}$ be well defined:
we choose a negative imaginary part,
which makes $e^{-ih\varpi}$ go to zero and
$e^{ih\varpi}$ to infinity.

Using the expression Eq.\ref{3.52} of $g$ and
${\widetilde V}^{(J)}_m=(g^{I_\varpi}_{J\;I_\varpi+m}
/E^{(J)}_m(\varpi))\,\sum\!_{_M}
(J,\varpi|_m^M
\ \xi^{(J)}_M$
where $(J,\varpi|_m^M$ is the inverse matrix
of $|J,\varpi)_M^m$,
Eq.\ref{4.16} is equivalent to
\beq
\lim_{\varpi\to\infty}
{\cal P}_{I_\varpi}\
e^{ih(J+m)/2}
\sqrt{e^{i\pi(J+m)} C_m^{(J)}(\varpi)} \>
\sum_M
(J,\varpi|_m^M
\xi^{(J)}_M
=
(2i)^m
e^{-ihm}
\xi^{(J)}_m.
\label{4.17}
\eeq

Let us prove now that the matrix $|J,\varpi)_M^m$
has a leading term proportional to the identity matrix
when $\varpi\to\infty$,
and consequently,
such is the leading term of its inverse matrix.
Recall Eq.\ref{3.35}
\beqa
&\vert J,\varpi)_M^m\,=
\sqrt{\hbox{$ {2J \choose J+M} $ }} \>e^{ihm/2}\times \nnn
&\sum_{(J-M+m-t)/2 \>\>\hbox{integer}}\, e^{iht(\varpi
+m)}
{J-M \choose (J-M+m-t)/2}\, {J+M \choose (J+M+m+t)/2}.
\label{4.18}
\eeqa
It is a polynomial in $e^{ih\varpi}$.
As $e^{ih\varpi}$ has an infinite limit,
the maximum value of $t$ dominates in the sum.
The boundaries
on t are given by the condition that
the q-factorials arguments be positive.
Thus, we get
\beq
\lim_{\varpi\to\infty}
e^{-ihJ\varpi}
\vert J,\varpi)_M^m\,=
\sqrt{\hbox{$ {2J \choose J+M} $ }}
e^{ihM(J+1/2)}\delta_{m,M}.
\label{4.19}
\eeq
This gives the limit of the inverse matrix directly.
In view of Eq.\ref{4.17}
we also need the limit of $C^{(J)}_m(\varpi)$.
Since  the complex exponential
with positive argument is dominant  in $\sin(h\varpi)$,
we get
\beq
\lim_{\varpi\to\infty}
e^{-ih2J\varpi}
C^{(J)}_m(\varpi)=
(-1)^{-J-m}e^{ihJ}{2J \choose J-m}
e^{ih(2J-1)m},
\eeq
which leads to Eq.\ref{4.16}.
\vskip 5mm

Now, we
take this limit $\varpi\to\infty$
in the braiding or fusing equations.
Begin with the fusing Eq.\ref{2.51}.
Let us write it down in term of three ${\widetilde V}$ and one $V$
$$
{\cal P}_{I_\varpi}\
{\widetilde V}^{(J_1)}_{m_1}
{\widetilde V}^{(J_2)}_{m_2}
=
\sum _{J_{12}= \vert J_1 - J_2 \vert} ^{J_1+J_2}
g_{J_1J_2}^{J_{12}}\
\left\{ ^{\quad J_1}_{I_\varpi+m_1+m_2}\,^{J_2}_{I_\varpi}
\right. \left |^{J_{12}}_{I_\varpi+m_1}\right\}
\times
$$
\beq
\sum _{\{\nu_{12}\}}
{\cal P}_{I_\varpi}\
{\widetilde V}^{(J_{12},\{\nu_{12}\})}_{m_1+m_2}
<\!\varpi_{J_{12}},{\{\nu_{12}\}} \vert
V ^{(J_1)}_{J_2-J_{12}} \vert \varpi_{J_2} \! >.
\label{4.21}
\eeq
The $\beta$ coefficients introduced by Eq.\ref{4.16} cancel out
thanks to the property Eq.\ref{3.53},
and, in the limit, we do get the fusing Eq.\ref{3.14}
of the $\xi$ operators,
provided that
\beq
\lim_{\varpi\to\infty}
\left\{ ^{\quad J_1}_{I_\varpi+m_1+m_2}
\,^{J_2}_{I_\varpi}\right. \left |^{\ J_{12}}_{I_\varpi
+m_1}\right\}
=
(J_1,m_1;J_2,m_2|J_{12}).
\label{4.22}
\eeq

Here again, we have to give an imaginary part to $\varpi$
so that the limits of the complex exponentials be well defined.
But the 6-j coefficients are only defined for positive
half-integer
spins.
So, we have to extend this definition to non integer $I_\varpi$
before going to the limit.
Such was not the case for the limit of
${\cal P}_{I_\varpi}{\widetilde V}^{(J)}_m$
considered above, since
everything was defined for any $\varpi$.
This gives us the condition that our extended
definition of the 6-j must be coherent with the limit of the
${\widetilde V}$, i.e. it must obey Eq.\ref{4.22}.

The expression of the 6-j coefficients
is given in Eq.\ref{2.54}.
The ambiguity of the extension lies in the range of summation.
For half-integer spins the boundaries are given by the condition
that the arguments of the q-factorial in the denominator
must not be negative integers.
For half integer $I_\varpi$ the initial definition
is strictly equivalent to the following one, obtained
by
the change of index
$\mu=z-2I_\varpi-J_{12}-m_1-m_2$
$$
\left\{ ^{\quad J_1}_{I_\varpi+m_1+m_2}
\,^{J_2}_{I_\varpi}\right. \left |^{\ J_{12}}_{I_\varpi
+m_1}\right\}
=
(-1)^{J_{12}-J_1-J_2}
\sqrt{\lfloor 2J_{12}+1 \rfloor
\lfloor 2I_\varpi
+2m_1+1 \rfloor}
\Delta(J_1,J_2,J_{12})\times
$$
$$
\Delta(J_1,I_\varpi,I_\varpi+m_1)
\Delta(I_\varpi+m_1+m_2,I_\varpi,J_{12})
\Delta(I_\varpi+m_1+m_2,J_2,I_\varpi+m_1)
\sum_{\mu\ {\rm integer}}
\! (-1)^\mu\times
$$
$$
\lfloor 2I_\varpi+J_{12}+m_1+m_2+\mu+1 \rfloor \! !
\bigg[
\lfloor 2I_\varpi -J_1-J_2+m_1+m_2+\mu \rfloor \! !
\lfloor J_{12}-J_1+m_2+\mu \rfloor \! !\times
$$
\beq
\lfloor J_{12}-J_2-m_1+\mu \rfloor \! !
\lfloor \mu \rfloor \! !
\lfloor J_1+J_2-J_{12}-\mu \rfloor \! !
\lfloor J_1+m_1-\mu \rfloor \! !
\lfloor J_2-m_2-\mu \rfloor \! !
\bigg]^{-1}.
\label{4.23}
\eeq

But, when we give an imaginary part to $I_\varpi$,
the q-factorials with a $I_\varpi$
have complex arguments and
yield no restriction. .
Such is the case of the last six factorials of the sum
with the definition
of Eq.\ref{2.54},
and,  of the first two  with the definition
of Eq.\ref{4.23}.
The first definition rapidly appears inadequate as it
yields no upper boundary for $z$.
The second definition (Eq.\ref{4.23}) leads to well defined
boundaries for the index $\mu$
and to a finite limit thanks to
\beq
{\lfloor  \alpha + 2I_\varpi \rfloor \! ! \over
\lfloor  \beta + 2I_\varpi \rfloor \! !} \sim
(2i \sin h)^{\beta-\alpha}
e^{ih(\alpha-\beta)2I_\varpi}
e^{ih(\alpha-\beta)(\alpha+\beta+1)/2}
\label{4.24}
\eeq
for $\varpi\to\infty$ with negative imaginary part,
similar to Eq.\ref{3.33} which was for positive
imaginary part.

We use this limit in the sum and in the $\Delta$ prefactors
of Eq.\ref{4.23},
and recognize
the expression of the Clebsch-Gordan coefficient
given in Eq.\ref{3.27}.
This justifies our extended definition of the 6-j.

Pictorially\footnote{for a better readability,
we omit the index $\varpi$ of $I_\varpi$
in the drawings}, this reads
\begin{equation}
\epsffile{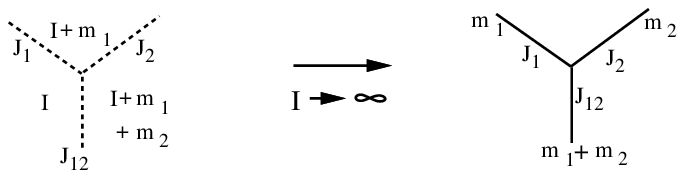}
\end{equation}
\vskip 5mm

We come now to the case of braiding.
It works like fusing.
From Eq.\ref{4.16} and the properties of the
$\beta$ coefficients Eq.\ref{3.53},
we see that we only have to prove
that when $J$ goes to infinity
with negative imaginary part
\beq
\lim_{\varpi\to\infty}
\left\{ ^{J_1}_{J_2}
\,^{I_\varpi+m}_{I_\varpi}\right.
\left |^{I_\varpi+m'_2}_{I_\varpi+m_1}\right\}
e^{i\pi (\Delta_{I_\varpi+m}+\Delta_{I_\varpi}
-\Delta_{I_\varpi+m'_2}-\Delta_{I_\varpi+m_1})}
=
(J_1,J_2)^{m'_1m'_2}_{m_1m_2}
\label{4.26}
\eeq
and
\beq
\lim_{\varpi\to\infty}
\left\{ ^{J_1}_{J_2}
\,^{I_\varpi+m}_{I_\varpi}\right.
\left |^{I_\varpi+m'_2}_{I_\varpi+m_1}\right\}
e^{-i\pi (\Delta_{I_\varpi+m}+\Delta_{I_\varpi}
-\Delta_{I_\varpi+m'_2}-\Delta_{I_\varpi+m_1})}
=
\overline{(J_1,J_2)}^{m'_1m'_2}_{m_1m_2}
\label{4.27}
\eeq
where $m\equiv m_1+m_2=m'_1+m'_2$ and
$(J_1,J_2)$ (resp.$\overline{(J_1,J_2)}$)
is the universal R-matrix $R$ (resp.$\overline R$),
following the conventions of ref\cite{G1} (see Eq.\ref{3.22}).
We only deal in details with the case of
the R-matrix $R$.
We compute its matrix element from its universal
form given in Eq.\ref{3.23}
$$
(J_1,J_2)^{m'_2m'_1}_{m_1m_2}
=
e^{-2ihm_1m_2}
{(-1)^n
(2i\sin(h))^n
e^{ihn(n+1)}
\over
\lfloor n \rfloor \! !}
e^{-ihnm_1}
e^{ihnm_2}\times
$$
\beq
\sqrt{
\lfloor J_1+m_1 \rfloor \! !
\lfloor J_1-m'_1 \rfloor \! !
\lfloor J_2-m_2 \rfloor \! !
\lfloor J_2+m'_2 \rfloor \! !
\over
\lfloor J_1-m_1 \rfloor \! !
\lfloor J_1+m'_1 \rfloor \! !
\lfloor J_2+m_2 \rfloor \! !
\lfloor J_2-m'_2 \rfloor \! !},
\label{4.28}
\eeq
for $n\equiv m'_2-m_2\ge 0$, and $0$ otherwise.
It is an upper triangular matrix.

To begin with, we examine the behavior of the
6-j coefficient when $\varpi$ goes to
infinity with a negative imaginary part.
This time, the suitable change of summation index
to define the 6-j for non-integer spins is
$x=2I_\varpi+J_1+J_2+m_1+m_2-z$
(basically, this is the same definition of the extension
as before,
which amounts to taking $z-2I_\varpi$ integer,
the extra integer shifts by $J_i+m_i$ being trivial).
Here, only the first and eighth factorials of the sum have
infinite arguments.
After using Eq.\ref{4.24}, we get
$$
\left\{ ^{J_1}_{J_2}
\,^{I_\varpi+m}_{\ I_\varpi}\left.
\right|^{I_\varpi+m'_2}_{I_\varpi+m_1}\right\}
\sim
e^{-ih2I_\varpi n}
e^{ih(m_1+m'_2+1)}
\sqrt{
\lfloor J_1+m_1 \rfloor \! !
\lfloor J_1-m_1 \rfloor \! !
\lfloor J_1+m'_1 \rfloor \! !
\lfloor J_1-m'_1 \rfloor \! !}
$$
$$
\times
\sqrt{
\lfloor J_2+m_2 \rfloor \! !
\lfloor J_2-m_2 \rfloor \! !
\lfloor J_2+m'_2 \rfloor \! !
\lfloor J_2-m'_2 \rfloor \! !}
(2i\sin(h))^n
e^{-ih/2(1+n)(m+m_1+m'_2+2)}\times
$$
\beq
\sum_{x\ {\rm integer}}
{e^{-4ihI_\varpi x}
(2i\sin(h))^{2x}
e^{-ihx(m+m_1+M'_2+2)}
(-1)^x
\over
\lfloor J_1-m_1-x \rfloor \! !
\lfloor J_1+m'_1-x \rfloor \! !
\lfloor J_2+m_2-x \rfloor \! !
\lfloor J_2-m'_2-x \rfloor \! !
\lfloor x \rfloor \! !
\lfloor n+x \rfloor \! !}.
\eeq
As $e^{-ihI_\varpi}\to 0$,
due to the factor $e^{-4ihI_\varpi x}$
the term of the sum with
$x$ minimal
will be dominant.
The lower boundary
for $x$ is given by
$\lfloor x \rfloor \! !$ and
$\lfloor n+x \rfloor \! !$:
for $n\ge 0$, the minimal allowed $x$ is 0, and
for $n\le 0$, it is $n$.
Hence, summarizing both cases,
the dominant behavior of the 6-j
is $e^{-ih2I_\varpi |n|}$.

Then, the extra factor
due to
$e^{\pm i\pi (\Delta_{I_\varpi+m}+\Delta_{I_\varpi}
-\Delta_{I_\varpi+m'_2}-\Delta_{I_\varpi+m_1})}$,
which is $e^{\pm ih2I_\varpi n}\times$ (finite term),
must be taken into account.
Altogether, it gives $e^{ih2I_\varpi(\pm n-|n|)}$,
i.e. $e^{-ih4I_\varpi|n|}$ when $n\le 0$,
and $1$ when $n\ge 0$, in the upper case$(+)$.
So, when $e^{-ih2I_\varpi} \to 0$, the limit is an upper
triangular matrix, as it should
(In the lower case$(-)$ it gives $1$ when $n\le 0$,
and $e^{-ih4I_\varpi|n|}$ when $n\ge 0$ and so, a lower
triangular matrix in the limit).

It is then straightforward to check that
the limit agrees with
Eq.\ref{4.28},
and this terminates the proof of
Eq.\ref{4.26}.

The pictorial representation of this is
\begin{equation}
\epsffile{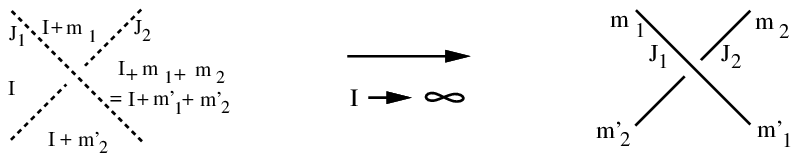}
\end{equation}
\vskip 5mm

We can use these limits on more complex braiding-fusing
identities of $\widetilde V$ fields,
thereby proving the same identities for $\xi$ fields.
For instance, we have the following limit
\begin{equation}
\epsffile{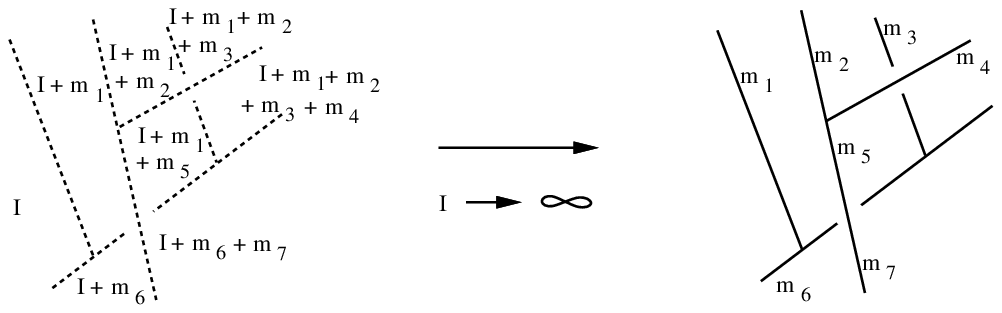}
\label{4.31}
\end{equation}
where we did not write the spins $J_i$ on the lines as they are not affected
by the limit,
and where the missing $m_i$ on the r.h.s can be deduced from
conservation of $\sum_im_i$ at each vertex.
Making this limit on both sides of identities in the shadow world
yields idendities in the normal world.

We can as well make a limit only on a part of the figure.
Eq.\ref{4.22} gives as well
\begin{equation}
{e^{-i\pi ((\Delta_{I+m_1+m_2}-\Delta_{J_{12}})}\over
e^{-i\pi
(\Delta_{I+m_1}-\Delta_{J_1}))}}\times\!
\begin{array}{c}
\epsffile{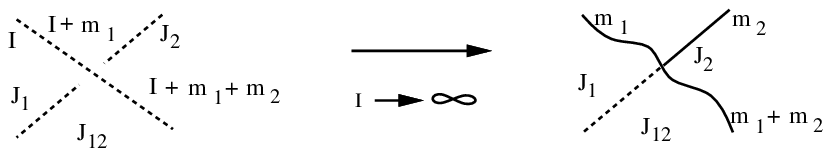}
\end{array}
\label{4.32}
\end{equation}
which gives on a composite figure
\begin{equation}
{e^{-i\pi (\Delta_{I+m_1+m_2+m_3+m_4}-\Delta_{J_c})}\over
e^{-i\pi (\Delta_{I+m_1}-\Delta_{J_a})}}
\times\!\!\!\!\!
\begin{array}{c}
\epsffile{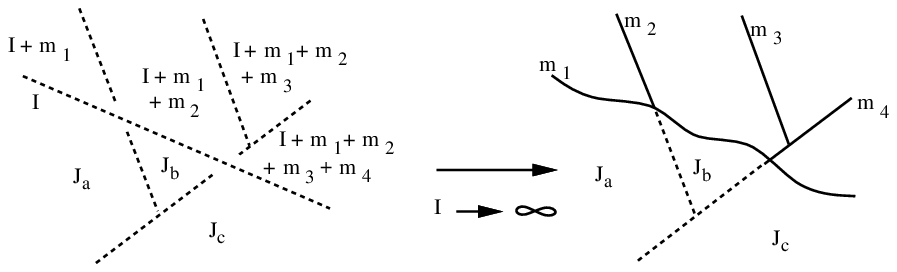}
\end{array}
\label{4.33}
\end{equation}
The extra phases in Eq.\ref{4.32} cancel out between two neighbouring
basic figures all along the wavy line,
and only the exterior ones remain as in Eq.\ref{4.33}.
Hence, they do not spoil equalities between transformed figures
since the exterior spins of both sides are the same.

To conclude with this part,
we show how this limit can be understood in the case of the 3D
representation by V3E polyhedra.
As an example,
we close the 2D figures \ref{4.31} and \ref{4.33} and get
the 3D representation
\begin{equation}
\epsffile{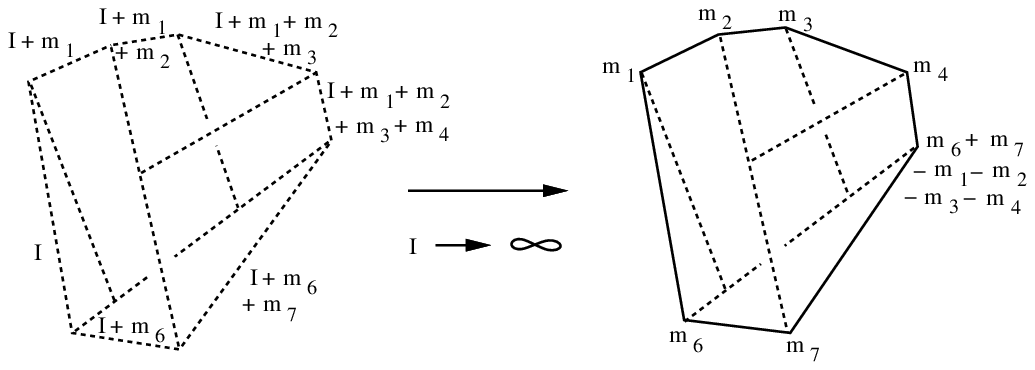}
\label{4.34}
\end{equation}
and
\begin{equation}
{e^{-i\pi (\Delta_{I+m_1+m_2+m_3+m_4}-\Delta_{J_c})}\over
e^{-i\pi (\Delta_{I+m_1}-\Delta_{J_a})}}
\times
\!\!\!\!\!
\begin{array}{c}
\epsffile{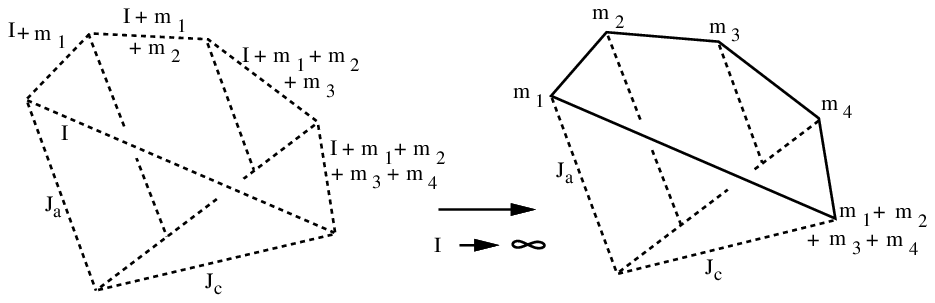}
\end{array}
\label{4.35}
\end{equation}
where again the spins on the internal lines, which are unchanged
by the limit have not been written down.

We see on these examples that the partial or global limits are not
fundamentally different.
On the V3E polyhedra, the normal world is obtained from the
shadow world by adding $I$ to all the edges of a
arbitrarily chosen closed loop,
and then letting $I$ go to infinity.
In this limit, only the differences between the values on the edges
remain finite and relevant; they go on the vertices.
When projecting in two dimensions,
the part of the drawing inside the projection of the closed
loop naturally becomes the normal world,
as the part outside becomes the shadow world.
Hence, depending on whether the closed loop is at the exterior
of the drawing or not,
there is only a normal world (3D in figure \ref{4.34}, 2D in
figure \ref{4.31})
or both worlds (3D in figure \ref{4.35}, 2D in figure \ref{4.33}),
but fundamentally, this is not different.

Following the remark of the end of part 4.1,
we note that we need to orient the $M_i$'s
at the vertices of the V3E polyhedra
(and then on the faces of the F3E polyhedra)
in order to know how to order the difference of the two
neighbouring spins in the infinite limit.

\section{  FUSION RULES FOR $\xi_{M\, \Mhat }^{(J\, \Jhat )}$
OPERATORS}
\markboth{5. fusion rules for general $\xi$ 
operators}
{5. fusion rules for general $\xi$  
operators}

In the previous sections, we have
been able to determine the coupling constants of
 the fusion rules for
the operators connected with the quantum group parameter $h$,
by assuming that  the general scheme of Moore and Seiberg
holds.
In this section, we deal with the most general operators
$V_{M\, \Mhat }^{(J\, \Jhat )}$, and $\xi_{M\, \Mhat }^{(J\, \Jhat )}$
that involve the two quantum group parameters $h$ and $\hhat$.
It is straightforward  to extend the full discussion
just given to this case,
 since the fusing and braiding matrices of the
$V_{M}^{(J)}$ (or equivalently of the $\xi_{M}^{(J)}$) with the
$V_{\Mhat }^{(\Jhat )}$ (or equivalently with  the
$\xi_{\Mhat }^{( \Jhat )}$) operators are simple phases.
We shall not do it explicitly to keep the length of this paper
within reasonable limits. We shall only determine the coupling
constants by using a short cut that only assumes that
the short-distance expansion of the
product of two fields $\xi$ do not depend on $\varpi$,  and
computes its matrix element between  the state  $|\varpi_0>$
and the appropriate highest-weight-state bra, in two equivalent
ways. In order to
illustrate the method, let us  first rederive the expression Eq.\ref{2.39}
for $g_{J_1J_2}^{J_{12}}$. We shall later on extend the method to the
fusion of the
$\xi_{M \Mhat}^{(J \Jhat )}$ fields.

   Let us compute $<\varpi_{J_3} \vert \xi_{M_1}^{(J_1)}(z) 
\xi_{M_2}^{(J_2)}(0) \vert \varpi_0 >$ in two different ways

(1) directly using the relation between $ \xi_M^{(J)} $ and
$V_M^{(J)}$ as well as equation (2.10)
$$<\varpi_{J_3} \vert \xi_{M_1}^{(J_1)}(z)
\xi_{M_2}^{(J_2)}(0) \vert \varpi_0 > =$$
\beq
E_{{J_2}-{J_3}}^{(J_1)}(\varpi_{J_3}) 
E_{-{J_2}}^{(J_2)}(\varpi_{J_2}) \vert J_1 ,\varpi_{J_3})_{M_1}^{{J_2}-{J_3}}
\vert J_2 ,\varpi_{J_2})_{M_2}^{-{J_2}}\>
z^{( \Delta _3 - \Delta _1 -\Delta _2 )};
\label{5.1}
\eeq

(2) using first the short distance expansion
\beq
 \xi_{M_1}^{(J_1)}(z) \xi_{M_2}^{(J_2)}(0) \simeq  
 \sum_{J_{12}} z^{{\Delta_{12}} - {\Delta_1} -{\Delta_2}}
[G_{M_1 M_2 {{M_1}+{M_2}}}^{J_1 \>  J_2 \quad  J_{12}} \  \xi_{{M_1}+{M_2}}^
{J_{12}}(0) +\cdots ] 
\label{5.2}
\eeq
where G is to be determined. Then
$$<\varpi_{J_3} \vert \xi_{M_1}^{(J_1)}(z) 
\xi_{M_2}^{(J_2)}(0) \vert \varpi_0 > =$$
\beq
G_{M_1 M_2 {{M_1}+{M_2}}}^{J_1 \>  J_2 \quad  J_3}\ 
 E_{-{J_3}}^{(J_3)}(\varpi_{J_3})\vert J_3 ,\varpi_{J_3})_{{M_1}+{M_2}}^{-{J_3}}
 \> z^{( \Delta _3 - \Delta _1 -\Delta _2 )} +\cdots .
\label{5.3} 
\eeq
One gets immediately
\beq
G_{M_1 M_2 {{M_1}+{M_2}}}^{J_1 \>  J_2 \quad  J_3} ={{
{E_{{J_2}-{J_3}}^{(J_1)}(\varpi_{J_3}) 
E_{-{J_2}}^{(J_2)}(\varpi_{J_2})}
\over{E_{-{J_3}}^{(J_3)}(\varpi_{J_3})}} }{                  
{\vert J_1 ,\varpi_{J_3})_{M_1}^{{J_2}-{J_3}} 
\vert J_2 ,\varpi_{J_2})_{M_2}^{-{J_2}}}\over
{\vert J_3 ,\varpi_{J_3})_{{M_1}+{M_2}}^{-{J_3}}}} 
\label{5.4}
\eeq
Applying  Eq.\ref{3.46}   with $m_{12}=-J_3 ,
\varpi =\varpi_{J_3} $ and noting that due to triangular
    inequalities the non symmetric 
6-j coefficient
$\left\{ ^{J_2}_{J_3}\,^{J_1}_0
\right. \left |^{\ J_3}_{J_3+m_1}\right\}$
is different from 0 only for $ m_1={J_2} -{J_3} $ we get
$${{\vert J_1 ,\varpi_{J_3})_{M_1}^{{J_2}-{J_3}} 
\vert J_2 ,\varpi_{J_2})_{M_2}^{-{J_2}}}\over{\vert J_3
,\varpi_{J_3})_{{M_1}+{M_2}}^{-{J_3}}}}=
e^{[ih({J_3}-{J_1}-{J_2})+i\pi ({J_3}-{J_1}-{J_2})]/2} \times  $$
\beq
\sqrt
{C_{{J_2}-{J_3}}^{(J_1)}(\varpi _{J_3})C_
{-{J_2}}^{(J_2)} (\varpi _{J_2})\over C_{-{J_3}}^{(J_3)} 
(\varpi_{J_3})}\> 
( J_1 M_1 J_2 M_2 \vert J_3)  
\label{5.5}
\eeq
As expected, this gives
\beq
G_{M_1 M_2 {{M_1}+{M_2}}}^{J_1 \>  J_2 \quad  J_{12}} =
 g_{{J_1}{J_2}}^{J_{12}}\>  (J_1 M_1 J_2 M_2 \vert J_{12}) 
\label{5.6}
\eeq
where  $ g_{{J_1}{J_2}}^{J_{12}} $ is given by Eq.\ref{3.55}.
It is worth to note that, instead of using the short distance 
expansion Eq.\ref{5.2}, it is consistent to use the exact Eq.\ref{3.14}
and we get the same result due to the fact that
$<\varpi _{J_3} \vert V_{M_1+M_2}^{(J_{12} ,\nu )} (0) 
\vert \varpi _0 > $ is different from 0 only for the highest 
weight operator $V_{M_1+M_2}^{(J_{12} ) } $ ($\nu =0$).

The same method applies exactly to the fusion of the
$\xi_{M \Mhat}^{(J \Jhat)}$ fields using now\cite{G1}
\beq
\xi_{M \Mhat}^{(J \Jhat)} =e^{4i\pi J\Jhat}
e^{-i\pi(M\Jhat +\Mhat J) }
\sum_{m,\hat m } \vert J,\varpi)_M^m
\hat {\vert} \Jhat,\widehat {\varpi} \hat ) _{\Mhat}^{\widehat m}
E^{(J \Jhat)}_{m \widehat m}(\varpi ) \>V^{(J \Jhat)}_{m \widehat m} 
\label{5.7}
\eeq
and
\beq
\vert J,\varpi +2\widehat {\ell} {\pi \over h} )^m_M =
e^{2i \pi \widehat {\ell}  m} \> e^{-2i \pi \widehat{\ell} (J-M)}
\vert J,\varpi )^m_M .
\label{5.8}
\eeq
The short distance expansion of the product of two
$\xi_{M \Mhat}^{J \Jhat}$ fields is written as
$$
\xi_{M_1  \Mhat_1 }^{(J_1  \Jhat_1) } (z)
  \xi_{M_2  \Mhat_2 }^{(J_2  \Jhat_2) } (0)
\simeq \sum_{J_3,\Jhat_3 }
z^{\Delta_{J_3\Jhat_3} -\Delta_{J_1\Jhat_1}-\Delta_{J_2\Jhat_2}}
\times
$$
\beq
[G^{J_1 \> \Jhat_1 \> J_2 \> \Jhat_2 \quad J_3 \quad \Jhat_3 \quad}
   _{M_1  \Mhat_1 M_2 \Mhat_2 M_1+M_2   \Mhat_1 +\Mhat_2 }
\xi^{(J_3 \quad \Jhat_3) \quad}_{M_1+M_2   \Mhat_1 +\Mhat_2 }
(0) +\cdots ]  .
\label{5.9}
\eeq
The final result reads
$$G^{J_1 \>  \Jhat_1 \> J_2 \> \Jhat_2 \quad J_3 \quad \Jhat_3 \quad} 
   _{M_1  \Mhat_1 M_2 \Mhat_2 M_1+M_2   \Mhat_1 +\Mhat_2 }
=g^{J_3  \Jhat_3 \quad }_{J_1 \Jhat_1 J_2 \Jhat_2}
(J_1 M_1 J_2 M_2 \vert J_3)
\hat ( \Jhat_1 \Mhat_1 \Jhat_2 \Mhat_2 \hat {\vert} \Jhat_3 \hat )\times
$$
\beq
e^{i \pi [M_1 \Jhat_2 -M_2 \Jhat_1 +\Mhat_1 J_2 -\Mhat_2 J_1
+(M_1 -M_2 )(\Jhat_3 -\Jhat_1 -\Jhat_2 )
+(\Mhat_1 -\Mhat_2 )(J_3 -J_1 -J_2 )]}
\label{5.10}
\eeq
with
\beq
g^{J_3  \Jhat_3  }_{J_1 \Jhat_1 J_2 \Jhat_2}
=g^{J_3}_{J_1 J_2} \widehat g^{\Jhat_3}_{\Jhat_1 \Jhat_2}
 h^{J_3  \Jhat_3  }_{J_1 \Jhat_1 J_2 \Jhat_2}
\label{5.11}
\eeq
where a careful calculation gives
\beq
h^{J_3  \Jhat_3  }_{J_1 \Jhat_1 J_2 \Jhat_2}=
(-1)^{p\widehat p +2p\Jhat_3 +2\widehat p J_3}
\prod_k \prod_{r,\widehat r \in X_k}
(r \sqrt{h/\pi} +\widehat r \sqrt{\pi /h} )^{\epsilon_k}
\label{5.12}
\eeq
where the $X_k's$ are the direct product of intervals
$[a_k,b_k]\otimes [{\widehat a}_k,{\widehat b}_k]$ and given by
         
$$
X_k= \left \{
\begin{array}{c}
\bigl [ 1, 2 J_1 \bigr ] \otimes [1,2\Jhat_1 ]  \\
\bigl [ 1, 2 J_1 -p\bigr ] \otimes [1,2\Jhat_1 -\widehat p ] \\
\bigl [ 1, 2 J_2 \bigr ] \otimes [1,2\Jhat_2 ]  \\
\bigl [ 1, 2 J_2 -p\bigr ] \otimes [1,2\Jhat_2 -\widehat p ] \\
\bigl [ 2, 2 J_3 +1 \bigr ] \otimes [2,2\Jhat_3 +1 ]  \\
\bigl [ 2, 2 J_3 +p+1 \bigr ] \otimes [2,2\Jhat_3 +\widehat p +1 ] \\
 \bigl [ 1,p\bigr ] \otimes \bigl [1,\widehat p \bigr ]
\end{array}
\right.
\>, \quad  \hbox{with} \quad
\epsilon_k =     
\left \{ 
\begin{array}{c} 
+1 \\ 
-1 \\ 
+1 \\
-1 \\
+1 \\
-1 \\
-1
\end{array} \right.
$$
and
\beq
p=J_1 +J_2 -J_3 ,
\widehat p=\Jhat _1 +\Jhat _2 -\Jhat _3 .
\eeq

We note that
G is symmetric under the exchange of 1 and 2.
This is in agreement with the consistency          
of the fusion rules and the braiding.

Using the properties of the $\Gamma $ function,
the factor h can be absorbed partially or completely.
A particular form which exhibits the symmetry between
 $(J_1 ,\Jhat_1)$, $(J_2 ,\Jhat_2)$ and 
$(-J_3 -1 ,-\Jhat_3 -1)$
is
\beq
g^{J_3  \Jhat_3  }_{J_1 \Jhat_1 J_2 \Jhat_2}
=(i/2)^{p + \widehat p }
{H_{p\widehat p}(J_1 ,\Jhat _1)H_{p\widehat p}(J_2 ,\Jhat _2)
H_{p\widehat p}(-J_3 -1,-\Jhat _3 -1)
\over{H_{p\widehat p}(p/2,\widehat p /2)}}
\label{5.14}
\eeq
with
$$
H_{p\widehat p}(J,\Jhat ) =
$$
\beq
\sqrt{\prod_{r=1}^p F(2\Jhat +1 +(2J -r+1)h/\pi )
   \prod_{\widehat r =1}^{\widehat  p}
    F(2J +1 + (2\Jhat - \widehat r +1 )\pi /h )}
\over
{\prod_{r=1}^p \prod_{\widehat r =1 }^{\widehat p }
[(2 J -r +1 )\sqrt{h/\pi} +(2\Jhat -\widehat r +1)\sqrt{\pi /h}]}
\label{5.15}
\eeq
and
$p=1+J_1 +J_2 + (-J_3-1)$
is symmetric.

The $H_{p\widehat p}(J,\Jhat ) $ function can be rewritten in a more
useful form by absorbing all the denominator factors
$$
H_{p\widehat p}(J,\Jhat ) =
e^{i \pi p \widehat p /2}
\prod_{\ell,\widehat \ell  \in \{ \gamma \} }
\left \{ F[R_\ell +(R_{\widehat \ell}+(1-\epsilon_{\widehat \ell})/2) \pi /h ]
(h/\pi)^{R_\ell-2J+p-1}
\right \} ^{\epsilon_{\widehat \ell}/2}
$$
\beq
\left \{ F[R_{\widehat \ell} +(\widehat R_\ell+(1-\epsilon_\ell)/2) \pi /h] 
(h/\pi)^{R_{\widehat \ell}-2\Jhat+\hat p-1} 
\right \} ^{\epsilon_\ell/2}
\label{5.16}
\eeq
where $\gamma$ is an arbitrary path in the plane $( R, \widehat R )$
 going from $( 2J -p + 1 ,2\Jhat - \widehat p + 1 )$
 to $(2J + 1 ,2\Jhat + 1 )$ and

$\widehat \ell$ labels the   elementary path
$( R_\ell ,R _{\widehat \ell}) \to
( R_\ell ,R _{\widehat \ell} + \epsilon_{\widehat \ell} )$

$\ell$   labels the elementary path
$( R_\ell ,R _{\widehat \ell} ) \to
( R_\ell +\epsilon_\ell ,R _{\widehat \ell} )$

\noindent
with $\epsilon_\ell$, $\epsilon_{\widehat \ell}$ = 0, $\pm1$.

This allows to relate two G by a product of elementary shifts 
which have a simple form. From
\beq
{{H_{p\widehat p}(J+1/2,\Jhat )} \over {H_{p\widehat p}(J,\Jhat )}}=
(h/\pi)^{-\widehat p /2}
\sqrt{
{F(2\Jhat +1 +(2J +1)h/\pi )} \over
{F(2\Jhat- \widehat p +1 +(2J -p+1)h/\pi )}}
\label{5.17}
\eeq

\beq
{{H_{p+1 \widehat p}(J,\Jhat )} \over {H_{p\widehat p}(J,\Jhat )}}=
\sqrt{
F(2\Jhat- \widehat p +1 +(2J -p)h/\pi )}
\label{5.18}
\eeq

\beq
{{H_{p+1 \widehat p}(p+1/2,\widehat p )} \over {H_{p\widehat p}(J,\Jhat )}}=
(h/\pi)^{-\widehat p /2}
\sqrt{
F( \widehat p +1 +(p+1)h/\pi )}
\label{5.19} 
\eeq
and related formula for $J \to J-1/2$, $ p \to \widehat p $ we get
$$
{{g^{J_3  \Jhat_3  }_{J_1 +1/2,  \Jhat_1, J_2 -1/2,  \Jhat_2}}
\over
{g^{J_3  \Jhat_3  }_{J_1  \Jhat_1 J_2  \Jhat_2}}}=
$$
\beq
\sqrt{
{F(2\Jhat_1 +1 +(2J_1 +1)h/\pi )F(2\Jhat_2 - \widehat p +1 +(2J_2 -p)h/\pi
)}\over
{F(2\Jhat_2 +1 +2J_2 h/\pi )F(2\Jhat_1 - \widehat p +1 +(2J_1
-p+1 )h/\pi)}},
\label{5.20}
\eeq

$$ 
{{g^{J_3  \Jhat_3  }_{J_1 +1/2,  \Jhat_1, J_2 +1/2,  \Jhat_2}}
\over 
{g^{J_3  \Jhat_3  }_{J_1  \Jhat_1 J_2  \Jhat_2}}}=
{i/ 2} (h\pi )^{-\widehat p /2}
\sqrt
{F(2\Jhat_1 +1 +(2J_1 +1)h/\pi )}\times
$$
\beq
\sqrt{
{F(2\Jhat_2 +1 +(2J_2 +1 )h/\pi )
F(-2\Jhat_3 -\widehat p  -1 -(2J_3 +p+1)h/\pi )}
\over
{F(\widehat p  +1 +(p+1)h/\pi )}}.
\label{5.21}
\eeq

All other relations can be deduced by inversion of these relations
or by the symmetry between $J_i ,h $ and $\Jhat_i ,\widehat h $
or the symmetry between $ J_1,J_2,-J_3-1$ (resp. $\Jhat_i$).

\section{CONCLUSION}
\markboth{6. Conclusion} {6. Conclusion}

As mentioned in the introduction, the earlier works only dealt
with leading-order fusions, explicitly.  In this case, one
only  retains the spin $J_1+J_2$ in fusing
operators with
 spins $J_1$ and $J_2$. We began the present study, when we
realized that,
beyond this approximation,
the OPA is not associative when it is considered too naively, that is
at the level of primaries. This difficulty appears at two levels.
First, in the product of three operators one sees that the
dominant behaviour depends upon the way the three world-sheet points
approach each other. Second, it is possible to define a limit such that
this naive analysis  gives   associativity conditions over
matrix-products of $F$ and $B$ matrices, but these relations are not
satisfied.  This fact is particularly clear for $\xi$ fields.
For them, OPE's at the level of primaries correspond to making
q-tensor product of the representations, and this is notoriously
non-associative. As a matter of fact, the 6-j symbols precisely
encode this non-associativity (see the defining relation Eq.\ref{3.13}),
as it is well known\footnote{this is also true for ordinary groups.}.
Specifically, the naive analysis leads to Eq.\ref{3.12} without
the fusing matrix $F$, or to Eq.\ref{3.13} without the 6-j symbol,
which are clearly wrong relations.
The general conclusion of the present work is that this difficulty is
solved when the primaries are included using the compact formulae of the
MS scheme, where all coefficients of the OPE are  expressed by the
$V$ matrix element in Eqs.\ref{2.52}, or \ref{3.15}. Then the
coupling constant $g_{JK}^L$ may be computed from the MS polynomial
equations, starting from the particular case where one of the operators
has $J=1/2$.

As was pointed out several times\cite{G1,G3,G5},
the $U_q(sl(2))$ quantum-group structure
remarkably comes out of the holomorphic OPA of Liouville theory.
It is interesting to think about a sort of reciprocal. Given the
quantum group quantities -- C.G. coefficients, 6-j symbols, universal
R matrix--, the tensor product of representations defines a
``multiplication'' of representations which is not associative:
this corresponds to the naive analysis mentioned above. Then, one may
consider introducing additional quantum numbers such that the
``product'' becomes  associative. A solution for this is given by the
indices $\{ \nu\}$ that caracterize the descendants, and then the
``product'' becomes the OPE of holomorphic fields we have displayed.
 Indeed, the only way to
have Eq.\ref{3.13} satisfied is that $F$ be equal to a 6-j symbol. This
forces the existence of the $V$'s and fixes their OPE.  Then some of the
$J$'s of the 6-j symbols have a natural interpretation as zero-modes
which are shifted by the $V$ fields as shown on Eq.A.12.
It is likely that this is the unique possibility, although a proof of this
fact would be beyond the scope of this article.
In any case, our present study indicates that the relationship
between $U_q(sl(2))$ and the holomorphic OPA of 2D gravity goes even
more deeply that previously thought.

One may forsee  future developments of the present work in several
directions.  The most interesting one  at this point concerns  the
strong-couplig regime $1<C<25$, with complex  $h$, where the present method
is the only available so far. In ref.\cite{G3} a unitary-truncation theorem
was derived for $C=7$, $13$, and $19$, by only
considering  leading-order
OPE's where it is sufficient to deal with primaries.  This discussion
may now be completed using the result of the present paper.
Moreover, the solutions of the MS relations just displayed should
allow us to  check that the associated non-critical string theories
are consistent  as such, namely,  that they satisfy
duality relations between crossed channels similar to the ones of the
Veneziano model.
Another point is that, now that the full bootstrap is at hand, one may
re-consider,  with a much greater insight,
  the extension to negative $J$, handled in ref.\cite{G3}
by  means of a symmetry principle between spins $J$ and $-J-1$.
Dealing with negative spins is unavoidable  since they are the
holomorphic components of exponentials of Liouville field
 with positive weights. These  must be introduced for proper  dressing by
2D gravity\cite{G5}.

At a more ambitious level, understanding the connection
between quantum group and 2D gravity more deeply is certainly a step toward
unravelling the non-commutative geometry of the latter theory.
\vskip 5mm

\noindent
{\bf Acknowledgements}

At an early stage of this work,
one of us (J.-L.G.) was visiting the
Mathematical Science Research Institute in Berkeley California.
He is grateful for the generous financial support,
the very stimulating surroundings  and the warm hospitality
from which he benefitted while participating
in the program "Strings in Mathematics".

This work was supported in part by the
European Twinning Program,
contrat $\#$540022.

\appendix
\section{APPENDIX A}
\markboth{Appendix}{Appendix}

In this appendix, some basic concepts of the present approach 
are recalled for completeness. The notations are identical 
to the ones used in the articles \cite{G1,G2,G3,G5} to which 
we refer for details. 

   We  recall some  basic properties of the  primary
fields  that came out as  holomorphic components of 
exponentials of the 
Liouville field. 
 Since one deals with functions
of a single  variable $\sigma - i\tau$
(see the beginning of part 2), one may work at $\tau=0$
without loss of generality
that is on the unit circle $u=e^{i\sigma}$.
These holomorphic components may thus be regarded as functions of 
$\sigma$.  
The starting point is that,   
for trivial Verma modules, 
there exist two equivalent free fields:
\beq
\phi_j(\sigma)=q^{(j)}_0+ p^{(j)}_0\sigma+i
\sum_{n\not= 0}e^{-in\sigma}\, p_n^{(j)}\bigl / n,
\quad j=1,\> 2,
\label{A.1}
\eeq
such that
(primes denote derivatives)
\beq
\Bigl [\phi'_1(\sigma_1),\phi'_1(\sigma_2) \Bigr ]=
\Bigl[\phi'_2(\sigma_1),\phi'_2(\sigma_2) \Bigr ]
=2\pi i\,  \delta'(\sigma_1-\sigma_2),
\qquad p_0^{(1)}=-p_0^{(2)},
\label{A.2}
\eeq
\beq
N^{(1)}\bigl (\phi'_1\bigr )^2+ \phi''_1/\sqrt \gamma=
N^{(2)}\bigl ( \phi'_2\bigr)^2+ \phi''_2/\sqrt \gamma.
\label{A.3}
\eeq
$N^{(1)}$ (resp. $N^{(2)}$)  denote  normal orderings
 with respect to the modes of $\phi_1$  (resp. of
$\phi_2$).
 Eq.\ref{A.3} defines the
stress-energy tensor  and the  coupling constant
$\gamma$ of  the quantum theory. The former  generates a
representation of the Virasoro algebra
with central charge $C=3+1/\gamma$. At an intuitive
level, the
correspondence between $\phi_1$ and $\phi_2$ may be
understood
from the fact that the Verma modules, which they  generate,
coincide
since   the highest weights only depend upon
$(p_0^{(1)})^2=(p_0^{(2)})^2$.
The chiral family is built up\cite{GN4,GN5,GN7,G1} from
the
 following operators
\beq
V_j =d_j\, N^{(j)}\bigl (e^{\sqrt{h /2\pi}\>
\phi_j}\bigr
), \quad
\Vhat_j ={\widehat d}_j\,
N^{(j)}\bigl (e^{\sqrt{\hhat /2\pi}\> \phi_j}\bigr ),
\quad
j=1,\>2,
\label{A.4}
\eeq
\beq
h={\pi \over 12}\Bigl(C-13 -
\sqrt {(C-25)(C-1)}\Bigr),\quad
\hhat={\pi \over 12}\Bigl(C-13
+\sqrt {(C-25)(C-1)}\Bigr).
\label{A.5}
\eeq
In Eq.\ref{A.4}, the zero modes are ordered, as is
standard in string vertices, so that the operators are
primary.
Explicitly one has
$$
  N^{(j)}\bigl( e^{\sqrt{h/2\pi}\, \phi_j} \bigr)
\equiv
e^{\sqrt{h/2\pi}\, q_0^{(j)}}\,e^{ \sqrt{h/2\pi}
p_0^{(j)}}
e^{-i\sigma h/4\pi}\times
$$
\beq
 \exp\left (\sqrt{h/2\pi}i\sum_{n<0}e^{-in\sigma}p_n^{(j)}/ n\right
)
\exp\left (\sqrt{h/2\pi}i\sum_{n>0}e^{-in\sigma}p_n^{(j)}/ n\right )
\label{A.6}
\eeq
with similar definitions for the hatted fields.
The relation between  $h$ or $\hhat$ and $C$ which is
equivalent
to
\beq
C=1+6({h\over\pi}+{\pi\over h}+2)=
1+6({\hhat\over\pi}+{\pi\over\hhat}+2),
\quad \hbox{with} \quad h\hhat=\pi^2,
\label{A.7}
\eeq
is such that $V_j$ and $\Vhat_j$ are  solutions of
the
 equations\cite{GN3,GN4}
$$
 -{d^2V_j(\sigma)\over d\sigma^2}+({h\over
\pi})\Bigl(\sum_{n<0}\,L_n\,e^{-in\sigma}
+{L_0\over 2}+({h\over 16\pi}-{C-1\over
24})\Bigr)V_j(\sigma)
$$
\beq
+({h\over
\pi})V_j(\sigma)\Bigl(\sum_{n>0}\,L_n\,e^{-in\sigma}
+{L_0\over 2}\Bigr)=0
\label{A.8}
\eeq
$$ -{d^2\Vhat_j(\sigma)\over d\sigma^2}+({\hhat\over
\pi})
\Bigl(\sum_{n<0}\,L_n\,e^{-in\sigma}
+{L_0\over 2}+({h\over 16\pi}-{C-1\over
24})\Bigr)\Vhat_j(\sigma)
$$
\beq
+({\hhat\over
\pi})\Vhat_j(\sigma)\Bigl(\sum_{n>0}\,L_n\,e^{-in\sigma}
+{L_0\over 2}\Bigr)=0
\label{A.9}
\eeq
These are  operator Schr\"odinger equations equivalent
to the decoupling of Virasoro
null-vectors\cite{GN4,GN5,GN7}.
Since there are two possible quantum modifications $h$
and
$\hhat$,
there are four solutions.
By operator product $V_j$, $j=1$, $2$, and $\Vhat
_j$,
$j=1$, $2$,
generate two infinite families of chiral
fields  $V_m^{(J)}$, $-J\leq m \leq J$,
and
$\Vhat_\mhat^{(\Jhat)}$, $-\Jhat\leq \mhat \leq \Jhat$;
  with $V_{-1/2}^{(1/2)}= V_1$,
$V_{1/2}^{(1/2)}= V_2$, and
$\Vhat_{-1/2}^{(1/2)}=
\Vhat_1$ ,
$\Vhat_{1/2}^{(1/2)}= \Vhat_2$.
The fields
$V_m^{(J)}$, $\Vhat_\mhat^{(\Jhat)}$,  are
 of the type ($1$, $2J+1$) and ( $2\Jhat+1$,$1$),
respectively,
in the BPZ
classification. For the zero-modes, it is
simpler\cite{G1}
to  define  the rescaled variables
\beq
\varpi=i p_0^{(1)} \sqrt{{2\pi\over h }}; \quad
\varpihat=i p_0^{(1)} \sqrt{{2\pi \over \hhat }};
\qquad \varpihat=\varpi\>{h\over \pi};
\quad \varpi=\varpihat \>{\hhat\over \pi}.
\label{A.10}
\eeq
The Hilbert space in which the operators $\psi$ and
$\psihat$
live,
 is a direct sum\cite{G1,G2,G3} of Verma modules 
${\cal H}(\varpi)$.  
They  are eigenstates of the quasi momentum $\varpi$, and
satisfy $L_n\vert \varpi>= 0$, $n>0$;
$(L_0~-~\Delta(\varpi))\vert \varpi>~=~0$.
The corresponding highest weights  $\Delta (\varpi)$
 may be rewritten as
\beq
\Delta(\varpi)\equiv {1\over
8\gamma}+{(p_0^{(1)})^2\over 2}
={h\over 4\pi}(1+{\pi\over h})^2-{h\over 4\pi}\varpi^2.
\label{A.11}
\eeq
The commutation relations Eq.\ref{A.2} are to be supplemented
by the
zero-mode
ones:
$$
 \bigl[ q^{(1)}_0,\, p^{(1)}_0\bigr]=
\bigl[ q^{(2)}_0,\, p^{(2)}_0\bigr]=i.
$$
It thus follows (see in particular Eq.\ref{A.6}), that the
fields $V$ and $\Vhat$  shift the quasi momentum
$p^{(1)}_0=-p^{(2)}_0$ by a fixed amount.
For an arbitrary c-number  function  $f$,  one has
\beq
V_m^{(J)}\>f(\varpi)=f(\varpi+2m)\>V_m^{(J)},\quad
 \Vhat_\mhat^{(\Jhat)}\>f(\varpi)=
f(\varpi+2\mhat\, \pi/h)\>\Vhat_\mhat^{(\Jhat)}.
\label{A.12}
\eeq
The fields $V$ and $\Vhat$ together with
their products may be naturally restricted to discrete
values
of $\varpi$. They thus live    in Hilbert
spaces
  of the form\footnote{
in the present paper,
we only deal with $V$ operators in detail,
so that $\nhat$ is kept equal to zero}
\beq
{\cal H}\equiv
\bigoplus_{n,\nhat=-\infty}^{+\infty}
{\cal H}(\varpi_0+n+\nhat\,\pi/h).
\label{A.13}
\eeq
where ${\cal H}(\varpi_0+n+\nhat\,\pi/h)$
are Verma modules. 
 $\varpi^0$ is a constant. The
$sl(2,C)$--invariant
vacuum corresponds to $\varpi_0=1+\pi/h$, \cite{G1}. With this 
choice, we use the notation ${\cal H}_{J\Jhat}$ 
instead of ${\cal H}(\varpi_0+n+\nhat\,\pi/h)$. For
$h$
real, Eq.\ref{A.13} shows that 
 the eigenvalues of $\varpi$ may be most naturally 
assumed to be 
real. Thus $(p^{(1)}_0)^\dagger =p^{(2)}_0$. As discussed
several times\cite{G1}, the natural hermiticity relation
is
$\bigl(\phi^{(1)}(\sigma)\bigr)^\dagger
=\phi^{(2)}(\sigma)$.
It follows from 
Eq.\ref{A.3}  that this is consistent with the usual
hermiticity relation $L_n^\dagger=L_{-n}$. 
As shown in \cite{G1}, one has
\beq
(\psi_m^{(J)})^\dagger=\psi_{-m}^{(J)}
\label{A.14}
\eeq
 The
most
general
$(2\Jhat+1, 2J+1)$
field $V_{m\,\mhat}^{(J\,\Jhat)}\sim
V_{m}^{(J)}\>\Vhat_{\mhat}^{(\Jhat)}$ has weight
\beq
 \Delta_{J\Jhat}
={C-1\over 24}-
{ 1 \over 24} \left((J+\Jhat+1) \sqrt{C-1}
-(J-\Jhat) \sqrt{C-25} \right)^2
\label{A.15}
\eeq
in agreement with Kac's formula.
It is  sometimes convenient to rewrite the operators 
$V_{M\,\Mhat}^{(J\,\Jhat)}$ as $V^{\mu,\nu;\>\muhat,\nuhat}$
where
\begin{equation}
2J=\mu+\nu,\quad  2m= \nu-\mu,\quad  2\Jhat=\muhat+\nuhat,
 \quad 2\mhat= \nuhat-\muhat.
\label{A.16}
\end{equation}
$V^{\mu,\nu;\>\muhat,\nuhat}(\sigma)$
is an operator whose normalization does not depend upon its
indices: its only non-vanishing matrix element is
\begin{equation}
< \varpi  \vert \,V^{\mu,\nu;\>\muhat,\nuhat}(\sigma)
\vert \varpi-\mu+\nu+(-\muhat+\nuhat)\pi/h >=
e^{i\sigma
(\varpi^2-(\varpi-\mu+\nu+(-\muhat+\nuhat)\pi/h)^2)h/4\pi}\>.
\label{A.17}
\end{equation}
  In Appendix E of  ref.\cite{G1} fields $\psi_{m\> \mhat} ^{(J\>
\Jhat)}$ were defined by the relation 
\begin{equation}
\psi^{(J\>\Jhat)}_{m\>\mhat}(\sigma)\equiv
C^{\mu,\nu;\>\muhat,\nuhat}\>
D^{\mu,\nu;\>\muhat,\nuhat}(\varpi)\>\>
 V^{\mu,\nu;\>\muhat,\nuhat}(\sigma),
\label{A.18}
\end{equation}
In the present article we use a condensed notation, by letting 
\beq
E^{(J\, \Jhat)}_{m\, \mhat}(\varpi) \equiv 
C^{\mu,\nu;\>\muhat,\nuhat}\>
D^{\mu,\nu;\>\muhat,\nuhat}(\varpi)
\label{A.19}
\eeq
where Eq.\ref{A.16} is to be used.  
According to Eqs.A.16 and E.3 of
\cite{G1}, $C^{\mu,\nu;\>\muhat,\nuhat}$ is given
by
\begin{equation}
C^{\mu,\nu;\>\muhat,\nuhat}=
C^{\mu,\nu}\Chat^{\muhat,\nuhat}\>\>
{\prod_{r=1}^{\mu+\nu}\prod_{\rhat=1}^{\muhat+\nuhat}
\bigl(r\sqrt{h/ \pi}+\rhat\sqrt{\pi/h}\bigr) \over
\Bigl [\prod_{s=1}^{\mu}\prod_{\shat=1}^{\muhat}
\bigl(s\sqrt{h/ \pi}+\shat\sqrt{\pi/h}\bigr)\,
\prod_{t=1}^\nu\prod_{\that=1}^\nuhat
\bigl(t\sqrt{h/ \pi}+\that\sqrt{\pi/h}\bigr)\Bigr ]},
\label{A.20} 
\end{equation}
\begin{equation}
 C^{\mu, \nu}=\prod_{r=1}^{\mu+\nu}
 \Gamma\bigl(1+rh/ \pi\bigr)\Bigl /\Bigl [
\prod_{s=1}^{\mu}\Gamma\bigl(1+sh/\pi\bigr)\,
\prod_{t=1}^\nu \Gamma\bigl(1+th/\pi\bigr)\Bigr ],
\label{A.21}
\end{equation}
with similar formulae for $\Chat^{\muhat,\nuhat}$. The $\varpi$
dependent part $D^{\mu,\nu;\>\muhat,\nuhat}(\varpi)$ is
given\footnote{ up to a few misprints}
  by Eqs.A.26a,b and E.4 of \cite{G1}:
\begin{equation}
D^{\mu,\nu;\>\muhat,\nuhat}(\varpi) =
D^{\mu,\nu}(\varpi)\>\Dhat^{\muhat,\nuhat}(\varpi) \>
M^{\mu,\nu;\>\muhat,\nuhat}(\varpi)
\label{A.22}
\end{equation}
\begin{equation}
 D^{\mu, \nu} (\varpi)=
\prod_{t=1}^{\mu-\nu} {\sqrt{\Gamma\bigl(-(\varpi-t+1)h/\pi
\bigr)}\over
\sqrt{\Gamma((\varpi-t) h/ \pi)}}
\prod_{r=1}^\mu \Gamma\bigl((\varpi-r) h/\pi )\bigr)
\prod_{s=1}^\nu \Gamma\bigl((-\varpi-s) h/\pi )\bigr),
\label{A.23}
\end{equation}
if $\mu>\nu$, and
\begin{equation}
 D^{\mu, \nu} (\varpi)=
\prod_{t=1}^{\nu-\mu}
{\sqrt{\Gamma\bigl((\varpi+t-1)h/\pi\bigr)}\over
\sqrt{ \Gamma((-\varpi-t) h/ \pi)}}
\prod_{r=1}^\mu \Gamma\bigl((\varpi-r) h/\pi )\bigr)
\prod_{s=1}^\nu \Gamma\bigl((-\varpi-s) h/\pi )\bigr),
\label{A.24}
\end{equation}
if $\nu>\mu$. Moreover, if
for instance, $\mu>\nu$, $\muhat>\nuhat$,
$$ M^{\mu,\nu;\>\muhat,\nuhat}(\varpi)=
$$
\beq
{{\prod_{t=1}^{\mu-\nu}\prod_{\that=1}^{\muhat-\nuhat}
\sqrt{-(\varpi-t+1)\sqrt{h/ \pi}+(\that-1)\sqrt{\pi/h}}
\sqrt{(\varpi-t)\sqrt{h/ \pi}-\that\sqrt{\pi/h}}}\over
{
\prod_{r=1}^{\mu}\prod_{\rhat=1}^{\muhat}
\bigl((\varpi-r)\sqrt{h/ \pi}-\rhat\sqrt{\pi/h}\bigr)
\>\prod_{s=1}^{\mu}\prod_{\shat=1}^{\nuhat}
\bigl(-(\varpi+s)\sqrt{h/ \pi}-\shat\sqrt{\pi/h}\bigr)}}.
\label{A.25}
\eeq

Next, an other basis of operators,
the $\xi$ fiels were introduced in ref.\cite{G1}
(see part 3 for the details).
In \cite{G1,G3} their braiding algebra
was completely determined. It was shown that it is given by 
the universal R-matrix of $U_q(sl(2))$ (see Eqs.\ref{3.21} -
\ref{3.23}).

\section {APPENDIX B}
In this section we  comment about the $z$-dependence of the 
fusion equations derived in section 2, and connect it   with the 
standard form where translation invariance   is manifest. 
 First consider Eq.\ref{2.15}. 
It follows from Eq.\ref{2.59} hat
the states  $|\varpi_1> $, and
$<\varpi_4 |$ are obtained from the limits
\beq
\lim_{z_1\to 0}   V_{-J_1}^{(J_1 )}(z_1)
\vert \varpi_0 >
= \vert \varpi_1 >, \quad
\lim_{z_4\to \infty }
(-z_4)^{2\Delta_4}
<-\varpi_0 \vert
V_{J_4+1, +1}^{(-J_4-1, -1 )}\! ( z_4)
=< \varpi_4 \vert.
\label{B.1}
\eeq
Thus we have 
$$<\varpi_4 | V_{\pm 1/2}^{(1/2)}(z_3) V_{m_2}^{(J_2)}(z_2)
|\varpi_1>=
$$
\beq
\lim_{z_4\to \infty, z_1\to 0} 
(-z_4)^{2\Delta_4} 
<-\varpi_0 | V_{J_4+1, 1}^{(-J_4-1, -1 )}(z_4) 
\> V_{\pm 1/2}^{(1/2)}(z_3) V_{m_2}^{(J_2)}(z_2) 
 V_{-J_1}^{(J_1 )}(z_1) 
\vert \varpi_0 > .
\label{B.2}
\eeq
Similarly one sees that 
$$<\varpi_1 |   V_{m_4}^{(J_4)}(z_4') V_{\pm 1/2}^{(1/2)}(z_3')  
|\varpi_2>=
$$
\beq
\lim_{z_1'\to \infty, z_2'\to 0}  
(-z_1')^{2\Delta_1}  
<-\varpi_0 | V_{J_1+1, 1}^{(-J_1-1, -1 )}(z_4')  
\> V_{m_4}^{(J_4)}(z_4') V_{\pm 1/2}^{(1/2)}(z_3')
 V_{-J_2}^{(J_2 )}(z_2')  
\vert \varpi_0 >  .
\label{B.3} 
\eeq 
Since $\vert \varpi_0 >  $ and $<-\varpi_0 |$ are 
the right and left $sl(2,C)$-invariant states, respectively, 
it follows that 
any matrix element  of primary fields 
$<-\varpi_0 | A_1 A_2 A_3 A_4 \vert \varpi_0 >  $, is of  the form 
$f/\Pi_{i,j} (z_i-z_j)^{\Delta_i+\Delta_j-\Delta}$, where 
f is invariant\footnote{it   may be regarded as a
function of the anharmonic (cross) ratio of the $z$'s} 
under projective (M\"obius) transformations of the 
$z$'s, and $\Delta= \sum_j \Delta_j/3$.
It is thus clear that the above $z$ and $z'$ variables 
should be related by a projective transformation. 
This is verified as follows. Eq.2.15 is such that   
 $z'_4=z_2$.  Thus the projective transformation 
 is characterized by the condition  $(\infty, z_2, 0) 
\to  (z_2, 0, \infty) $, and  one has
\beq
z'_k=z_2(1-z_2/z_k).  
\label{B.4}
\eeq
This immediately gives $z_3'=z_2(1-z_2/z_3)$ in agreement with
Eq.2.15.  Moreover there  is 
the Jacobian factor 
$J= \prod_k (dz_k'/dz_k)^{\Delta_k}$ which arises  due to the 
transformation law Eq.2.6 of the primary fields.  
Since  $dz'/dz=(z_2/z)^2$, one has  
$$J= 
z_2^{2(\Delta_1+\Delta_3+\Delta_4)} \prod_{l=1,3,4}
z_l^{-2\Delta_l} 
\sim_{z_1\to 0}  
(-z_1')^{2\Delta_1} (z_4)^{-2\Delta_4}
z_2^{2(\Delta_3+\Delta_4-\Delta_1)} \> z_3^{-2\Delta_3}
$$
The first two terms on the right-hand side coincide with 
the factors in Eqs.\ref{B.2}, and \ref{B.3} that respectively
 ensure the
finiteness of the limits   $z_4\to \infty$, and 
$z_1'\to \infty$.  
 The other factors are equal to the  first  two terms 
of the right-hand side  of Eq.\ref{2.15}. 

Finally, we show how invariance under $z$ translations may be
recovered. Assume, as usual that Eq.\ref{2.22} is true for
arbitrary matrix element. Then we may write
$$ V_{\pm 1/2}^{(1/2)}(z_3) V_{m_2}^{(J_2)}(z_2)
 = \left ( {z_2\over z_3}\right ) ^{2\Delta_{1/2}}
\sum_{\epsilon=\pm 1} f_{\pm 1,\, \epsilon}(J_2,m_2;\varpi)
\times
$$
\beq
\sum_{\{\nu\}}
   V_{m_2\pm 1/2}^{(J_\epsilon, \{\nu \})}(z_2)
 <\varpi_\epsilon , \{\nu\}| V_{-\epsilon /2}^{(1/2)}
\left (z_2(1-{z_2\over z_3})\right )
|\varpi_2>,
\label{B.5}
\eeq
where $\varpi$ is the zero-mode  operator. This formula is
invariant under the transformation $z\to \lambda z$ which
preserves the points $0$ and $\infty$ asociated with the 
matrix element we started from. Covariance under $z$ translations
($z\to  z+a$), is achieved by performing the M\"obius
transformation $ z \to \zeta=-z_2 z_3 /z$ on the three operators
$V_{\pm 1/2}^{(1/2)}(z_3)$, $ V_{m_2}^{(J_2)}(z_2)$, 
$V_{m_2\pm 1/2}^{(J_\epsilon, \{\nu \})}(z_2)$, and rescaling the
argument of $V_{-\epsilon /2}^{(1/2)}$ in the matrix element 
so that $z_2(1-z_2/ z_3)\to z_3-z_2$. 
 Taking account of the 
transformation factor of Eq.\ref{2.6}, one finds 
$$ V_{\pm 1/2}^{(1/2)}(-z_2) V_{m_2}^{(J_2)}(-z_3) 
 =  
\sum_{\epsilon=\pm 1} f_{\pm 1,\, \epsilon}(J_2,m_2;\varpi) 
\times   
$$ 
\beq 
\sum_{\{\nu\}}   
   V_{m_2\pm 1/2}^{(J_\epsilon, \{\nu \})}(-z_3) 
 <\varpi_\epsilon , \{\nu\}| V_{-\epsilon /2}^{(1/2)} 
\left (z_3- z_2)\right ) 
|\varpi_2>. 
\label{B.6} 
\eeq 
Up to a change of sign in the definition of the $z$ variable,
this takes  the usual form.

\section {APPENDIX C}
This section is devoted to the explicit derivation of 
Eq. \ref{3.8}, that is 
$$\sum_{\eta,\, \mu,  (\eta+\mu=n)} |1/2,\varpi)_\alpha^\eta | J,\,
\varpi+2\eta)_M^\mu\>
N\!\!\left | ^{1/2}_\eta \,  ^{ J;}_{\mu ;}\, ^{J+\epsilon/2}_n  
;\varpi \right |= 
$$
\beq
(1/2,\alpha;J,M|J+\epsilon/2)\>  | J+\epsilon/2,\,
\varpi)_{M+\alpha}^n,
\label{C.1}
\eeq
The case $\epsilon=1$ is already discussed in ref.\cite{G1} where
the general fusion coefficients to leading order  were determined.
We shall, nevertheless rediscuss this case first for completeness,
and since the case $\epsilon=-1$ is exactly similar.  Letting, 
$n=m+1/2$, one sees that 
we have to deal with the expression 
\beq 
|1/2,\varpi)_\alpha^{-1/2} | J,\, \varpi-1)_M^m \> 
{\lfloor \varpi+ J+m\rfloor \over \lfloor \varpi \rfloor}
+|1/2,\varpi)_\alpha^{1/2} | J,\, \varpi+1)_M^{m-1} \>  
{\lfloor \varpi- J+m-1\rfloor \over \lfloor \varpi \rfloor}.
\label{C.2}
\eeq
The first thing is to get rid of the $\lfloor \varpi \rfloor$ 
denominators. This is done by using the following recurrence
relation satisfied by the coefficients $| J,\, \varpi)_M^m$:
\beq  
| J,\, \varpi+1)_M^{m-1}= e^{-ih/2} 
{\lfloor \varpi+ J+m\rfloor \over \lfloor J-m+1 \rfloor} 
| J,\, \varpi-1)_M^m- e^{ih(M-1/2)} 
{ \lfloor \varpi \rfloor \over \lfloor J-m+1 \rfloor}
| J,\, \varpi)_M^m.
\label{C.3}
\eeq
This relation was established in ref.\cite{G1} in order to derive  the
braiding matrix of the $\xi$ fields (Eqs.A.35-39) from the 
one of the  $\psi$ fields.

Next, we proceed by  only considering  Eq.\ref{C.2} for $\alpha=-1/2$,
since this other value is treated in exactly the same way.
Let us  transform this relation  by 
using the explicit expressions for $|1/2,\varpi)_\alpha^{\pm 1/2}$, 
that is,  
\beq
|1/2,\varpi)_{\pm 1/2}^{\pm 1/2} =e^{ih(\varpi\pm 1/2)},\quad
|1/2,\varpi)_{\pm 1/2}^{\mp 1/2} =e^{-ih\varpi /2}. 
\label{C.6}
\eeq
One gets
$${e^{-ih(\varpi+1)/2}\over \lfloor J-m+1\rfloor}
\left\{  e^{ih(J-m+1)}\lfloor \varpi+J+m\rfloor 
| J,\, \varpi-1)_M^m \right. 
$$
$$  \left. - e^{ihM} 
\lfloor \varpi-J+m-1\rfloor | J,\, \varpi)_M^m \right \}.
$$
Eq.A.28 shows that 
the coefficients $| J,\, \varpi)_M^m$ are of the 
form $\sum_t e^{iht(\varpi+m)} A_M^t(J,m)$. 
 Accordingly, the 
Fourier expansion  of the last equation is 
$$ {e^{-ih(\varpi+1)/2}\over \lfloor J-m+1\rfloor}
\sum_t e^{iht(\varpi+m)} e^{ih(M+J+1-t-m)/2} 
\left \{ \left \lfloor {3+3J-M-t-m\over 2}\right \rfloor 
A_M^{t-1}(J,m)\right.  
$$
$$ \left.  +
\left \lfloor {1+J+M+t+m\over 2}\right \rfloor 
A_M^{t+1}(J,m)\right\}.  
$$
Using the explicit expression of $A_M^t(J,m)$, one sees that this
last expression  is equivalent to 
$$e^{ih(-\varpi+M+J-m+1/2)/2} \sqrt{ \lfloor J-M+1\rfloor \over 
\lfloor 2J+1\rfloor} \times 
\sum_t e^{iht(\varpi+m-1/2)} A_{M-1/2}^{t-1/2}(J+1/2,m-1/2). 
$$
Thus we finally get 
$$|1/2,\varpi)_{-1/2}^{-1/2} | J,\, \varpi-1)_M^m \>
{\lfloor \varpi+ J+m\rfloor \over \lfloor \varpi \rfloor}
+|1/2,\varpi)_{-1/2}^{1/2} | J,\, \varpi+1)_M^{m-1} \>
{\lfloor \varpi- J+m-1\rfloor \over \lfloor \varpi \rfloor}
$$
\beq
= \sqrt{ \lfloor J-M+1\rfloor \over 
\lfloor 2J+1\rfloor} e^{ih(J+M)/2} 
| J+1/2,\, \varpi)_{M-1/2}^{m-1/2}. 
\label{C.7}
\eeq
This is the desired relation since (see, e.g. ref.\cite{G1})  
$$(1/2,-1/2 ;J,M|J+1/2)= e^{ih(J+M)/2}  
\sqrt{ \lfloor J-M+1\rfloor \over 
\lfloor 2J+1\rfloor}. 
$$
Let us next turn to the case $\epsilon=-1$. Instead of
Eq.\ref{C.2}, 
one has to deal with
$${i\over 2 \sin (h) \sqrt {\lfloor 2J\rfloor \lfloor 2J+1\rfloor}} 
\left \{ 
|1/2,\varpi)_\alpha^{-1/2} | J,\, \varpi-1)_M^m \>
{\lfloor  J+m\rfloor \over 
\lfloor \varpi \rfloor}\right. 
$$
\beq
\left. +|1/2,\varpi)_\alpha^{1/2} | J,\, \varpi+1)_M^{m-1} \>
{\lfloor - J+m-1\rfloor \over \lfloor \varpi \rfloor} 
\right \}.
\label{C.8}
\eeq
Again, we consider the case $\alpha=-1/2$, and make use of
Eq.\ref{C.3}. The last expression becomes
$${-i e^{-ih(\varpi+1)/2}\over 
2\sin (h) \sqrt {\lfloor 2J\rfloor \lfloor 2J+1\rfloor}}
\left\{  e^{-ih(J+m)}
| J,\, \varpi-1)_M^m 
   - e^{ihM}
 | J,\, \varpi)_M^m \right \}.
$$
Its Fourier expansion is found to be
$$-{e^{-ih(\varpi+1)/2} 
 \over \sqrt {\lfloor 2J\rfloor \lfloor 2J+1\rfloor}} 
\sum_t e^{iht(\varpi+m)} e^{ih(M-J-t-m)/2}
 \left \lfloor {J+M+t+m\over 2}\right \rfloor
A_M^{t}(J,m). 
$$
Using the explicit expression for $A_M^{t}(J,m)$, one verifies
finally that this leads to 
$$ {i\over 2 \sin (h) \sqrt {\lfloor 2J\rfloor \lfloor
2J+1\rfloor}} 
\left \{ 
|1/2,\varpi)_\alpha^{-1/2} | J,\, \varpi-1)_M^m \> 
{\lfloor  J+m\rfloor \over 
\lfloor \varpi \rfloor}\right.   
$$ 
$$\left. +|1/2,\varpi)_\alpha^{1/2} | J,\, \varpi+1)_M^{m-1} \>
{\lfloor - J+m-1\rfloor \over \lfloor \varpi \rfloor}
\right \}
$$
\beq = -e^{ih(-J+M-1/2)/2} 
\sqrt{\lfloor J+M\rfloor \over \lfloor 2J+1\rfloor } 
| J-1/2,\, \varpi)_{M+1/2}^{m-1/2}.
\label{C.9}
\eeq
This is the desired result since 
$$ (1/2,-1/2 ;J,M|J-1/2)= -e^{ih(-J+M-1)/2}
\sqrt{\lfloor J+M\rfloor \over \lfloor 2J+1\rfloor }
$$

One may check the case $\alpha=1/2$ in the same way, using the fact
that 
$$(1/2,1/2 ;J,M|J+1/2)= e^{-ih(J-M)/2}
\sqrt{ \lfloor J+M+1\rfloor \over
\lfloor 2J+1\rfloor}
$$
\beq 
(1/2,1/2 ;J,M|J-1/2)= e^{ih(J+M+1)/2}
\sqrt{\lfloor J-M\rfloor \over \lfloor 2J+1\rfloor }
\label{C.10}
\eeq

\typeout{*****************************************}
\typeout{If the figures do not print,   
 create first a postscript file }
\typeout{by using something like dvips -f cgr.dvi > cgr.ps.}  
\typeout{If nothing works, ask for a hard copy by e-mail to }
\typeout{ gervais@physique.ens.fr. Good luck !} 
\typeout{*****************************************}

\begin{thebibliography}{99}
\markboth{ References}{ References}

\bibitem{V} E. Verlinde, \np B300, 1988, 360.

\bibitem{TK} A. Tsuchia, Y. Kanie,  
{Advanced Study in Pure Mathematics}, {\bf 16}, 
(1987), 297.

\bibitem{K} T. Khono, {\sl Ann. Inst. Fourier} (Grenoble) 
{\bf 37, N:4}, (1987), 139.

\bibitem{F} J. Fr\"ohlich, 1987  Carg\`ese Lectures, 
in {\sl Non-Perturbative Quantum Field Theory}, 
G. 't Hooft editor, Plenum Press (1988).
 
\bibitem{FFK} G. Felder, J. Fr\"ohlich, J. Keller,  
\cmp 124, 1989, 646.

\bibitem{FK} J. Fr\"ohlich, C. King, {\sl Int. J. Mod. Phys.
} {\bf A4} (1989) 5321.

\bibitem{R} K.H. Rehren, \cmp 116, 1988, 675. 

\bibitem{RS} K.H. Rehren, B. Schroer, \np B312, 1989, 715.

\bibitem{GN4} J.-L. Gervais, A. Neveu,
\np B238, 1984, 125, \hfill \break 
\np B238, 1984, 396.

\bibitem{MS} G. Moore, N. Seiberg, \cmp 123, 1989, 77.

\bibitem{MR} G. Moore, N. Reshetikhin, \np B238, 1989, 557.

\bibitem{P} V. Pasquier, \cmp 118, 1988, 355.

\bibitem{B} O. Babelon,
\pl   B215,  1988, 523.


\bibitem{AGS} L. Alvarez-Gaum\'e, C. Gomez, G. Sierra, 
\np B319, 1989. 155 , \pl B220, 1989, 142, \np B[FS], 1990, 347.

\bibitem{W} E. Witten, \np B330, 1990, 205.

\bibitem{T} I. Todorov, {\sl Proceedings of the 
VIII Clausthal-Zellerfeld Workshop on Mathematical 
Physics} Springer Lecture Notes, Springer; 
L. Hadjivanov, R. Paunov, I. Todorov, \np B356, 1991, 387.

\bibitem{FGP} P. Furlan, A. Ganchev, V. Petkova, 
\np B343, 1990, 205.

\bibitem{Ga} K. Gaw\c edski, \np (Proc. Suppl.)18B,
1990, 78.

\bibitem{G1} J.-L. Gervais,  \cmp 130, 1990, 257.


\bibitem{G3} J.-L. Gervais, \cmp 138, 1991, 301.

\bibitem{GN3} J.-L. Gervais, A. Neveu,  \np B224,
1983, 329.

\bibitem{GN5} J.-L. Gervais, A. Neveu,  \np B257[FS14],
1985, 59.

\bibitem{GN7} J.-L. Gervais, A. Neveu, \np B264, 1986, 557.

\bibitem{GoS} C. Gomez, G. Sierra, \np B352, 1991, 791.

\bibitem{KR} A. Kirilov, N. Reshetikhin, {\sl Infinite Dimensional
Lie Algebras and Groups, Advanced Study in
Mathematical Physics} {\bf vol. 7},
 Prooceedings of the 1988 Marseille
Conference, V. Kac editor, p. 285, World scientific.


\bibitem{CG} E. Cremmer, J.-L. Gervais,
 \cmp 144, 1992, 279.


\bibitem{G5} J.-L. Gervais, ``Quantum group derivation
of 2D gravity-matter coupling'' Invited talk at
the Stony Brook meeting {\sl String and Symmetry 1991}.
``Gravity-Matter couplings from
Liouville Theory'',  preprint LPTENS 91-22,
hep-th/9205034, Nucl. Phys. to be published.

\bibitem{G2} J.-L. Gervais, \pl B243, 1990, 85.
 

\bibitem{DisGr} G. Ponzano, T. Regge, {\sl in} 
``Spectroscopic and group theoretical 
methods in Physics'', F. Bloch ed., 
North Holland, Amesterdam, 1968. 
B. Hasslacher, M. Perry, \pl 103B, 1981, 21,
S. M. Lewis, \pl 122B, 1983, 265. 
 


\bibitem{B2} O. Babelon,
\cmp 139, 1991, 619.

\bibitem{GS} J.-L. Ger\-vais, B. Sakita, \np B34,
1971, 477.

\bibitem{HHM} Bo-yu Hou, Bo-yuan Hou, 
Zhong-qi Ma, {\sl Comm. Theor. Phys.}, {\bf 13},  181, 
(1990); {\sl Comm. Theor. Phys.}, {\bf 13}, 1990, 341.


\end{thebibliography}
\end{document}